\documentclass{article}
\usepackage{amsfonts}

\title{Affine symmetry in mechanics of collective and
internal modes. Part I. Classical models}

\author{ J. J. S{\l}awianowski, V. Kovalchuk, A. S\l awianowska,\\
B. Go\l ubowska, A. Martens, E. E. Ro\.zko, Z. J. Zawistowski\\
Institute of Fundamental Technological Research,\\
Polish Academy of Sciences,\\
21 \'{S}wi\c{e}tokrzyska str., 00-049 Warsaw, Poland\\
e-mails: jslawian@ippt.gov.pl, vkoval@ippt.gov.pl,\\
aslawian@ippt.gov.pl, bgolub@ippt.gov.pl,\\ 
amartens@ippt.gov.pl, erozko@ippt.gov.pl, \\
zzawist@ippt.gov.pl }

\begin{document}

\maketitle
\begin{abstract}

Discussed is a model of collective and internal degrees of freedom
with kinematics based on affine group and its subgroups. The main
novelty in comparison with the previous attempts of this kind is
that it is not only kinematics but also dynamics that is
affinely-invariant. The relationship with the dynamics of
integrable one-dimensional lattices is discussed. It is shown that
affinely-invariant geodetic models may encode the dynamics of
something like elastic vibrations.

\end{abstract}

\noindent {\bf Keywords:} collective modes, affine invariance,
integrable lattices, nonlinear elasticity.

\section*{Introduction}
In some of our earlier papers including rather old ones
\cite{Frac-Ser-JJS91,Gol02,Gol03,Gol04,Gol04_2,Mart02,Mart03,Mart04,AKS-JJS91,JJS73_2,JJS75_1,JJS75_2,JJS82_2,JJS82_1,JJS86,JJS88_1,JJS02_2,JJS04,JJS04_2,JJS-VK03,JJS-VK04_2}
we have discussed the concept of affinely-rigid body, i.e.,
continuous, discrete, or simply finite system of material points
subject to such constraints that all affine relations between its
elements are frozen during any admissible motion. For example, all
material straight lines remain straight lines in the course of
evolution, and their parallelism is also a constant, non-violated
property. Unlike this, the metrical features, like distances and
angles, need not be preserved. In other words, such a body is
restricted in its behaviour to rigid translations, rigid
rotations, and homogeneous deformations. Models of this kind may
be successfully applied in a very wide spectrum of physical
problems like nuclear dynamics \cite{Bohr-Mot75} (droplet model of
the atomic nuclei), molecular vibrations, macroscopic elasticity
\cite{Erin62,Erin68,AKS-JJS91,JJS73_2,JJS75_1,JJS75_2,JJS82_2,JJS82_1,JJS86,JJS88_1,JJS02_2,JJS04,JJS04_2}
(in situations when the length of excited waves is comparable with
the size of the body), in the theory of microstructured bodies
\cite{JJS82_1} (micromorphic continua), in geophysics
\cite{Bog85,Chan69} (the theory of the shape of Earth), and even
in large-scale astrophysics (vibrating stars, vibrating
concentrations of the cosmic substratum, like galaxies or
concentrations of the interstellar dust).

From the purely mathematical point of view such a model provides
an interesting example of a system with the group-theoretical
background of the geometry of degrees of freedom
\cite{Arn78,Her68,Her70,JJS73_1,JJS82_1,JJS88_1,JJS91_3}. It is an
affine generalization of the usual rigid top with the orthogonal
group replaced by the linear one (isometries replaced by affine
transformations). Let us remind also that there is an interesting
formulation of the general non-constrained continuum mechanics
based on the infinite-dimensional "Lie group" of all
diffeomorphisms or volume-preserving diffeomorphisms
\cite{Abr-Mars78,Arn78,Binz91,Ebin77,Ebin_Mars70,Mars81,Mars-Hugh83}.
This theory is rather complicated (although heuristically very
fruitful) because of serious mathematical problems with
infinite-dimensional groups. The mechanics of an affinely-rigid
body is a simple compromise between rigid-body mechanics and such
a continuum theory, because admitting deformative degrees of
freedom it is simultaneously based on the finite-dimensional
framework.

Let us stress, however, that, in spite of its non-questionable
physical applicability and formally interesting features, the
referred mechanics of affinely-rigid body is in a sense
disappointing from the point of view of the mathematical theory of
Lie group motivated systems. The point is that in the latter
theory it is not only kinematics (finally, geometry of degrees of
freedom) but also dynamics that is ruled by the underlying group.
Due to the isotropy of the physical space, Lagrangian of a free
rigid top, i.e., its kinetic energy, is invariant under all left
regular translations (all spatial rotations); the same is valid,
of course, for the resulting equations of motion (Euler
equations). If the material structure of the top is isotropic
(spherical inertial tensor), then the model is also invariant
under right regular translations. When formulating the theory of
ideal incompressible fluids in terms of the group of all
volume-preserving diffeomorphisms, one obtains an
infinite-dimensional Hamiltonian system invariant under all right
regular translations. This is due to the fact that in the usual
Euler description of fluid its Lagrangian coordinates are not very
essential, and the fluid particles have a rather limited
individuality. Summarizing, in these theories one deals with
Lagrangians or Hamiltonians based on left-, right-, or even
two-side invariant metric tensors on the Lie group used as a
configuration space. It is never the case in the above-quoted
model of affinely-rigid body. This brings about the question as to
the hypothetic affine counterpart of left- and right-invariant
geodetic models on the orthogonal group and their potential
perturbations. This interest is at least academically motivated.
But at the same time, from the physical point of view, such models
look rather esoteric. In any case, the previously mentioned
applications of affine collective modes are dynamically
well-established, because they are based on the d'Alembert
principle in theory of constrained systems. There are, however,
some indications that physical applicability is not a priori
excluded.

This problem has to do with the very philosophy of the origin of
collective and internal degrees of freedom. We say that a "large"
system of material points (continuous, denumerable, or just finite
admitted) has collective modes when there exists a "small" number
of parameters $q^{1},\ldots,q^{n}$ that are dynamically relevant,
i.e., satisfy an approximately autonomous system of evolution
equations, if for our purposes the kinematical information about
the system, encoded in them, is sufficient, and (very important!)
if they depend on individual particles in a non-local way. The
latter means that positions and velocities of all particles enter
the $q^{i}$-variables on essentially equal footing, with the same
strength, order of magnitude, so to speak. This is, of course, a
rough, qualitative introduction of the term, but there is no place
here to develop a rigorous mathematical description. As a
mathematical model we can realize some quotient manifolds of
multiparticle state spaces or their submanifolds (e.g.,
representatives of cosets). On the contrary, internal degrees of
freedom are described in terms of fibre bundles over the physical
space, space-time, or the configuration space. They give an
account of phenomena which are either essentially non-extended in
space, or perhaps cannot be described in terms of composed systems
because their spatial details are unapproachable to our
experimental abilities. For example, from the point of view of
contemporary science, spin systems seem to be based on essentially
internal quantities \cite{Lan-Lif58,Trim97}. In any case, spin
media do not look like the Cosserat continuous limit of discrete
systems of molecular "gyroscopes". The latter model works
successfully in the theory of Van der Waals crystals and granular
media.

Apparently, the most natural and intuitive origin of collective
modes, e.g., of some microstructure variables, is based on the
mechanism of constraints and the d'Alembert principle. Collective
motion is then "large", whereas non-collective one is "small" and
merely reduced to some vibrations about the appropriate constraint
submanifold. The collective kinetic energy, i.e., dynamical metric
element, is obtained from the restriction of the total one to the
constraints surface (the first fundamental quadratic form). This
corresponds to the classical relationship between kinetic energy
and inertia \cite{Arn78,Cap89,Cap00,Syn60}. In this case, as a
rule, the collective kinetic energy is invariant under a proper
subgroup of a group underlying geometry of the constraints
submanifold. But one can also realize another mechanism, namely,
such one that the hidden non-collective motion is just large, and
that the emerging collective modes have to do with the averaged
behaviour of hidden modes, i.e., with the time dependence of some
relatively slowly-varying mean values. Then it is quite natural to
expect that the collective Lagrangian will be based on a kinetic
energy whose underlying dynamical metric tensor will be
non-interpretable in terms of the restriction of the usual
multi-particle metric tensor of the kinetic energy to the
constraints manifold (i.e., to the first fundamental form of
constraints). Similarly, equations of motion need not be derivable
from the usual d'Alembert principle based on the original spatial
metric. Therefore, the relationship between kinetic energy and
inertia may become rather non-classical, to some extent exotic in
comparison with the usual requirements (cf., e.g., the discussion
by Capriz and Trimarco \cite{Cap89,Cap00,Trim97}). In such
situations the only reasonable procedure is to postulate the
kinetic term of the Lagrangian on the basis of some natural and
physically justified postulates. Let us mention two examples from
the two completely opposite scales of the physical phenomena,
namely, the atomic nuclei and vibrating-rotating stars (by the
way, the neutron stars are in a sense exotic and gigantic nuclei
with $Z=0$ and enormous $A$). As objects more close to the Earth
one can think, e.g., kinetic bodies as discussed by Capriz, and
various non-standard microstructure elements like gas bubbles,
voids, and defects in solids \cite{Cap89,Cap00,Kos-Zor67}. Though
bubbles and voids can be hardly treated as constrained pieces of a
substance or systems of material points.

Situation is even much more complicated, when one deals with
essentially internal degrees of freedom, like, e.g., spin systems
\cite{Trim97}. Then, although we have some guiding hints from the
theory of extended systems, any choice of Lagrangian, Hamiltonian,
or equations of motion is based on some rather hypothetic
postulates, first of all, on certain invariance requirements.

There is also another point worth of mentioning. Namely, usually
in variational theories of analytical mechanics, Lagrangian
consists of the kinetic and potential parts. The first one has to
do with inertia, constraints, metric structure, whereas the other
one describes true interactions. But even in traditional problems
of analytical mechanics there are approaches where the structure
of interactions is encoded in an appropriate metric structure,
i.e., in a kind of kinetic term. There is a well-known example,
namely, the Jacobi-Maupertuis variational principle. If $ds$ is
the usual metric (arc) element of the configuration space, and $V$
is the potential energy, then one uses a modified metric
\cite{Arn78}
\[
d\sigma_{V}=\sqrt{E-V}ds,
\]
where $E$ denotes a fixed energy value. This is so-called
isoenergetic dynamics, based on the homogeneous "Lagrangian"
\[
\mathcal{L}=\sqrt{E-V}\sqrt{g_{ij}\frac{dq^{i}}{d\lambda}
\frac{dq^{j}}{d\lambda}},
\]
$\lambda$ denoting an arbitrary parameter (not time). This
variational principle, based on the metric element $d\sigma_{V}$,
gives trajectories with the energy value $E$, but without the
time-dependence. There are also spatiotemporal forms of this
principle, where the time variable occurs as one of coordinates
$q^{i}$, and there is no restriction to the fixed energy value.

In a slightly different context, in certain problems we will
follow the idea of encoding the interaction structure in an
appropriately postulated kinetic energy form, i.e., metric tensor
on the configuration space.

As mentioned, we concentrate below on models with kinematics (and
dynamics) ruled by the linear group GL$(n,\mathbb{R})$, or, more
rigorously, affine group GAf$(n,\mathbb{R})$ (physically $n=2,3$).
Of course, the usual rigid body in $n$ dimensions is ruled by
SO$(n,\mathbb{R})$, or, if translations are taken into account, by
the isometry group E$(n,\mathbb{R})=$
SO$(n,\mathbb{R})\times_{s}\mathbb{R}^{n}$. But there are also
other possibilities of finite-dimensional collective modes, e.g.,
SL$(n,\mathbb{R})$ (or
SL$(n,\mathbb{R})\times_{s}\mathbb{R}^{n}$), i.e., incompressible
affinely-rigid body, or, just conversely, the Weyl group
$\mathbb{R}^{+}$SO$(n,\mathbb{R})$ generated by rotations and
translations (the shape of the body is preserved, but not
necessarily its size). In some future we are going to investigate
systems ruled by the projective group in $n$ dimensions,
Pr$(n,\mathbb{R})\simeq$ SL$(n+1,\mathbb{R})$, cf., e.g.,
\cite{JJS-VK04}. This is quite a natural extension of
affinely-rigid body, when the system of material straight-lines is
preserved but their parallelism may be violated. Another
interesting model would be given by the Euclidean-conformal group
CO$(n,\mathbb{R})$. Let us mention that there was also some very
interesting attempt by unjustly forgotten German physicist
Westpfahl \cite{West67}, who invented the idea of using the
unitary group U$(3)$ as a basis for collective modes in three
dimensions, quite independently of later applications of unitary
symmetry in elementary particle physics.

Finally, it is quite often so that the complexification idea leads
to physically interesting results. It is not excluded that
complexifying the physical space $\mathbb{R}^{n}$ to
$\mathbb{C}^{n}$ and replacing the real groups GL$(n,\mathbb{R})$,
U$(n)$ by GL$(n,\mathbb{C})$ we could obtain some interesting
models of collective or internal degrees of freedom
\cite{Woj86,Zhel78,Zhel83}. The idea is particularly tempting,
because GL$(n,\mathbb{R})$ and U$(n)$ are two different (and
qualitatively opposite) real forms of the same complex group
GL$(n,\mathbb{C})$. But, of course, such exotic ideas are rather
far from realization and they are mentioned here only because of
their obvious conceptual relationship with the usual and
generalized models of affinely-rigid bodies.

The group space is a particular model of systems with kinematics
and dynamics ruled by a Lie group. In general, the microstructure
or collective configuration space (the manifold $\mathcal{M}$ in
the sense of Capriz book \cite{Cap89} and related papers) is a
homogeneous quotient space $G/H$. Here $G$ is a fundamental group
of the model, and $H$ is an appropriate subgroup of $G$, not
necessarily normal one, thus $G/H$ need not inherit the group
structure from $G$ \cite{Cap89,Cap00}.

\section{Dynamical systems based on Lie groups}

Dynamical systems based on Lie groups and their homogeneous spaces
are widely used as a model of internal and collective degrees of
freedom \cite{Cap89,Cap00,Cap-Mar(eds)03,Cap-Mar03,Mar01,Mar03}.
They present also interest by themselves from the purely
mathematical point of view. They are realistic and quite often
they possess rigorous analytical solutions in terms of special
functions and power series; this is probably due to the analytical
structure of Lie groups. The first step of analysis is the theory
of left- and right-invariant geodetic systems, when the Lagrangian
and total energy are identical with the kinetic energy expression
based on an appropriate Riemannian structure of $G$.

For simplicity let us use the language of linear groups; by the
way, nonlinear groups are exceptional in applications, and the
most known examples are the universal covering groups
$\overline{{\rm GL}(n,\mathbb{R})}$, $\overline{{\rm
SL}(n,\mathbb{R})}$ of the indicated linear groups. For any curve
$\mathbb{R}\ni t\mapsto g(t)\in G$ its tangent vectors
$\dot{g}(t)\in T_{g(t)}G$ may be transported to the Lie algebra
$G^{\prime}=T_{e}G$ with the help of right or left
$g(t)^{-1}$-translations, resulting in quantities
$\Omega(t):=\dot{g}(t)g(t)^{-1}$,
$\hat{\Omega}(t):=g(t)^{-1}\dot{g}(t)$. In this way the tangent
and cotangent bundles $TG$, $T^{\ast}G$ may be, in two canonical
ways, identified with the Cartesian products: $TG\simeq G\times
G^{\prime}$, $T^{\ast}G \simeq G \times G^{\prime\ast}$. It is
clear that the left and right regular translations $g\mapsto
L_{k}(g)=kg$, $g\mapsto R_{k}(g)=gk$ transform quasi-velocities
either according to the adjoint rule or trivially:
\begin{eqnarray}
L_{k}&:&\quad \Omega\mapsto {\rm Ad}_{k}\Omega=k\Omega k^{-1},
\qquad
\hat{\Omega}\mapsto \hat{\Omega},\nonumber\\
R_{k}&:&\quad \Omega\mapsto \Omega,\quad \hat{\Omega}\mapsto {\rm
Ad}_{k^{-1}}\hat{\Omega}=k^{-1}\hat{\Omega}k.\nonumber
\end{eqnarray}

Left-invariant geodetic systems on $G$ are based on kinetic
energies, which are quad\-ratic forms of $\hat{\Omega}$ with
constant coefficients. If $G$ is non-Abelian, then $\hat{\Omega}$
is a non-holonomic quasi-velocity and the corresponding Riemannian
structure on $G$ is curved. Similarly, right-invariant kinetic
energies are quadratic forms of $\Omega$ with constant
coefficients. As a canonical example of left-invariant systems we
can realize the free rigid body in $n$ dimensions, $G=$
SO$(n,\mathbb{R})$ (if we neglect translational motion). If the
rigid body is spherical (its inertial tensor is completely
degenerate), then $T$ is also right-invariant, and the underlying
metric tensor on $G$ is proportional to the Killing tensor. Such a
pattern may be followed in all semisimple Lie groups
\cite{Arn78,Mars-Rat94,Mars-Rat99,Rat82}. Quite a different
example is provided by the theory of the ideal fluids
\cite{Arn78}. The configuration space is identified with SDiff
$\mathbb{R}^{3}$
--- the infinite-dimensional group of all volume-preserving diffeomorphisms of
$\mathbb{R}^{3}$ (provided that we discuss the physical
three-dimensional case). If we admitted the fluid to be
compressible, we would have to use the full group Diff
$\mathbb{R}^{3}$ of all diffeomorphisms. The functional of kinetic
energy is invariant under right regular translations in SDiff
$\mathbb{R}^{3}$. What concerns left regular translations, it is
invariant only under the six-dimensional isometry group  of
$\mathbb{R}^{3}$. The reason for this relatively poor
left-hand-side invariance is that the kinetic energy expression
depends in an essential way on the spatial metric tensor. At the
same time, from the point of view of the material space, the
particles of fluid have a rather limited individuality, and that
is why the kinetic energy form of incompressible fluid is
invariant under the huge group of sufficiently smooth and
volume-preserving "permutations" of particles, i.e., under SDiff
$\mathbb{R}^{3}$.

In some of our earlier papers
\cite{Gol01,Gol02,Gol03,Gol04,Gol04_2,Mart02,Mart03,Mart04,AKS-JJS91,JJS73_2,JJS75_1,JJS75_2,JJS82_2,JJS82_1,JJS86,JJS88_1,JJS02_2,JJS04,JJS04_2,JJS-VK03,JJS-VK04_2}
we discussed the object called "affinely-rigid body", i.e., the
system of material points "rigid" in the sense of affine geometry,
i.e., all affine relationship between constituents being kept
fixed during any admissible motion. Such a model is geometrically
interesting in itself and has a wide range of applications in
macroscopic elasticity, mechanics of micromorphic continua with
internal degrees of freedom, molecular vibrations, nuclear
dynamics, vibrations of astrophysical objects, and the theory of
the shape of Earth \cite{Bog85,Chan69}. Analytically, the
configuration space of $n$-dimensional affinely-rigid body may be
identified with the semi-direct product
GL$(n,\mathbb{R})\times_{s}\mathbb{R}^{n}$, or simply
GL$(n,\mathbb{R})$ when we neglect translational degrees of
freedom.

The kinetic energy of an extended affinely-rigid body in Euclidean
space may be calculated in the usual way, by summation of kinetic
energies of its constituents. Velocity vectors are squared with
the use of the fixed metric tensor of the physical space. The
resulting metric tensor of the configuration space is flat, and it
is not invariant either under left or right regular translations,
except two subgroups isomorphic with the Euclidean group
SO$(n,\mathbb{R})\times_{s}\mathbb{R}^{n}$. Because of this the
resulting geodetic model, although kinematically based on the
group manifold, dynamically is incompatible with it. Besides, it
is physically non-realistic and useless, because geodetics are
straight lines in L$(n,\mathbb{R})\times_{s}\mathbb{R}^{n}$,
therefore, in certain directions the body would suffer a
non-limited extension or squeezing. It is impossible to avoid such
non-physical catastrophic phenomena without introducing some
potential term.

The very taste and mathematical machinery of systems with
group-manifold degrees of  freedom consist in the invariance of
geodetic models under the total group of regular translations.
This motivates the search for left- or right-invariant kinetic
energies, i.e., Riemannian structures on
GL$(n,\mathbb{R})\times_{s}\mathbb{R}^{n}$ or GL$(n,\mathbb{R})$.
The first step is purely mathematical: the very construction and
some primary analysis. Later on some hypotheses are formulated
concerning the physical applicability of such apparently exotic
"non-d'Alembertian" models.

\section{Kinematics and Poisson brackets}

Let us remind briefly the basic ideas concerning the extended
affinely-rigid body in a flat Euclidean space
\cite{AKS-JJS91,JJS73_2,JJS75_1,JJS75_2,JJS82_2,JJS82_1,JJS86,JJS88_1,JJS02_2,JJS04}.
It is convenient to use the standard terms of continuum mechanics,
although the model applies also to discrete or finite systems of
material points (provided there exist at least $n+1$ material
points in $n$-dimensional space). Two Euclidean spaces are used,
namely, the material space $(N,U,\eta)$ and the physical space
$(M,V,g)$; the symbols $N$, $M$ denote the underlying sets, $U$
and $V$ are their linear spaces of translations, and $\eta\in
U^{\ast}\otimes U^{\ast}$, $g\in V^{\ast}\otimes V^{\ast}$ are
metric tensors. We put $\dim N=\dim M=n$. The points of $N$ are
labels of material points. The configuration space $Q$ of
affinely-rigid body in $M$ is given by AfI$(N,M)$, i.e., the
manifold of affine isomorphisms of $N$ onto $M$. Obviously, it is
an open submanifold of AfI$(N,M)$ --- the affine space of all
affine mappings of $N$ into $M$ (including non-invertible ones).
In some configuration $\Phi\in Q$ the material point $a\in N$
occupies the spatial position $x=\Phi(a)\in M$. The co-moving,
i.e., Lagrangian, mass distribution within the body will be
described by the constant (time-independent) positive measure
$\mu$ on $N$; it may be $\delta$-like (concentrated at single
points), continuous with respect to the Lebesgue measure, or
mixed. Cartesian (Lagrange) coordinates $a^{K}$ in $N$  are chosen
in such a way that their origin is placed at the centre of mass
$\mathcal{C}$, i.e.,
\[
\int a^{K}d\mu(a)=0.
\]
The manifold AfI$(N,M)$ may be identified with the Cartesian
product $M\times$ LI$(U, V)$, where LI$(U, V)$ denotes the
manifold of all linear isomorphism of $U$ onto $V$; it is an open
submanifold of the linear space L$(U,V)$. The first factor refers
to translational motion, i.e., to the centre of mass position
$x=\Phi(\mathcal{C})$. The linear part of $\Phi$,
$\varphi=L[\Phi]=D\Phi\in$ LI$(U,V)$, describes the relative
(internal) motion. Analytically, when some Cartesian coordinates
in $M$ are used, motion is described by the dependence of Euler
(current) coordinates on Lagrangian (material) ones and on the
time variable:
\[
\Phi(t,a)^{i}=\varphi^{i}{}_{K}(t)a^{K}+x^{i}(t).
\]
In practical calculations it is often technically convenient,
although may be geometrically misleading, to identify both $U$ and
$V$ with $\mathbb{R}^{n}$ and $Q$ with semi-direct product
GAf$(n,\mathbb{R})\simeq$
GL$(n,\mathbb{R})\times_{s}\mathbb{R}^{n}$. Another natural model
of $Q$ is $M\times$ F$(V)$, where F$(V)$ denotes the manifold of
all linear frames in $V$. By the way, F$(V)$ as a model of
internal (relative-motion) degrees of freedom is essentially
identical with LI$(U,V)$ if we put $U=\mathbb{R}^{n}$ and use the
natural isomorphism between linear mappings $\varphi\in$
LI$(\mathbb{R}^{n},V)$ and co-moving frames $e\in$ F$(V)$ frozen
into the body and attached at the centre of mass. This must be
done when the body is infinitesimal and the relative motion is
replaced by the dynamics of essentially internal degrees of
freedom. Then $\mathbb{R}^{n}$ becomes the micromaterial space of
internal motion.

Inertia of the body is described by two constant quantities,
namely, the total mass and the second-order moment of internal
inertia $J\in U\otimes U$, i.e.,
\[
m:=\int_{N}d\mu(a),\qquad J^{KL}:=\int_{N}a^{K}a^{L}d\mu(a)
\]
(cf., e.g.,
\cite{AKS-JJS91,JJS73_2,JJS75_1,JJS75_2,JJS82_2,JJS82_1,JJS86,JJS88_1,JJS02_2,JJS04});
$J$ is symmetric and positively-definite.

Summing up the kinetic energies of constituents,
\[
T = \frac{1}{2}g_{ij}\int\frac{\partial\Phi^{i}}{\partial
t}(t,a)\frac{\partial\Phi^{j}}{\partial t}(t,a)d\mu(a),
\]
one obtains:
\begin{equation}\label{2.1}
T=T_{\rm tr}+T_{\rm int}=
\frac{m}{2}g_{ij}\frac{dx^{i}}{dt}\frac{dx^{i}}{dt}+
\frac{1}{2}g_{ij}\frac{d\varphi^{i}{}_{A}}{dt}
\frac{d\varphi^{i}{}_{B}}{dt}J^{AB};
\end{equation}
the symbols "tr" and "int" refer, obviously, to the translational
and internal (relative) terms.

The phase space of our system may be identified with the manifold
$P:=M\times{\rm LI}(U,V)\times V^{\ast}\times{\rm L}(V,U)$. The
factor $V^{\ast}$ refers to translational canonical momentum,
whereas L$(V,U)$ to the internal one, in the sense of the obvious
pairing between $\pi\in$ L$(V,U)$ and generalized internal
velocity $\xi\in$ L$(U,V)$: $\langle\pi,\xi\rangle={\rm
Tr}(\pi\cdot\xi)={\rm Tr}(\xi\cdot\pi)$. Cartesian coordinates in
$M$ generate parametrization $p_{i}$, $p^{A}{}_{i}$ of canonical
momenta. For Lagrangians of the form $L=T-V(x,\varphi)$ Legendre
transformation
\begin{equation}\label{2.2}
p_{i}=mg_{ij}\frac{dx^{j}}{dt},\qquad
p^{A}{}_{i}=g_{ij}\frac{d\varphi^{j}{}_{B}}{dt}J^{BA}
\end{equation}
leads to the following kinetic term of the Hamiltonian:
\[
\mathcal{T}=\frac{1}{2m}g^{ij}p_{i}p_{j}+
\frac{1}{2}g^{ij}p^{A}{}_{i}p^{B}{}_{j}\widetilde{J}_{AB},
\]
where, obviously, $g^{ij}$ are components of the reciprocal
contravariant metric of $g$, and $\widetilde{J}\in U^{\ast}\otimes
U^{\ast}$ is reciprocal to $J$,
$\widetilde{J}_{AC}J^{CB}=\delta_{A}{}^{B}$, do not confuse it
with $J$ with the $\eta$-lowered indices. This kinetic term (and
its underlying flat metric on $Q$) is invariant under Abelian
additive translations in $Q=M\times$ LI$(U,V)$; those in the
second term are meant in the sense
\begin{equation}\label{2.3}
{\rm LI}(U,V)\ni\varphi\mapsto\varphi+\alpha,\qquad \alpha\in{\rm
L}(U,V).
\end{equation}
Therefore, without the interaction term (for $L=T$), the
Hamiltonian generators $p_{i}$, $p^{A}{}_{i}$ are constants of
motion. However, as mentioned above, such geodetic models for
deformable bodies are physically non-interesting, because they
predict unlimited expansion, contraction, and passing through
singular configurations with $\det\varphi=0$. The latter, although
non-acceptable in continuum mechanics, may be to some extent
admissible in mechanics of discrete bodies. If we once decide that
the internal configuration space is given by LI$(U,V)$, then the
above transformation group is only local. At the same time, even
for purely geodetic systems, as mentioned, there is no invariance
under geometrically interesting affine groups of left or right
affine regular translations in $Q$. Even if, at the present stage,
models with affinely-invariant kinetic energy might seem rather
academic, they present some interest at least from the purely
mathematical point of view. Besides, some physical applications
seem to be possible in hydrodynamics, astrophysics, nuclear
dynamics, and in certain elastic problems. It is interesting that
even without any genuine interactions, on the purely geodetic
level such models may predict bounded and stable elastic
vibrations of incompressible bodies. It is so as if the
interaction was encoded in the very kinetic energy, i.e.,
configuration metric, so as it is, e.g., in Jacobi-Maupertuis
variational principle. To formulate such models we must introduce
and partially remind certain geometric objects.

Affine velocity in laboratory representation, i.e., expressed in
terms of space-fixed frames, is defined as
\[
\Omega:=\frac{d\varphi}{dt}\varphi^{-1 }\in{\rm L}(V),\qquad
\Omega^{i}{}_{j}=\frac{d\varphi^{i}{}_{K}}{dt}\left(\varphi^{-1}\right){}^{K}{}_{j}.
\]
The corresponding co-moving object, related to the body-fixed
frame, is given by
\[
\hat{\Omega}:=\varphi^{-1}\frac{d\varphi}{dt}\in{\rm L}(U),\qquad
\hat{\Omega}^{A}{}_{B}=\left(\varphi^{-1}\right){}^{A}{}_{i}\frac{d\varphi^{i}{}_{B}}{dt}.
\]
Obviously, $\Omega=\varphi\hat{\Omega}\varphi^{-1}$,
$\Omega^{i}{}_{j}=
\varphi^{i}{}_{A}\hat{\Omega}^{A}{}_{B}\left(\varphi^{-1}\right){}^{B}{}_{j}$.
These are Lie-algebraic objects corresponding to the structure of
$Q$ as the group space of a Lie group. They provide an affine
counterpart of the rigid-body angular velocities, and in fact
reduce to them when $\varphi$ is confined to the manifold of
isometries of $(U,\eta)$ onto $(V,g)$; then they become
skew-symmetric respectively with respect to $\eta$ or $g$.

The object $\Omega$ may be represented in terms of continua as a
gradient of the Euler velocity field, namely, the material point
passing the fixed spatial point $y$ has the translational
velocity:
\[
{}^{E}v(y)^{i}=\frac{dx^{i}}{dt}+\Omega^{i}{}_{j}(y^{j}-x^{j}),
\]
i.e., simply $\Omega^{i}{}_{j}y^{j}$ in the instantaneous rest
frame of the centre of mass, placed also at the instantaneous
position of this centre in $M$. Similarly, $d\varphi^{i}{}_{B}/dt$
has to do with the gradient of the Lagrange velocity field,
because the instantaneous velocity of the $a$-th particle $(a\in
N)$ is given by
\[
{}^{L}v(a)^{i}=\frac{dx^{i}}{dt}+\frac{d\varphi^{i}{}_{K}}{dt}a^{K}
\]
(concerning the standard concepts of continuum mechanics consult,
e.g., \cite{Erin62,Erin68,Lan-Lif58_1}). In certain problems it is
also convenient to express the centre of mass translational
velocity $v^{i}=dx^{i}/dt$ in co-moving terms, i.e.,
$\hat{v}^{A}=\left(\varphi^{-1}\right){}^{A}{}_{i}v^{i}$.

It is very convenient to introduce the canonical affine spin, also
in two representations, the spatial and co-moving ones $\Sigma\in$
L$(V)$, $\hat{\Sigma}\in$ L$(U)$. In terms of coordinates they are
given by the following formulas:
$\Sigma^{i}{}_{j}=\varphi^{i}{}_{A}p^{A}{}_{j}$,
$\hat{\Sigma}^{A}{}_{B}=p^{A}{}_{i}\varphi^{i}{}_{B}$. As
previously, $\Sigma=\varphi\hat{\Sigma}\varphi^{-1}$,
$\Sigma^{i}{}_{j}=\varphi^{i}{}_{A}\hat{\Sigma}^{A}{}_{B}
\left(\varphi^{-1}\right){}^{B}{}_{j}$.

They are purely Hamiltonian quantities defined on the phase space;
without any precisely defined Lagrangian or Hamiltonian we cannot
relate them to generalized velocities. It is seen, however, that
they are dual objects to affine velocities, i.e., they are
non-holonomic canonical momenta conjugate to them in the sense of
following pairing:
\[
\langle\Sigma,\Omega\rangle=\langle\hat{\Sigma},\hat{\Omega}\rangle:={\rm
Tr}(\Sigma\Omega)={\rm
Tr}(\hat{\Sigma}\hat{\Omega})=p^{A}{}_{i}v^{i}{}_{A},
\]
where $v^{i}{}_{A}$ are generalized velocities  of internal
(relative) motion. This canonical isomorphism between Lie algebras
GL$(U)^{\prime}=$ L$(U)$, GL$(V)^{\prime}=$ L$(V)$ and their duals
simplifies remarkably all formulas and considerations.

It is clear that quantities $\Sigma^{i}{}_{j}$ are Hamiltonian
generators of GL$(V)$ acting on LI$(U,V)$ through the left
translations:
\begin{equation}\label{2.4}
\varphi\mapsto A\varphi,\qquad \varphi\in{\rm LI}(U,V),\quad A\in
{\rm GL}(V).
\end{equation}
Similarly, $\hat{\Sigma}^{A}{}_{B}$ generate right regular
translations in the internal configuration space:
\begin{equation}\label{2.5}
\varphi\mapsto\varphi B,\qquad \varphi\in{\rm LI}(U,V),\quad B\in
{\rm GL}(U).
\end{equation}

In continuum mechanics these mappings are referred to,
respectively, as spatial and material transformations; in this
case they include rotations and homogeneous deformations.
Obviously, to use correctly such terms we must be given metric
tensors in $V$ and $U$. Then the $g$-antisymmetric part of
$\Sigma$ and the $\eta$-antisymmetric part of $\hat{\Sigma}$
generate, respectively, spatial and material  rigid rotations; the
symmetric parts generate deformations.

The doubled antisymmetric parts are referred to as spin $S$ and
vorticity $V$ \cite{Dys68},
\begin{equation}\label{2.6}
S^{i}{}_{j}=\Sigma^{i}{}_{j}-g^{ik}g_{jm}\Sigma^{m}{}_{k},\qquad
V^{A}{}_{B}=\hat{\Sigma}^{A}{}_{B}-\eta^{AC}\eta_{BD}\hat{\Sigma}^{D}{}_{C}.
\end{equation}

\noindent {\bf Attention!} There is an easy mistake possibility:
if motion is not metrically-rigid, then $V$ is not a co-moving
representation of $S$, i.e.,
\[
S^{i}{}_{j}\neq\varphi^{i}{}_{A}V^{A}{}_{B}\left(\varphi^{-1}\right){}^{B}{}_{j}.
\]

Just as translational velocity, the canonical linear momentum may
be expressed in co-moving terms according to the following rule:
$\hat{p}_{A}=p_{i}\varphi^{i}{}_{A}$.

The objects $\Omega$ and $\Sigma$ are invariant under material
transformations, but the spatial action of $A\in$ GL$(V)$
transforms them according to the adjoint rule, i.e.,
$\Omega\mapsto A\Omega A^{-1}$, $\Sigma\mapsto A\Sigma A^{-1}$. On
the contrary, $\hat{\Omega}$ and $\hat{\Sigma}$ are invariant
under GL$(V)$ but experience the inverse adjoint rule under $B\in$
GL$(U)$, i.e., $\hat{\Omega}\mapsto B^{-1}\hat{\Omega}B$,
$\hat{\Sigma}\mapsto B^{-1}\hat{\Sigma}B$. This formally agrees
with formulas for systems with configuration spaces identical with
Lie groups, but one must stress that there are some subtle
differences due to the fact that LI$(U,V)$ is not a Lie group (may
be identified with it, but there is an infinity mutually
equivalent identifications).

The translational or orbital affine momentum with respect to some
point $\mathcal{O}\in M$ is defined as follows:
\[
\Lambda(\mathcal{O})^{i}{}_{j}:=x^{i}p_{j},
\]
where $x^{i}$ are Cartesian coordinates of the
$\mathcal{O}$-radius vector of the current position of the centre
of mass in $M$. The total affine momentum with respect to
$\mathcal{O}$ is given by
\[
I(\mathcal{O})^{i}{}_{j}:=\Lambda(\mathcal{O})^{i}{}_{j}+\Sigma^{i}{}_{j}.
\]
$\Lambda(\mathcal{O})$ and $I(\mathcal{O})$ depend explicitly on
the choice of $\mathcal{O}$. Unlike this, $\Sigma$ is objective
(in a fixed Galilean reference frame). There is a complete analogy
with the properties of angular momentum, the doubled
$g$-antisymmetric part of the above objects. The quantity
$I(\mathcal{O})$ is a Hamiltonian generator of the group of affine
transformations of $M$ preserving $\mathcal{O}$
($\mathcal{O}$-centred affine subgroup).

Poisson brackets of $\Sigma$-quantities follow directly from the
standard ones for $x^i$, $p_{i}$, $\varphi^{i}{}_{A}$,
$p^{A}{}_{i}$. The non-vanishing ones are simply given by the
structure constants of linear group,
\[
\{\Sigma^{i}{}_{j},\Sigma^{k}{}_{l}\}=
\delta^{i}{}_{l}\Sigma^{k}{}_{j}-\delta^{k}{}_{j}\Sigma^{i}{}_{l},
\quad \{\Sigma^{i}{}_{j},\hat{\Sigma}^{A}{}_{B}\}=0,
\]
\[
\{\hat{\Sigma}^{A}{}_{B},\hat{\Sigma}^{C}{}_{D}\}=
\delta^{C}{}_{B}\hat{\Sigma}^{A}{}_{D}-
\delta^{A}{}_{D}\hat{\Sigma}^{C}{}_{B}
\]
(similarly for $\Lambda$, $I$). There are also non-vanishing
Poisson brackets related to the left or right affine groups
GAf$(M)$, GAf$(N)$. Here belong the above ones and besides, those
involving linear momenta,
\[
\{\hat{\Sigma}^{A}{}_{B},\hat{p}_{C}\}=\delta^{A}{}_{C}\hat{p}_{B},
\qquad \{I^{i}{}_{j},p_{k}\}=\{\Lambda^{i}{}_{j},p_{k}\}=
\delta^{i}{}_{k}p_{j}.
\]
If $F$ is any function depending only on the configurations
variables, then, obviously,
\[
\{F,\Sigma^{i}{}_{j}\}=\varphi^{i}{}_{A}\frac{\partial
F}{\partial\varphi^{j}{}_{A}},\qquad
\{F,\Lambda^{i}{}_{j}\}=x^{i}\frac{\partial F}{\partial
x^{j}},\qquad
\{F,\hat{\Sigma}^{A}{}_{B}\}=\varphi^{i}{}_{B}\frac{\partial
F}{\partial\varphi^{i}{}_{A}}.
\]
Geometric meaning of the last formulas is clear, because the
differential operators used on their right-hand sides are
identical with vector fields on $Q$ generating the action of
one-parameter subgroups of GAf$(M)$ and GAf$(N)$.

As mentioned, the above Poisson brackets follow directly from the
standard  definition \cite{Arn78,Gold50,Gut71}
\[
\{F,G\}:=\frac{\partial F}{\partial q^{\alpha}}\frac{\partial
G}{\partial p_{\alpha}}-\frac{\partial F}{\partial
p_{\alpha}}\frac{\partial G}{\partial q^{\alpha}},
\]
where $q^{\alpha}$ are generalized coordinates and $p_{\alpha}$
are their conjugate canonical momenta. In our model $q^{\alpha}$
are given by $x^{i}$, $\varphi^{i}{}_{A}$, and $p_{\alpha}$ by
$p_{i}$, $p^{A}{}_{i}$. In applications it is sufficient to
remember that $\{q^{\alpha},q^{\beta}\}=0$,
$\{p_{\alpha},p_{\beta}\}=0$,
$\{q^{\alpha},p_{\beta}\}=\delta^{\alpha}{}_{\beta}$, that Poisson
bracket is bilinear (over constant reals $\mathbb{R}$),
skew-symmetric, i.e., $\{F,G\}$$=-\{G,F\}$, satisfies the Jacobi
identity $\{\{F,G\},H\}+\{\{G,H\},F\}+\{\{H,F\},G\}=0$, and
finally that
\[
\{F,H(G_{1},\ldots,G_{k})\}=
\sum^{k}_{p=1}H_{,p}(G_{1},\ldots,G_{k})\{F,G_{p}\},
\]
where commas before indices denote the partial derivatives. The
formerly-quoted Poisson brackets together with the above rules are
sufficient for all calculations concerning equations of motion and
their analysis.

To define a non-dissipative (Hamiltonian) dynamical model, we must
be given some Lagrangian $L(q,\dot{q})$, perform the Legendre
transformation, $p_{\alpha}=\partial L/\partial \dot{q}^{\alpha}$,
invert it, i.e., solve with respect to generalized velocities
$\dot{q}^{\alpha}$, and substitute the result to the energy
function $E=\dot{q}^{\alpha}\partial
L/\partial\dot{q}^{\alpha}-L$. In this way one obtains the
Hamilton function $H(q,p)$. Equations of motion may be then
expressed in terms of Poisson brackets,
\[
\frac{dF}{dt}=\{F,H\},
\]
where $F$ runs over some finite family of basic functions, e.g., $(p_{i},\Sigma^{i}{}_{j},x^{i},\varphi^{i}{}_{A})$,
$(\hat{p}_{A},\hat{\Sigma}^{A}{}_{B},x^{i},\varphi^{i}{}_{A})$, or
something else. The basic dynamical laws are given by the balance
equations for the linear momentum and affine spin either in
laboratory or co-moving representation (one could use equivalently
the linear momentum and the total affine momentum, however, the
previous choice is more convenient). The procedure based on
Poisson brackets and canonical formalism is very often more easy
and computationally less embarrassing than the one directly using
the Euler-Lagrange equations.

\noindent {\bf Remark:} Legendre transformation may be also
expressed in terms of non-holonomic objects, moreover, this is
often more convenient and effective than the use of generalized
velocities. Expressing Lagrangian in terms of
$(v^{i},\Omega^{i}{}_{j})$ or
$(\hat{v}^{A},\hat{\Omega}^{A}{}_{B})$ instead of
$(\dot{x}^{i},\dot{\varphi}^{i}{}_{A})$, we can describe the
Legendre transformation as follows:
\[
p_{i}=\frac{\partial L}{\partial v^{i}},\qquad
\Sigma^{i}{}_{j}=\frac{\partial L}{\partial
\Omega^{j}{}_{i}},\qquad {\rm or}\qquad \hat{p}_{A}=\frac{\partial
L}{\partial\hat{v}^{A}},\qquad
\hat{\Sigma}^{A}{}_{B}=\frac{\partial L}{\partial
\hat{\Omega}^{B}{}_{A}}.
\]

When dealing with the Hamiltonian form of equations of motion, we
need often Poisson brackets involving deformation tensors and
certain by-products of the inertial tensor, like, e.g., the
Eulerian quadrupole of the mass distribution.

Obviously, for systems with affine degrees of freedom the Green
and Cauchy deformation tensors $G\in U^{\ast}\otimes U^{\ast}$,
$C\in V^{\ast}\otimes V^{\ast}$ are respectively given by the
following expressions: $G=\varphi^{\ast}g$,
$C=\left(\varphi^{-1}\right)^{\ast}\eta$, i.e., in analytical
terms $G_{AB}=g_{ij}\varphi^{i}{}_{A}\varphi^{j}{}_{B}$,
$C_{ij}=\eta_{AB}\left(\varphi^{-1}\right){}^{A}{}_{i}\left(\varphi^{-1}\right){}^{B}{}_{j}$.
Their inverses $\widetilde{G}\in U\otimes U$, $\widetilde{C}\in
V\otimes V$ are defined by
$\widetilde{G}^{AC}G_{CB}=\delta^{A}{}_{B}$,
$\widetilde{C}^{ik}C_{kj}=\delta^{i}{}_{j}$, and one must be
careful to avoid mistaking $\widetilde{G}^{AB}$,
$\widetilde{C}^{ij}$ with $\eta^{AC}\eta^{BD}G_{CD}$,
$g^{ik}g^{jl}C_{kl}$. Therefore, the usual convention of the
upper- and lower-case indices may be misleading. Analytically,
$\widetilde{G}^{AB}=\left(\varphi^{-1}\right){}^{A}{}_{i}
\left(\varphi^{-1}\right){}^{B}{}_{j}g^{ij}$,
$\widetilde{C}^{ij}=\varphi^{i}{}_{A}\varphi^{j}{}_{B}\eta^{AB}$.
When there is no deformation, i.e.,  $\varphi\in$
LI$(U,\eta;V,g)$, then $G=\eta$, $C=g$. The corresponding
deformation measures vanishing in the non-deformed state, i.e.,
Lagrange and Cauchy deformation tensors $E\in U^{\ast}\otimes
U^{\ast}$, $e\in V^{\ast}\otimes V^{\ast}$ are given by (see,
e.g., \cite{Erin62,Erin68}):
\[
E:=\frac{1}{2}(G-\eta),\qquad e:=\frac{1}{2}(g - C).
\]
One uses also their contravariant versions $E^{AB}$, $e^{ij}$;
unlike $\widetilde{G}^{AB}$, $\widetilde{C}^{ij}$ they are defined
via the $\eta$- and $g$-raising of indices.

\noindent {\bf Remark:} $G$ is independent of $\eta$ and may be
defined even if the material space is purely affine, amorphous.
Similarly, $C$ is independent of $g$ and is well-defined even if
the physical space is metric-free. Therefore, the literally meant
term "deformation" is better expressed by $E$, $e$ than $G$, $C$.
However, in many formulas $G$, $C$ are more natural and
convenient. Deformation tensors behave under the action of
isometries in a very peculiar way, namely, for any $A\in$
O$(V,g)$,  $B\in$ O$(U, \eta)$, we have:
\[
G[A\varphi]_{KL}=G[\varphi]_{KL},\qquad G[\varphi
B]_{KL}=G[\varphi]_{CD}B^{C}{}_{K}B^{D}{}_{L},
\]
\[
C[A\varphi]_{ij}=C[\varphi]_{ab}\left(A^{-1}\right){}^{a}{}_{i}
\left(A^{-1}\right){}^{b}{}_{j},\quad C[\varphi
B]_{ij}=C[\varphi]_{ij}.
\]
By the way, the last two formulas are valid for any $A\in$
GL$(V)$, $B\in$ GL$(U)$. The first two equations (invariance
rules) imply the Poisson-bracket rules 
\[
\{G_{KL},S^{i}{}_{j}\}=0,\quad
\{C_{ij},V^{A}{}_{B}\}=0, 
\]
and similarly for $E_{KL}$, $e_{ij}$.

Deformation invariants are important mechanical quantities. They
are scalar measures of deformation, basic stretchings, which do
not contain any information concerning the orientation of
deformation (its principal axes) in the physical or material
space. They may be chosen in various ways, but in an
$n$-dimensional space exactly $n$ of them may be functionally
independent. The particular choice of $n$ basic invariants depends
on the considered problem and on the computational details. When
non-specified, they will be denoted by $\mathcal{K}_{a}$,
$a=\overline{1,n}$. Let us define mixed tensors $\hat{G}\in
U\otimes U^{\ast}$, $\hat{C}\in V\otimes V^{\ast}$, $\hat{E}\in
U\otimes U^{\ast}$, $\hat{e}\in V\otimes V^{\ast}$, namely,
\[
\hat{G}^{A}{}_{B}:=\eta^{AC}G_{CB},\qquad
\hat{C}^{i}{}_{j}:=g^{ik}C_{kj},\qquad
\hat{E}^{A}{}_{B}:=\eta^{AC}E_{CB},\qquad
\hat{e}^{i}{}_{j}:=g^{ik}e_{kj}.
\]
A class of possible and geometrically natural choices of
$\mathcal{K}_{a}$ is given by the following expressions: ${\rm
Tr}(\hat{G}^{k})$, ${\rm Tr}(\hat{C}^{k})$, ${\rm
Tr}(\hat{E}^{k})$, ${\rm Tr}(\hat{e}^{k})$, $k=\overline{1,n}$. In
certain problems it is convenient to use the following
eigenequations:
\[
\det\left[\hat{G}^{A}{}_{B}-\lambda\delta^{A}{}_{B}\right]=0,\quad
\det\left[\hat{C}^{i}{}_{j}-\lambda\delta^{i}{}_{j}\right]=0,
\]
\[
\det\left[\hat{E}^{A}{}_{B}-\lambda\delta^{A}{}_{B}\right]=0,\quad
\det\left[\hat{e}^{i}{}_{j}-\lambda\delta^{i}{}_{j}\right]=0.
\]
These are $n$-th order algebraic (polynomial) equations with
respect to $\lambda$. Their solutions provide one of possible
choices of basic invariants. Another, very convenient one is given
by coefficients at $\lambda^{p}$, $p=\overline{0,(n-1)}$
\cite{Erin62,Erin68} (the coefficient at $\lambda^{n}$ is standard
and equals one). Deformation invariants are non-sensitive with
respect to spatial and material isometries, i.e., for any $A\in$
O$(V,g)$, $B\in$ O$(U,\eta)$ we have $\mathcal{K}_{a}[A\varphi
B]=\mathcal{K}_{a}[\varphi]$. This implies the obvious Poisson
brackets:
$\{\mathcal{K}_{a},S^{i}{}_{j}\}=\{\mathcal{K}_{a},V^{A}{}_{B}\}=0$.

In certain computational problems, but also in theoretical
analysis, it is very convenient to use quantities
$Q^{a}=\sqrt{\lambda_{a}}$, where $\lambda_{a}$ are solutions of
the above eigenequations, or $q^{a}=\ln Q^{a}$ (i.e.,
$Q^{a}=\exp(q^{a})$). The eigenvalues of $\hat{C}$ equal
$\left(\lambda_{a}\right)^{-1}=(Q^{a})^{-2}=\exp(-2q^{a})$.

Any function $F$ on the configuration space which depends on $\varphi$ only through the
deformation invariants is doubly isotropic, i.e., satisfies $F(A\varphi B)=F(\varphi)$ for any
$A\in$ O$(V,g)$, $B\in$ O$(U,\eta)$, $\varphi\in$ LI$(U,\eta;V,g)$. All such functions have
vanishing Poisson brackets with spin and vorticity, i.e.,
$\{F,S^{i}{}_{j}\}=\{F,V^{A}{}_{B}\}=0$. In certain formulas we need the spatial inertial
quadrupole, $J[\varphi]^{ab}=\varphi^{a}{}_{K}\varphi^{b}{}_{L}J^{KL}$. It is related to
$J^{KL}$ just as $\widetilde{C}$ is to $\eta$. When the body is inertially isotropic,
$J[\varphi]$ becomes proportional to the inverse Cauchy deformation tensor. Unlike the
co-moving internal tensor $J\in U\otimes U$, $J[\varphi]\in V\otimes V$ is
configuration-dependent, thus variable in time.

\section{Traditional d'Alembert model}

At least for the comparison with more exotic (although
geometrically and perhaps physically interesting) suggestions we
must start with a brief reporting and extension of the traditional
model based on the d'Alembert principle. As shown in
\cite{JJS74_2,JJS75_1,JJS75_2,JJS75_3,JJS75_4,JJS82_2},
Lagrangians of the form $L=T-V(x,\varphi)$ with $T$ given by
(\ref{2.1}) lead to the following dynamical laws:
\begin{equation}\label{3.1}
\frac{dp_{i}}{dt}=-\frac{\partial V}{\partial x^{i}}=Q_{i},\qquad
\frac{d\Sigma^{i}{}_{j}}{dt}=\Omega^{i}{}_{m}\Sigma^{m}{}_{j}-
\varphi^{i}{}_{A}\frac{\partial
V}{\partial\varphi^{j}{}_{A}}=\Omega^{i}{}_{m}\Sigma^{m}{}_{j}+
Q^{i}{}_{j},
\end{equation}
expressed in terms of Cartesian coordinate systems. This is the
balance for fundamental Hamiltonian generators. It becomes a
closed dynamical system when considered together with the Legendre
transformation (\ref{2.2}) or its equivalent description
\begin{equation}\label{3.2}
p_{i}=mg_{ij}\frac{dx^{j}}{dt},\qquad
\Sigma^{i}{}_{j}=g_{jk}\Omega^{k}{}_{m}J[\varphi]^{mi}.
\end{equation}
Substituting these expressions to the dynamical balance for
$p_{i},\Sigma^{i}{}_{j}$ one obtains some reformulation of the
Euler-Lagrange equations. Similarly, some form of canonical
Hamilton equations is obtained when the balance (\ref{3.1}) is
unified with the inverse Legendre transformation, i.e.,
\[
\frac{dx^i}{dt}=\frac{1}{m}g^{ij}p_{j},\qquad
\Omega^{i}{}_{j}=\widetilde{J}[\varphi]_{jk}\Sigma^{k}{}_{m}g^{mi},
\]
where, obviously,
$J[\varphi]^{ik}\widetilde{J}[\varphi]_{kj}=\delta^{i}{}_{j}$ (do
not confuse $\widetilde{J}[\varphi]$ with $g$-shift of indices of
$J[\varphi]$).

Obviously, the general balance form may  be used for dissipative
non-Largan\-gian models. Simply  the covariant force $Q_{i}$ and
the generalized internal force $Q^{i}{}_{j}$ (affine moment of
forces, hyperforce) must involve appropriately defined dissipative
forces (in the case of affinely-constrained continuum one can also
consider the mutual coupling of mechanical phenomena with
discretized thermal effects).

As shown in the mentioned papers, the above equations of motion
may be formulated in various equivalent forms adapted to the kind
of considered problems. For example, instead of the canonical
(Hamilton) form, one can write them down in purely kinematical
velocity-based terms, i.e.,
\begin{equation}\label{3.3}
m\frac{d^{2}x^{i}}{dt^{2}}=F^{i},\qquad
\varphi^{i}{}_{A}\frac{d^{2}\varphi^{j}{}_{B}}{dt^{2}}J^{AB}=N^{ij},
\end{equation}
where contravariant forces $F^{i}$ and hyperforce $N^{ij}$ (affine
dynamical moment) may depend on all possible arguments, i.e., $t$,
$x^{i}$, $\varphi^{i}{}_{A}$, $dx^{i}/dt$,
$d\varphi^{i}{}_{A}/dt$. Obviously, for potential models they
depend only on generalized coordinates and possibly on the time
variable $t$ itself, and then
\begin{equation}\label{3.4}
F^{i}=g^{ij}Q_{j}=-g^{ij}\frac{dV}{dx^{j}},\qquad
N^{ij}=Q^{i}{}_{k}g^{kj}=-\varphi^{i}{}_{A}\frac{\partial
V}{\partial\varphi^{k}{}_{A}}g^{kj}.
\end{equation}

\noindent {\bf Remark:} In spite of the tradition based on
Riemannian geometry and relativity theory we shall refrain from
the graphical identification of symbols $F^{i}$, $N^{ij}$
respectively with $Q^{i}$, $Q^{ij}$. In our treatment this would
be just confusing, because we shall use various prescriptions for
shifting the tensorial indices, i.e., various isomorphisms between
contravariant and covariant objects.

As mentioned, the above equations of the motion (\ref{3.3}) may be
derived directly in Newtonian terms, basing merely on the
d'Alembert principle and its underlying spatial metric $g$ in $M$.
The primary quantities of this approach are the monopole and
dipole moments of the distributions of linear kinematical momentum
and forces within the body. These quantities, just as all
high-order multipoles may be defined for any unconstrained system
of material points, does not matter finite, discrete, or
continuous. Let $\Phi(t,a)^{i}$ denote as previously Cartesian
coordinates of the current position of the $a$-th material point
at the time instant $t$, $x^{i}(t)$ be the current position of the
centre of mass, and $\mathcal{F}^{i}(t,x,\Phi(t,a),dx/dt,
(\partial\Phi/\partial t)(t,a);a)$ be the density of forces per
unit mass. As mentioned, the affine constraints are not yet
assumed.

The monopoles are simply the total quantities: the total
kinematical momentum $k^{i}$ (do not confuse it at this stage with
the canonical one $p_{i}$) and the total force $F^{i}$ affecting
the centre of mass motion, i.e.,
\[
k^{i}=\int\frac{\partial\Phi^{i}}{\partial t}(t,a)d\mu(a),\qquad
F^{i}=\int\mathcal{F}^i\left(t,x,\Phi(t,a),\frac{dx}{dt},
\frac{\partial\Phi}{\partial t}(t,a);a\right)d\mu(a).
\]
The dipole moments with respect to the centre of mass current
position are referred to as kinematical affine spin $K^{ij}$ (do
not confuse it at this stage with the canonical one
$\Sigma^{i}{}_{j}$) and the affine moment of forces $N^{ij}$ (not
to be confused with its potential version $Q^{i}{}_{j}$). They are
given respectively by the following expressions:
\begin{eqnarray}
K^{ij}&=&\int\left(\Phi^{i}(t,a)-x^{i}\right)
\left(\frac{\partial\Phi^{j}}{\partial
t}(t,a)-\frac{dx^{j}}{dt}\right)d\mu(a),
\nonumber\\
N^{ij}&=&\int\left(\Phi^{i}(t,a)-x^{i}\right)
\mathcal{F}^{j}\left(t,x,\Phi(t,a),\frac{dx}{dt},
\frac{\partial\Phi}{\partial t}(t,a);a\right)d\mu(a).\nonumber
\end{eqnarray}

The dipoles may be also referred to some space-fixed centre
$\mathcal{O}\in M$, e.g., the origin of Cartesian coordinates in
$M$. The difference is that "the lever arm" $(\Phi^{i}-x^{i})$ is
then replaced by $\Phi^{i}$ itself, and its velocity
$\left(\partial\Phi^{i}/\partial t-dx^i/dt\right)$ by
$(\partial\Phi^{i}/\partial t)(t,a)$. The resulting dipoles will
be denoted respectively by $K(\mathcal{O})^{ij}$,
$N(\mathcal{O})^{ij}$. For the sake of uniformity, it may be also
convenient to denote the previous dipoles by $K(\rm cm)^{ij}$ and
$N(\rm cm)^{ij}$ instead of $K^{ij}$, $N^{ij}$. We shall also use
affine moments of the centre of mass characteristics with respect
to the origin $\mathcal{O}$. Thus, the translational (orbital)
affine moment of kinematical linear momentum and translational
affine moment of forces are as follows:
\[
K_{\rm tr}(\mathcal{O})^{ij}=
x^{i}k^{j}=mx^{i}\frac{dx^{j}}{dt},\qquad N_{\rm
tr}(\mathcal{O})^{ij}= x^{i}F^{j}.
\]
The doubled skew-symmetric parts of the above quantities, i.e.,
\[
S^{ij}=K^{ij}-K^{ji}, \ L_{\rm tr}(\mathcal{O})^{ij}=K_{\rm
tr}(\mathcal{O})^{ij}-K_{\rm tr}(\mathcal{O})^{ji}, \
\mathcal{J}(\mathcal{O})^{ij}=K(\mathcal{O})^{ij}-K(\mathcal{O})^{ji},
\]
\[
\mathcal{N}^{ij}=N^{ij}-N^{ji},\ \mathcal{N}_{\rm
tr}(\mathcal{O})^{ij}=N_{\rm tr}(\mathcal{O})^{ij}-N_{\rm
tr}(\mathcal{O})^{ji}, \
\mathcal{N}(\mathcal{O})^{ij}=N(\mathcal{O})^{ij}-N(\mathcal{O})^{ji},
\]
represent the kinematical angular momentum and the moment of
forces (torque). They also occur in three versions concerning,
respectively, the internal motion (thus $S$ is spin), motion of
the centre of mass with respect to $\mathcal{O}$, and the total
motion with respect to $\mathcal{O}$.

If now we assume that the motion is affine, then the above
expressions simplify to
\[
K^{ij}=\varphi^{i}{}_{A}\frac{d\varphi^{j}{}_{B}}{dt}J^{AB}, \quad
K(\mathcal{O})^{ij}=K_{\rm
tr}(\mathcal{O})^{ij}+K^{ij}=mx^{i}\frac{dx^{j}}{dt}+
\varphi^{i}{}_{A}\frac{d\varphi^{j}{}_{B}}{dt}J^{AB},
\]
\[
N(\mathcal{O})^{ij}=N_{\rm
tr}(\mathcal{O})^{ij}+N^{ij}=x^{i}F^{j}+N^{ij}.
\]
Obviously, $F$ and $N$ become now functions of $t$, $x^{i}$,
$dx^{i}/dt$, $\varphi^{i}{}_{A}$, $d\varphi^{i}{}_{A}/dt$. Let us
stress that, just as it was the case with the kinetic energy, the
above additive splitting into translational and internal parts is
based on the assumption that the current centre of mass has
permanently Lagrangian coordinates $a^{K}=0$. This is consistent
because barycenters are invariants of affine transformations.

By summation of elementary time rates of work over the body
constituents, one can show that in the affine motion the total
rate is given by
\[
\mathcal{P}=g_{ij}\frac{dx^{i}}{dt}F^{j}+g_{ij}\Omega^{i}{}_{k}N^{kj}.
\]
Let us remind however that, besides of active generalized forces
$F$, $N$ controlling affine modes of motion, there are also hidden
structural forces keeping affine constraints, i.e., reactions.
Their density $\mathcal{F}_{R}$ does not vanish, however, their
monopole and dipole moments $F_{R}$, $N_{R}$ do because, according
to the d'Alembert principle, the reaction time rate of work
vanishes for any constraints-compatible virtual velocities, i.e.,
for any possible $dx^{i}/dt$, $\Omega$:
\[
\mathcal{P}_{R}=g_{ij}\frac{dx^{i}}{dt}F_{R}{}^{j}
+g_{ij}\Omega^{i}{}_{k}N_{R}{}^{kj}=0.
\]
Therefore, the effective reaction-free equations of motion are
obtained from the primary non-constrained system by calculating
the monopole and dipole moments.

The above derivation is quite general and valid for all kinds of
forces, including non-potential and dissipative ones. It relies
only on the metric structure $g$ in $M$ and on the d'Alembert
principle. Obviously, if equations of motion follow from the
Lagrangian $L=T-V(x,e)$, $T$ given by (\ref{2.1}), then the above
analysis implies equations (\ref{3.4}).

Similarly, one can easily show that
\begin{equation}\label{3.5}
K^{ij}=\Sigma^{i}{}_{m}g^{mj},\qquad k^{i}=g^{ij}p_{j}.
\end{equation}
But these relationships become false when Lagrangian depends on
velocities not only through the kinetic energy $T$ but also
through some generalized potential $V$, e.g., when magnetic or
gyroscopic external forces are present. This is one of reasons we
avoid denoting $K^{ij}$ by $\Sigma^{ij}$ or $K^{i}{}_{j}$ by
$\Sigma^{i}{}_{j}$.

Kinematical quantities $k^{i}$, $K^{ij}$ are intuitive because of
their direct operational interpretation in terms of positions and
velocities. At the same time they are lowest-order multipoles
(monopoles and dipoles) of the distribution of kinematical linear
momentum within the body, and it is difficult to over-estimate the
application of multipoles and all moment quantities in practical
problems of mechanics and field theory (cf, e.g., all
Galerkin-type procedures \cite{JJS-AKS93}). On the other hand,
their canonical counterparts $p_{i}$, $\Sigma^{i}{}_{j}$ have a
very deep geometrical interpretation as Hamiltonian generators of
fundamental transformation groups. Because of this, they are very
often important constants of motions. In mechanics of
affinely-rigid body, equations of motion are equivalent to the
balance laws for $p_{i},\Sigma^{i}{}_{j}$ or, in a sense
equivalently, to the ones for $k^{i},K^{ij}$, because Lagrangians
of non-dissipative models, or at least Lagrangians of
non-dissipative background dynamics, establish some link between
these concepts. Similarly, in rigid-body mechanics equations of
motion are equivalent to the balance for
$p_{i},(\Sigma^{i}{}_{j}-g^{ik}g_{jm}\Sigma^{m}{}_{k}$) or for
$k^{i},S^{ij}$.

Equations of motion (\ref{3.3}) may be written in several mutually
equivalent balance forms. Let us quote some of them based on
kinematical quantities like $k^{i},K^{ij}$ or their co-moving
representation $\hat{k}^{A},\hat{K}^{AB}$, where, obviously,
$k^{i}=\varphi^{i}{}_{A}\hat{k}^{A}$,
$K^{ij}=\varphi^{i}{}_{A}\varphi^{j}{}_{B}\hat{K}^{AB}$. The
co-moving components $\hat{F}^{A},\hat{N}^{AB}$ of generalized
forces are given by analogous expressions, thus,
$F^{i}=\varphi^{i}{}_{A}\hat{F}^{A}$,
$N^{ij}=\varphi^{i}{}_{A}\varphi^{j}{}_{B}\hat{N}^{AB}$.

The dynamical balance expressed in terms of kinematical
(non-canonical) quantities in spatial (Eulerian) representation
reads:
\begin{equation}\label{3.6}
\frac{dk^{i}}{dt}=F^{i},\qquad
\frac{dK^{ij}}{dt}=\frac{d\varphi^{i}{}_{A}}{dt}
\frac{d\varphi^{j}{}_{B}}{dt}J^{AB}+N^{ij}.
\end{equation}
For non-dissipative potential systems with Lagrangians
$L=T-V(x,\varphi)$, it reduces to (\ref{3.1}), because then
(\ref{3.4}) holds. Let us observe that even in the
interaction-free case, when $N=0$, the balance for $K$ is not a
conservation law due to the first non-dynamical term on its
right-hand side. One can write that
\[
\frac{dK^{ij}}{dt}=N^{ij}+2\frac{\partial T_{\rm int}}{\partial
g_{ij}}.
\]
On the Hamiltonian level, this means that the non-conservation of
$K$ even in geodetic motion is due to the fact that the kinetic
energy depends explicitly on the spatial metric tensor. Affine
symmetry of degrees of freedom is broken and reduced to the
Euclidean one.

The system (\ref{3.6}) may be written in the following form:
\begin{equation}\label{3.7}
\frac{dk^{i}}{dt}=F^{i},\qquad
\frac{dK(\mathcal{O})^{ij}}{dt}=m\frac{dx^{i}}{dt}\frac{dx^{j}}{dt}
+\frac{d\varphi^{i}{}_{A}}{dt}\frac{d\varphi^{j}{}_{B}}{dt}J^{AB}
+N(\mathcal{O})^{ij},
\end{equation}
as a balance for the kinematical linear momentum and the total
affine momentum with respect to some space-fixed origin
$\mathcal{O}\in M$.

If the body is rigid in the usual metrical sense, i.e., all
distances between its constituents are constant, then the
d'Alembert principle implies that the second subsystems in
(\ref{3.6}), (\ref{3.7}) are to be replaced by their
skew-symmetric parts, thus,
\[
\frac{dS^{ij}}{dt}=\mathcal{N}^{ij},\qquad
\frac{d\mathcal{J}(\mathcal{O})^{ij}}{dt}=\mathcal{N}(\mathcal{O})^{ij}.
\]
To these equations the rigidity condition, i.e.,
$\eta_{AB}=g_{ij}\varphi^{i}{}_{A}\varphi^{j}{}_{B}$, may be
automatically substituted without paying any attention to reaction
forces responsible for the metrical rigidity.

The above balance laws for kinematical angular momenta become
conservation laws in the interaction-free case, and even under
weaker, realistic conditions that $N$ or $N(\mathcal{O})$ is
symmetric. This is very natural on the Hamiltonian level, because
$S$, $\mathcal{J}(\mathcal{O})$ are then directly related to their
canonical counterparts. The latter are Hamiltonian generators of
the isometry group of ($M,g$), thus, according to the Noether
theorem, they are constants of motion in geodetic problems,
because gyroscopic kinetic energy is isometry-invariant.

Another very convenient balance form of equations of motion is
obtained when generalized velocities $d\varphi^{i}{}_{A}/dt$ are
expressed through the non-holonomic quantities $\Omega^{i}{}_{j}$,
then
\begin{equation}\label{3.8}
\frac{dk^{i}}{dt}=F^{i},\qquad
\frac{dK^{ij}}{dt}=\Omega^{i}{}_{m}K^{mj}+N^{ij}.
\end{equation}
Expressing our balance in co-moving (material) terms we obtain
\begin{equation}\label{3.9}
\frac{d\hat{k}^{A}}{dt}=-\hat{k}^{B}\widetilde{J}_{BC}\hat{K}^{CA}+\hat{F}^{A},
\qquad
\frac{d\hat{K}^{AB}}{dt}=-\hat{K}^{AC}\widetilde{J}_{CD}\hat{K}^{DB}+\hat{N}^{AB},
\end{equation}
or, using non-holonomic velocities,
\[
m\frac{d\hat{v}^{A}}{dt}=-m\hat{\Omega}^{A}{}_{B}\hat{v}^{B}+\hat{F}^{A},
\qquad J^{AC}\frac{d\hat{\Omega}^{B}{}_{C}}{dt}=-
\hat{\Omega}^{B}{}_{D}\hat{\Omega}^{D}{}_{C}J^{CA}+\hat{N}^{AB}.
\]
It is a nice feature of the co-moving representation that all
non-dynamical terms are built only of expressions
$\hat{k}^{A},\hat{K}^{AB}$ or $\hat{v}^{A},\hat{\Omega}^{A}{}_{B}$
without any direct using of mixed quantities like
$\varphi_{A}^{i},d\varphi^{i}{}_{A}/dt$. The second (internal)
subsystems are exactly affine counterparts of gyroscopic Euler
equations and exactly reduce to them when the rigid-body
constraints are imposed. The relationship between two co-moving
forms is based on the equation
$\hat{K}^{AB}=\hat{\Omega}^{B}{}_{C}J^{CA}$ following directly
from the definition of $K$. There is some relationship between
this formula and Legendre transformation for Lagrangians
$L=T-V(x,\varphi)$. Namely, one can show that the internal part of
(\ref{3.2}) may be equivalently written in the following form:
\begin{equation}\label{3.10}
\hat{\Sigma}^{A}{}_{B}=
\hat{K}^{AC}G_{CB}=G_{BC}\hat{\Omega}^{C}{}_{D}J^{DA},
\end{equation}
where $G\in U^{\ast}\otimes U^{\ast}$ denotes as previously the
Green deformation tensor. Therefore, the canonical affine spin is
obtained from the kinematical one by the $G$-lowering of the
second index. As we saw, there was a similar formula (\ref{3.5})
in the spatial representation, i.e.,
\begin{equation}\label{3.11}
\Sigma^{i}{}_{j}=K^{im}g_{mj}.
\end{equation}
It is seen that there is an easy possibility of confusion. Namely
a superficial analogy with the last formula might suggest us to
use the $\eta$-shifting of indices for establishing the Legendre
link between $K$ and $\Sigma$. However, for any reasonable
Lagrangian $\hat{\Sigma}^{A}{}_{B}\neq\hat{K}^{AC}\eta_{CB}$
except the special case of metrically-rigid motion. This is an
additional reason for avoiding ambiguous symbols like
$\hat{\Sigma}^{AB}$ or $\hat{K}^{A}{}_{B}$. More generally, if
some tensor objects in $V$ are related to each other by the
$g$-shifting of indices, then the corresponding co-moving objects
in $U$ are interrelated by the $G$-shifting. And conversely, if
two tensors in $U$ are interrelated by the $\eta$-shift of
indices, then their spatial counterparts in $V$ are obtained from
each other by the $C$-shifting, where $C\in V^{\ast}\otimes
V^{\ast}$ is the Cauchy deformation tensor. The contravariant
inverses of $G$ and $C$ are carefully denoted by $\widetilde{G}
\in U\otimes U$ and $\widetilde{C} \in V\otimes V$, i.e.,
$\widetilde{G}^{AC}G_{CB}=\delta^{A}{}_{B}$,
$\widetilde{C}^{ik}C_{kj}=\delta^{i}{}_{j}$. The notation
$G^{AB}$, $C^{ij}$ would be misleading because of the possible
confusion with the objects $\eta^{AC}\eta^{BD}G_{CD}$,
$g^{ik}g^{jm}C_{km}$ obtained from $G$, $C$ by the usual $\eta$-
or $g$-metrical operations on indices.

For Lagrangian systems $L=T-V(x,\varphi)$ generalized forces $Q$,
$N$ are interrelated by (\ref{3.4}), thus, just as in
(\ref{3.11}), $Q^{i}{}_{j}=N^{im}g_{mj}$. But in the co-moving
representation, in analogy to (\ref{3.10}), we have
$\hat{Q}^{A}{}_{B}=\hat{N}^{AC}G_{CB}$,
$\hat{Q}^{A}{}_{B}\neq\hat{N}^{AC}\eta_{CB}$. This has to do with
different $\varphi$-transformation properties of $Q$, $N$, i.e.,
$Q^{i}{}_{j}=\varphi^{i}{}_{A}\hat{Q}^{A}{}_{B}(\varphi^{-1})^{B}{}_{j}$,
$N^{ij}=\varphi^{i}{}_{A}\varphi^{j}{}_{B}\hat{N}^{AB}$; similarly
for $\Sigma^{i}{}_{j}$, $K^{ij}$, $\hat{\Sigma}^{A}{}_{B}$,
$\hat{K}^{AB}$.

As seen from equations (\ref{3.8}), (\ref{3.9}) even in the
interaction-free case neither $K^{ij}$ nor $\hat{K}^{AB}$ are
constants of motion. The same concerns their canonical
counterparts $\Sigma^{i}{}_{j}$, $\hat{\Sigma}^{A}{}_{B}$. The
reason is that the kinetic energy is not invariant under spatial
and material affine transformations (except translations, of
course). At the same time, purely geodetic Hamiltonian models with
$L=T$ are physically useless because, except of rest-states, all
their trajectories (straight lines in $M\times$ LI$(U,V)$) escape
to infinity. In particular, the body may expand to infinity and
contract in finite time to a point. The metric on $Q=M\times$
LI$(U,V)$ underlying the kinetic energy (\ref{2.1}) is unable to
encode realistic interactions and predict elastic vibrations in
purely geodetic terms.

\section{Dynamical affine invariance}

Basing on the motivation presented in previous sections, we shall
now consider some models which are ruled by affine groups not only
on the kinematical but also on the dynamical level. In particular,
we shall discuss left- and right-invariant Riemann metrics on
linear and affine groups or rather, more precisely, on their
free-action homogeneous spaces. We concentrate on geodetic models,
when there is no potential term and the structure of interactions
is encoded in an appropriately chosen metric tensor on the
configuration space.

Let us begin with the internal sector, when translational degrees
of freedom are frozen and the configuration space reduces to
$Q_{\rm int}=M \times$ LI$(U, V)$, or equivalently to F$(V)$ (when
for simplicity we put $U=\mathbb{R}^{n}$). According to the
transformation rules for $\Omega$, $\hat{\Omega}$, the most
general metric tensor on $Q_{\rm int}$ invariant under the action
of GL$(V)$ through (\ref{2.4}) is that underlying the kinetic
energy form given by
\begin{equation}\label{4.1}
T_{\rm int}=\frac{1}{2}\mathcal{L}^{B}{}_{A}{}^{D}{}_{C}
\hat{\Omega}^{A}{}_{B}\hat{\Omega}^{C}{}_{D},
\end{equation}
where coefficients $\mathcal{L}$ are constant and symmetric in
bi-indices $({}^{B}{}_{A})$, $({}^{D}{}_{C})$. This quadratic form
is also assumed to be non-degenerate, although not necessarily
positively definite. As $\hat{\Omega}$ is a non-holonomic
velocity, i.e., it is not a time derivative of any system of
generalized coordinates, the underlying metric on $Q_{\rm int}$ is
curved.

Quite similarly, the most general kinetic energy invariant under
material affine transformations (\ref{2.5}) has the following
form:
\begin{equation}\label{4.2}
T_{\rm int}=\frac{1}{2}\mathcal{R}^{j}{}_{i}{}^{l}{}_{k}
\Omega^{i}{}_{j}\Omega^{k}{}_{l},
\end{equation}
where $\mathcal{R}$ is also constant and symmetric in bi-indices
$({}^{j}{}_{i})$, $({}^{l}{}_{k})$. The underlying metric tensor
on $Q_{\rm int}$ is also curved, i.e., essentially Riemannian.

In general, (\ref{4.1}) is not right, i.e., materially, invariant
under GL$(U)$ acting through (\ref{2.5}), and (\ref{4.2}) is not
invariant under GL$(V)$ acting through (\ref{2.4}) in the physical
space, i.e., on the left. The exceptional situation of
simultaneous spatial and material invariance leads us to
\[
T_{\rm int}=\frac{A}{2}\Omega^{i}{}_{j}\Omega^{j}{}_{i}
+\frac{B}{2}\Omega^{i}{}_{i}\Omega^{j}{}_{j}=
\frac{A}{2}\hat{\Omega}^{K}{}_{L}\hat{\Omega}^{L}{}_{K}+
\frac{B}{2}\hat{\Omega}^{K}{}_{K}\hat{\Omega}^{L}{}_{L},
\]
where $A$, $B$ are some constants. Using invariant terms, we can
say that such $T_{\rm int}$ is a linear combination of two basic
second-order Casimir invariants, i.e.,
\begin{equation}\label{4.3}
T_{\rm int}=\frac{A}{2}{\rm Tr}(\Omega^{2})+\frac{B}{2}({\rm
Tr\Omega})^{2}=\frac{A}{2}{\rm
Tr}(\hat{\Omega}^{2})+\frac{B}{2}({\rm Tr\hat{\Omega}})^{2}.
\end{equation}
Such $T_{\rm int}$ is never positively-definite. The reason is
that the maximal semisimple subgroups SL$(V)$, SL$(U)$ (their
determinants equal to unity) are non-compact, thus, the quadratic
form ${\rm Tr}(\Omega^{2})={\rm Tr}(\hat{\Omega}^{2})$ has the
hyperbolic signature ($n(n+1)/2\ +$, $n(n-1)/2\ -$), where the
positive contribution corresponds to the "non-compact" dimensions
and the negative one to the "compact" dimensions in GL$(V)$,
GL$(U)$.

By the way, the above quadratic forms reduce to the Killing form
(Killing scalar products) on L$(V)$, L$(U)$
\cite{Her68,Her70,Kob-Nom63} when $A=2n$, $B=-2$. As L$(V)$,
L$(U)$ are non-semisimple, in this special unhappy case the scalar
product (kinetic energy) is degenerate, thus, non-applicable in
usual mechanical problems. The singularity consists of
dilatational Lie algebras $\mathbb{R}{\rm Id}_{V}$,
$\mathbb{R}{\rm Id}_{U}$. More generally, the same holds when
$A=-Bn$. Paradoxically enough, non-degenerate forms (\ref{4.3})
($A\neq-Bn$) may be mechanically useful in spite of their
non-definiteness.

The usual d'Alembert model (\ref{2.1}) invariant under additive
translations (\ref{2.3}) is the special case of general models of
the following form:
\begin{equation}\label{*32}
T_{\rm int}=\frac{1}{2}\mathcal{A}^{K}{}_{i}{}^{L}{}_{j}
\frac{d\varphi^{i}{}_{K}}{dt}\frac{d\varphi^{j}{}_{L}}{dt},
\end{equation}
where $\mathcal{A}$ is constant and symmetric in bi-indices
$({}^{K}{}_{i})$, $({}^{L}{}_{j})$. The peculiarity of (\ref{2.1})
within this class is that $\mathcal{A}$ factorizes, i.e.,
$\mathcal{A}^{K}{}_{i}{}^{L}{}_{j}=g_{ij}J^{KL}$, and is invariant
under the left action of SO$(V,g)$ and the right action of SO$(U,
\widetilde{J})$ (in particular, SO$(U, \eta)$, when the inertia is
isotropic, i.e., $J=\mu\eta$). It is clear that the
$\mathcal{A}$-based models of $T_{\rm int}$ are never invariant
under GL$(V)$, GL$(U)$. The underlying metric on LI$(U, V)$ is
flat.

Let us now consider the translational sector of motion. The only
model of translational kinetic energy invariant under GAf$(M)$
(affine group of $M$) is as follows:
\begin{equation}\label{4.4}
T_{\rm tr}=\frac{m}{2}C_{ij}\frac{dx^{i}}{dt}\frac{dx^{j}}{dt}
=\frac{m}{2}\eta_{AB}\hat{v}^{A}\hat{v}^{B}.
\end{equation}
It looks like the usual kinetic energy, however, there is a very
essential difference. Namely, in the above expression the velocity
vector is not squared with the help of the constant and absolutely
fixed spacial metric  $g \in V^{\ast}\otimes V^{\ast}$. Instead of
it, the Cauchy deformation tensor $C$ is used as an instantaneous
metric tensor of $M$. Being a function of the internal
configuration $\varphi \in$ LI$(U, V)$, it depends on time. It is
so as if the instantaneous internal configuration created a
dynamical metric in an essentially amorphous affine space $M$. In
this sense the model is an oversimplified toy simulation of
general relativity. At the same time it is clear that $T_{\rm tr}$
is not invariant under GL$(U)$ because the fixed material metric
$\eta$ restricts the symmetry to O$(U,\eta)$ ($\eta$-rotations of
$U$).

If we wish the translational kinetic energy to be
GL$(U)$-invariant, then the only reasonable model is the usual
one, based on the fixed $M$-metric $g$, i.e.,
\begin{equation}\label{4.5}
T_{\rm tr}=\frac{m}{2}g_{ij}\frac{dx^{i}}{dt}\frac{dx^{j}}{dt}
=\frac{m}{2}G_{AB}\hat{v}^{A}\hat{v}^{B}.
\end{equation}
It is impossible to construct any model of $T_{\rm tr}$ and
$T=T_{\rm tr}+T_{\rm int}$, which would be purely affine both in
$M$ and $N$; in one of these spaces some metric structure must be
assumed. Therefore, although $T_{\rm int}$ alone may be affine
simultaneously in $M$ and $N$, there are no reasons to stick to
such models, the more so they are never positively-definite. These
problems have to do with the non-existence of a doubly (left- and
right-) invariant pseudo-Riemannian structure on the affine group
GAf$(n,\mathbb{R}) \simeq$ GL$(n,
\mathbb{R})\times_{s}\mathbb{R}^{n}$. Any doubly-invariant twice
covariant tensor field on this group is degenerate. Therefore, it
is reasonable to concentrate on kinetic energies which are affine
in $M$ and $\eta$-metrical in $N$ or, conversely, $g$-metrical in
$M$ and affine in $N$. The corresponding geodetic models in $Q$
have a maximal possible symmetry, being at the same time really
true geodetic problems (non-singular underlying metric). Such
models are special cases of (\ref{4.1}), (\ref{4.2}), (\ref{4.4}),
(\ref{4.5}), thus, we start with some statements concerning the
general case.

For the model (\ref{4.1}), (\ref{4.4}) affine in space and
matrical in the material, Legendre transformation has the form:
\begin{equation}\label{4.6}
\hat{\Sigma}^{A}{}_{B}=\mathcal{L}^{A}{}_{B}{}^{C}{}_{D}\hat{\Omega}^{D}{}_{C},
\quad \hat{p}_{A}=m\eta_{AB}\hat{v}^{B},
\end{equation}
where, obviously, the second equation may be rewritten as
\begin{equation}\label{4.7}
p_{i}=m C_{ij}v^{j}.
\end{equation}
The corresponding geodetic Hamiltonian is given by
$\mathcal{T}=\mathcal{T}_{\rm tr}+\mathcal{T}_{\rm int}$, where
\[
\mathcal{T}_{\rm
tr}=\frac{1}{2m}\eta^{AB}\hat{p}_{A}\hat{p}_{B}=\frac{1}{2m}\widetilde{C}^{ij}p_{i}p_{j},
\qquad \mathcal{T}_{\rm
int}=\frac{1}{2}\widetilde{\mathcal{L}}^{A}{}_{B}{}^{C}{}_{D}
\hat{\Sigma}^{B}{}_{A}\hat{\Sigma}^{D}{}_{C},
\]
and the symmetric bimatrix $\widetilde{\mathcal{L}}$ is reciprocal
to $\mathcal{L}$, i.e.,
\[
\widetilde{\mathcal{L}}^{A}{}_{B}{}^{K}{}_{L}\mathcal{L}^{L}{}_{K}{}^{C}{}_{D}
=\delta^{A}{}_{D}\delta^{C}{}_{B}.
\]

For the model (\ref{4.2}), (\ref{4.5}), metrical in space and
affine in the material, Legendre transformation may be represented
as follows:
\begin{equation}\label{4.8}
\Sigma^{i}{}_{j}=\mathcal{R}^{i}{}_{j}{}^{k}{}_{l}\Omega^{l}{}_{k},
\qquad p_{i}=mg_{ij}v^{j},
\end{equation}
where, in analogy to the $\mathcal{L}$-case, the second subsystem
may be rewritten as
\[
\hat{p}_{A}=mG_{AB}\hat{v}^{B}.
\]
Inverting the Legendre transformation, we obtain geodetic
Hamiltonian $\mathcal{T}=\mathcal{T}_{\rm tr}+\mathcal{T}_{\rm
int}$, where, dually to the $\mathcal{L}$-formulas,
\[
\mathcal{T}_{\rm
tr}=\frac{1}{2m}g^{ij}p_{i}p_{j}=\frac{1}{2m}\widetilde{G}^{AB}\hat{p}_{A}\hat{p}_{B},
\qquad  \mathcal{T}_{\rm
int}=\frac{1}{2}\widetilde{\mathcal{R}}^{i}{}_{j}{}^{k}{}_{l}\Sigma^{j}{}_{i}
\Sigma^{l}{}_{k}.
\]
Obviously again $\widetilde{\mathcal{R}}$ denotes the inverse
bimatrix of $\mathcal{R}$, i.e.,
$\widetilde{\mathcal{R}}^{a}{}_{b}{}^{k}{}_{l}\mathcal{R}^{l}{}_{k}{}^{j}{}_{i}
=\delta^{a}{}_{i}\delta^{j}{}_{b}$.

Let us admit non-geodetic models of the form $L=T-V(x,\varphi)$,
$H=\mathcal{T}+V(x,\varphi)$, where $V(x,\varphi)$ is a usual
potential energy depending only on the indicated configuration
variables. Then the balance equations for $\mathcal{L}$-models
(affine in space, metrical in the material) read:
\begin{equation}\label{4.9}
\frac{dp_{i}}{dt}=Q_{i},\qquad
\frac{d\Sigma^{i}{}_{j}}{dt}=-\frac{1}{m}\widetilde{C}^{ik}p_{k}p_{j}+
Q^{i}{}_{j},
\end{equation}
with the same meaning of generalized forces as previously, i.e.,
\[
Q_{i}=-\frac{\partial V}{\partial x^{i}},\qquad
Q^{i}{}_{j}=-\varphi^{i}{}_{A}\frac{\partial V}{\partial
\varphi^{j}{}_{A}}.
\]
When taken together with the Legendre transformation, the above
balance laws are equivalent to the Hamilton canonical equations.
They may be also generalized so as to include some
non-Hamiltonian, e.g., dissipative terms on their right-hand
sides. It is seen that in the purely geodetic case (when
$Q_{i}=0$, $Q^{i}{}_{j}=0$) the canonical linear momentum is
conserved, but the affine spin is not so due to the first term on
the right-hand side of the balance law for $\Sigma$. The reason is
that $\Sigma^{i}{}_{j}$ generate linear transformations of
internal degrees of freedom; these transformations do not affect
translational variables. Therefore, the full affine symmetry is
broken, and $\Sigma$ is not a constant of motion. But one can
reformulate the balance laws (\ref{4.9}), just as in the
d'Alembert model, by introducing the total canonical affine
momentum $I(\mathcal{O})$, related to some fixed origin
$\mathcal{O} \in M$,
\[
I(\mathcal{O})^{i}{}_{j}:=
\Lambda(\mathcal{O})^{i}{}_{j}+\Sigma^{i}{}_{j}=x^{i}p_{j}+\Sigma^{i}{}_{j}.
\]
The first term refers to the translational motion, the second one
to the relative (internal) motion. Something similar may be done
for generalized forces,
\[
Q_{\rm tot}(\mathcal{O})^{i}{}_{j}:=Q_{\rm
tr}(\mathcal{O})^{i}{}_{j}+ Q^{i}{}_{j}=x^{i}Q_{j}+Q^{i}{}_{j}.
\]
Then the system of balance equations (\ref{4.9}) may be written in
the following equivalent form:
\begin{equation}\label{4.10}
\frac{dp_{i}}{dt}=Q_{i},\qquad
\frac{dI(\mathcal{O})^{i}{}_{j}}{dt}=Q_{\rm
tot}(\mathcal{O})^{i}{}_{j}.
\end{equation}
It is seen that in the geodetic case one obtains conservation laws
for $p_{i}$, $I(\mathcal{O})^{i}{}_{j}$, i.e., for the system of
generators of the spatial affine group GAf$(M)$, just as expected.

Let us observe some funny feature of our geodetic equations.
Namely, the canonical linear momentum is a constant of motion, but
the translational velocity is not. This is because the Legendre
transformation (\ref{4.7}) implies that the translational motion
is influenced by internal phenomena. Except some special solutions
even the direction of translational velocity is non-constant and
depends on what happens with "internal" degrees of freedom. This
is a kind of "drunk missile" effect. Something similar occurs in
the dynamics of defects in solids \cite{Kos-Zor67}. It is also
non-excluded that the non-conservation of velocity might be an
over-simplified model of certain specially-relativistic phenomena
(internal motion results in changes of internal energy, and
therefore, in the rest mass pulsations; but the latter ones
influence the effective inertia, and therefore, the translational
motion). One can show that the time-rate of translational velocity
may be expressed as follows:
\[
m\frac{dv^{a}}{dt}=-
\widetilde{C}^{aj}\frac{dC_{jb}}{dt}v^{b}+F^{a}=mv^{b}(\Omega^{a}{}_{b}
+\widetilde{C}^{ad}\Omega^{m}{}_{d}C_{mb})+F^{a},
\]
where the contravariant force $F$ is given by the following
expression: $F^{a}=\widetilde{C}^{ab}Q_{b}$ (because of the
formerly mentioned reasons, we avoid denoting $F^{a}$ as $Q^{a}$).
It is explicitly seen that $v$ is variable even in the purely
geodetic motion. The $\mathcal{L}$-based geodetics in $M \times$
LI$(U, V)$ do not project onto straight lines in $M$.

Let us now consider the $\mathcal{R}$-based models (\ref{4.2}),
(\ref{4.5}), metrical in space and affine in the material. As
mentioned, they are somehow related to the
Arnold-Ebin-Marsden-Binz approach to the dynamics of ideal fluids
\cite{Abr-Mars78,Arn78,Binz91,Ebin77,Ebin_Mars70,Mars81,Mars-Hugh83}.
Our Poisson brackets imply that the balance form of equations of
motion may be expressed as follows:
\begin{equation}\label{4.11}
\frac{dp_{a}}{dt}=Q_{a},\qquad \frac{d
\hat{\Sigma}^{A}{}_{B}}{dt}=\hat{Q}^{A}{}_{B} ,
\end{equation}
again with
\[
Q_{a}=-\frac{\partial V}{\partial x^{a}}, \qquad
\hat{Q}^{A}{}_{B}=-\frac{\partial V}{\partial
\varphi^{i}{}_{A}}\varphi^{i}{}_{B}=(\varphi^{-1})^{A}{}_{i}
Q^{i}{}_{j}\varphi^{j}{}_{B}
\]
in the potential case. In geodetic models $p_{a}$,
$\hat{\Sigma}^{A}{}_{B}$ are conserved quantities as explicitly
seen from the balance equations and expected on the basis of
invariance properties. Indeed, the $\mathcal{R}$-model of $T$ is
invariant under the Abelian group of translations in $M$.
Therefore, $p_{a}$ are constants of motion as Hamiltonian
generators of this group. Similarly, as seen from our Poisson
brackets, $\hat{\Sigma}^{A}{}_{B}$ Poisson-commute with $p_{a}$
and $\Sigma^{i}{}_{j}$, therefore, with the total geodetic
Hamiltonian $\mathcal{T}$. This is due to the explicitly obvious
invariance of the $\mathcal{R}$-based $T$ under the group of
material linear transformations GL$(U)$. Surprisingly enough, the
co-moving components of linear momentum,
$\hat{p}_{A}=p_{i}\varphi^{i}{}_{A}$, are not constants of motion.
But this is also clear because the material space $N$ has in our
model a distinguished point, i.e., the Lagrangian position of the
centre of mass. Because of this, translations in $N$ fail to be
symmetries and their Hamiltonian generators $\hat{p}_{A}$ are
non-conserved. According to the structure of Legendre
transformation, translational velocity
$v^{a}=dx^{a}/dt=g^{ab}p_{b}$ is also a constant of motion, just
as $p$ itself. Therefore, in geodetic $\mathcal{R}$-models, the
geodesic curves in $M \times$ LI$(U, V)$ project to $M$ onto
straight lines swept with constant velocities (uniform motions).
This means that there is no "drunk missile effect" and
contravariant representation of the translational balance takes on
the usual form:
\[
m\frac{dv^{a}}{dt}=m\frac{d^{2}x^{a}}{dt^{2}}=F^{a}, \qquad
F^{a}=g^{ab}Q_{b}.
\]

As previously, the balance laws (\ref{4.11}) become a closed
system of equations of motion when considered jointly with the
Legendre transformation (\ref{4.8}). Let us observe that the
structure of (\ref{4.11}) is in a sense less "aesthetical" than
that of (\ref{4.9}) because it is more non-homogeneous. The point
is that in (\ref{4.9}) both subsystems are written in terms of
spatial objects, whereas in (\ref{4.11}) one uses the mixed
representation: spatial for the translational motion and material
for the internal one. Obviously, (\ref{4.11}) may be done
symmetric, dual to (\ref{4.9}), by substituting
$\hat{p}_{A}=p_{i}\varphi^{i}{}_{A}$. But this immediately makes
the translational equation more complicated.

There is also another problem. The simplicity of our balance laws
(conservation laws in the geodetic case) is rather illusory. The
point is that, as mentioned above, the total system of equations
of motion consists of the balance laws and Legendre
transformation. The balance (\ref{4.9}), (\ref{4.10}) looks simple
in Euler (spatial) representation, but the corresponding Legendre
transformation is simple in Lagrangian (material) representation
(\ref{4.6}). And quite conversely, the internal part of
(\ref{4.11}) is simple in the co-moving terms, but its Legendre
transformation is simple when expressed in the spatial (Eulerian)
form (\ref{4.8}).

One can easily show that the internal parts of Legendre
transformations (\ref{4.6}), (\ref{4.8}) may be respectively
expressed as follows:
\begin{equation}
\Sigma^{i}{}_{j}=\check{\mathcal{L}}^{i}{}_{j}{}^{k}{}_{l}
\Omega^{l}{}_{k},\qquad \label{4.12}
\hat{\Sigma}^{A}{}_{B}=\hat{\mathcal{R}}^{A}{}_{B}{}^{C}{}_{D}
\hat{\Omega}^{D}{}_{C},\label{4.13}
\end{equation}
where
\[
\check{\mathcal{L}}^{i}{}_{j}{}^{k}{}_{l}
=\varphi^{i}{}_{A}(\varphi^{-1})^{B}{}_{j}\varphi^{k}{}_{C}
(\varphi^{-1})^{D}{}_{l}\mathcal{L}^{A}{}_{B}{}^{C}{}_{D},
\]
\[
\hat{\mathcal{R}}^{A}{}_{B}{}^{C}{}_{D}
=(\varphi^{-1})^{A}{}_{i}\varphi^{j}{}_{B}(\varphi^{-1})^{C}{}_{k}
\varphi^{l}{}_{D}\mathcal{R}^{i}{}_{j}{}^{k}{}_{l}.
\]
Obviously, this form is rather complicated because the
coefficients at $\Omega$ and $\hat{\Omega}$ are non-constant; they
depend on the internal configuration $\varphi$. Simplicity of the
balance laws is incompatible with simplicity of Legendre
transformations.

As mentioned, when translational degrees of freedom are taken into
account, there are no sensible models which would be affine
simultaneously in space and in the material. The highest symmetry
of mathematical interest and at the same time physically
reasonable is that affine in space and rotational in the material,
and conversely, Euclidean in space and (centro-)affine in the
body. The latter model is an over-simplified discretization of
dynamical systems on diffeomorphisms group as used in
hydrodynamics and elasticity.

In materially isotropic $\mathcal{L}$-models the quantity
$\mathcal{L}^{A}{}_{B}{}^{C}{}_{D}$ is a linear combination of
tensors $\eta^{AC}\eta_{BD}$, $\delta^{A}{}_{D}\delta^{C}{}_{B}$,
$\delta^{A}{}_{B}\delta^{C}{}_{D}$. Similarly, in spatially
isotropic $\mathcal{R}$-models the tensor
$\mathcal{R}^{i}{}_{j}{}^{k}{}_{l}$ is a linear combination of
terms $g^{ik}g_{jl}$, $\delta^{i}{}_{l}\delta^{k}{}_{j}$,
$\delta^{i}{}_{j}\delta^{k}{}_{l}$. Therefore, (\ref{4.1}),
(\ref{4.2}) take on, respectively, the following forms:
\begin{eqnarray}
T_{\rm
int}&=&\frac{I}{2}\eta_{KL}\hat{\Omega}^{K}{}_{M}\hat{\Omega}^{L}{}_{N}\eta^{MN}+
\frac{A}{2}\hat{\Omega}^{K}{}_{L}\hat{\Omega}^{L}{}_{K}+
\frac{B}{2}\hat{\Omega}^{K}{}_{K}\hat{\Omega}^{L}{}_{L},\label{4.14}\\
T_{\rm
int}&=&\frac{I}{2}g_{ik}\Omega^{i}{}_{j}\Omega^{k}{}_{l}g^{jl}+
\frac{A}{2}\Omega^{i}{}_{j}\Omega^{j}{}_{i}+
\frac{B}{2}\Omega^{i}{}_{i}\Omega^{j}{}_{j},\label{4.15}
\end{eqnarray}
where the constants $I$, $A$, $B$ are generalized internal inertia
scalars. It is clear that if $I=0$, then these expressions become
identical. The $I$-terms break the centro-affine symmetry in $U$
and $V$, and restrict it to the metrical one, respectively, in the
sense of metric tensors $\eta$ or $g$. The first term in
(\ref{4.14}), just as (\ref{4.4}), may be expressed in terms of
the Cauchy deformation tensor, i.e.,
\begin{equation}\label{4.16}
\frac{I}{2}C_{ij}\frac{d\varphi^{i}{}_{A}}{dt}
\frac{d\varphi^{j}{}_{B}}{dt}\eta^{AB}.
\end{equation}

Let us observe that the isotropic inertial tensor $I\eta^{AB}$ in
(\ref{4.16}) might be replaced by the general one,
\begin{equation}\label{4.17}
\frac{1}{2}C_{ij}\frac{d\varphi^{i}{}_{A}}{dt}
\frac{d\varphi^{j}{}_{B}}{dt}J^{AB}=
\frac{1}{2}\eta_{KL}\hat{\Omega}^{K}{}_{A}\hat{\Omega}^{L}{}_{B}J^{AB}.
\end{equation}
This expression is structurally similar to the d'Alembert formula
(\ref{2.1}). The difference is that the fixed metric $g$ is
replaced by the $\varphi$-dependent Cauchy tensor $C$. There is
not only formal similarity but also some asymptotic correspondence
between (\ref{4.17}) and (\ref{2.1}). Obviously, for the general
$J$, (\ref{4.17}) is not metrically isotropic and its internal
symmetry is reduced to O$(U, \eta) \cap {\rm O}(U,
\widetilde{J})$. The $I$-terms in (\ref{4.14}), (\ref{4.15}) are
positively definite if $I>0$. Moreover, the total expressions
(\ref{4.14}), (\ref{4.15}) are positively definite for some open
range of $(I,A,B)\in\mathbb{R}^{3}$. Roughly speaking, the
absolute values of $A$, $B$ must be "sufficiently small" in
comparison with $I$.

The internal part of Legendre transformation (\ref{4.6}) for
$\mathcal{L}$-models becomes now (i.e., for (\ref{4.14}))
\begin{equation}\label{4.18}
\hat{\Sigma}^{K}{}_{L}=I\eta^{KM}\eta_{LN}\hat{\Omega}^{N}{}_{M}+
A\hat{\Omega}^{K}{}_{L}+B\delta^{K}{}_{L}\hat{\Omega}^{M}{}_{M}.
\end{equation}
This may be alternatively written as follows:
$\Sigma^{i}{}_{j}=I\widetilde{C}^{ib}C_{ja}\Omega^{a}{}_{b}+A\Omega^{i}{}_{j}+
B\delta^{i}{}_{j}\Omega^{m}{}_{m}$, it is the very special case of
(\ref{4.12}).

The inverse Legendre transformation has the same structure, i.e.,
\begin{equation}\label{4.19}
\hat{\Omega}^{K}{}_{L}=\frac{1}{\widetilde{I}}\eta^{KM}\eta_{LN}\hat{\Sigma}^{N}{}_{M}
+\frac{1}{\widetilde{A}}\hat{\Sigma}^{K}{}_{L}+
\frac{1}{\widetilde{B}}\delta^{K}{}_{L}\hat{\Sigma}^{M}{}_{M},
\end{equation}
where $\widetilde{I}=\left(I^{2}-A^{2}\right)/I$,
$\widetilde{A}=\left(A^{2}-I^{2}\right)/A$,
$\widetilde{B}=-\left(I+A\right)\left(I+A+nB\right)/B$. When
written in Eulerian (spatial) terms, this formula becomes
\[
\Omega^{i}{}_{j}=\frac{1}{\widetilde{I}}\widetilde{C}^{ib}C_{ja}\Sigma^{a}{}_{b}
+\frac{1}{\widetilde{A}}\Sigma^{i}{}_{j}+
\frac{1}{\widetilde{B}}\delta^{i}{}_{j}\Sigma^{m}{}_{m},
\]
with the same as previously meaning of modified inertial
coefficients $\widetilde{I}$, $\widetilde{A}$, $\widetilde{B}$.

Similarly, for $\mathcal{R}$-models based on (\ref{4.15}) the
internal sector of Legendre transformation has the following form
\begin{equation}\label{4.20}
\Sigma^{i}{}_{j}=Ig^{im}g_{jn}\Omega^{n}{}_{m}+A\Omega^{i}{}_{j}+
B\delta^{i}{}_{j}\Omega^{m}{}_{m},
\end{equation}
which is inverted as
\begin{equation}\label{4.21}
\Omega^{i}{}_{j}=\frac{1}{\widetilde{I}}g^{im}g_{jn}\Sigma^{n}{}_{m}
+\frac{1}{\widetilde{A}}\Sigma^{i}{}_{j}
+\frac{1}{\widetilde{B}}\delta^{i}{}_{j}\Sigma^{m}{}_{m}.
\end{equation}
The co-moving representation of these formulas is given by the
following expressions:
\begin{eqnarray}
\hat{\Sigma}^{K}{}_{L}&=&I\widetilde{G}^{KM}G_{LN}\hat{\Omega}^{N}{}_{M}
+A\hat{\Omega}^{K}{}_{L}+B\delta^{K}{}_{L}\hat{\Omega}^{M}{}_{M},
\nonumber\\
\hat{\Omega}^{K}{}_{L}&=&\frac{1}{\widetilde{I}}\widetilde{G}^{KM}G_{LN}\hat{\Sigma}^{N}{}_{M}
+\frac{1}{\widetilde{A}}\hat{\Sigma}^{K}{}_{L}+
\frac{1}{\widetilde{B}}\delta^{K}{}_{L}\hat{\Sigma}^{M}{}_{M}.\nonumber
\end{eqnarray}
The general balance laws (\ref{4.9}), (\ref{4.11}) considered
jointly with these Legendre transformations (including the obvious
translational sector) provide the complete system of equations of
motion (naturally, the definitions of $\Omega$, $\hat{\Omega}$ are
to be substituted). These equations, due to the very special
structure of (\ref{4.14}), (\ref{4.15}) are relatively readible
and effective. At the same time one can show that for
incompressible bodies even in the purely geodetic case ($Q_{i}=0$,
$Q^{i}{}_{j}=0$) there exists an open set of solutions which are
bounded in the internal configuration space LI$(U,V)$, so the
elastic vibrations may be encoded in the very kinetic energy
(Riemann structure) without the explicite use of forces.

After substituting the above inverse of Legendre transformations
to kinetic energies (\ref{4.14}), (\ref{4.15}), we obtain the
following geodetic Hamiltonians of internal motion:
\begin{eqnarray}
\mathcal{T}_{\rm
int}&=&\frac{1}{2\widetilde{I}}\eta_{KL}\hat{\Sigma}^{K}{}_{M}
\hat{\Sigma}^{L}{}_{N}\eta^{MN}+\frac{1}{2\widetilde{A}}\hat{\Sigma}^{K}{}_{L}
\hat{\Sigma}^{L}{}_{K}+\frac{1}{2\widetilde{B}}\hat{\Sigma}^{K}{}_{K}
\hat{\Sigma}^{L}{}_{L},\label{4.22}\\
\mathcal{T}_{\rm
int}&=&\frac{1}{2\widetilde{I}}g_{ik}\Sigma^{i}{}_{j}
\Sigma^{k}{}_{l}g^{jl}+\frac{1}{2\widetilde{A}}\Sigma^{i}{}_{j}
\Sigma^{j}{}_{i}+\frac{1}{2\widetilde{B}}\Sigma^{i}{}_{i}
\Sigma^{j}{}_{j}.\label{4.23}
\end{eqnarray}
In certain problems it may be convenient to write down the first
formula in spatial terms; similarly, the second one may be
expressed with the use of co-moving representation. Therefore, we
obtain, respectively, the following expressions:
\begin{eqnarray}
\mathcal{T}_{\rm
int}&=&\frac{1}{2\widetilde{I}}C_{kl}\Sigma^{k}{}_{m}
\Sigma^{l}{}_{n}\widetilde{C}^{mn}+\frac{1}{2\widetilde{A}}\Sigma^{k}{}_{l}
\Sigma^{l}{}_{k}+\frac{1}{2\widetilde{B}}\Sigma^{k}{}_{k}
\Sigma^{l}{}_{l},\nonumber\\
\mathcal{T}_{\rm
int}&=&\frac{1}{2\widetilde{I}}G_{KL}\hat{\Sigma}^{K}{}_{M}
\hat{\Sigma}^{L}{}_{N}\widetilde{G}^{MN}+\frac{1}{2\widetilde{A}}\hat{\Sigma}^{K}{}_{L}
\hat{\Sigma}^{L}{}_{K}+\frac{1}{2\widetilde{B}}\hat{\Sigma}^{K}{}_{K}
\hat{\Sigma}^{L}{}_{L}.\nonumber
\end{eqnarray}
The corresponding velocity-based formulas (\ref{4.14}),
(\ref{4.15}) for kinetic energy may be written in an analogous
way. Simply $1/\widetilde{I}$, $1/\widetilde{A}$,
$1/\widetilde{B}$ in the last expressions are to be replaced by
$I$, $A$, $B$, and simultaneously one must substitute
$\Omega^{k}{}_{l}$, $\hat{\Omega}^{K}{}_{L}$ instead of
$\Sigma^{k}{}_{l}$, $\hat{\Sigma}^{K}{}_{L}$.

For certain purposes it is convenient to rewrite geodetic
Hamiltonians (\ref{4.22}), (\ref{4.23}) in an alternative form:
\begin{eqnarray}
\mathcal{T}_{\rm int}&=&\frac{1}{2\alpha}{\rm
Tr}(\hat{\Sigma}^{2})+ \frac{1}{2\beta}({\rm
Tr}\hat{\Sigma})^{2}-\frac{1}{4\mu}{\rm Tr}(V^{2}),\label{4.24}
\\
\mathcal{T}_{\rm int}&=&\frac{1}{2\alpha}{\rm Tr}(\Sigma^{2})+
\frac{1}{2\beta}({\rm Tr}\Sigma)^{2}-\frac{1}{4\mu}{\rm
Tr}(S^{2}),\label{4.25}
\end{eqnarray}
where $\alpha:=I+A$, $\beta:=-(I+A)(I+A+nB)/B$,
$\mu:=(I^{2}-A^{2})/I$, and $V$, $S$ denote, respectively, the
vorticity and spin given by (\ref{2.6}). It is clear that the only
distinction between expressions (\ref{4.24}) and (\ref{4.25}) is
that in their third terms, thus, we can rewrite them concisely
like
\begin{eqnarray}
\mathcal{T}_{\rm int}&=&\frac{1}{2\alpha}C(2)+
\frac{1}{2\beta}C(1)^{2}-\frac{1}{4\mu}{\rm
Tr}(V^{2}),\label{4.26}
\\
\mathcal{T}_{\rm int}&=&\frac{1}{2\alpha}C(2)+
\frac{1}{2\beta}C(1)^{2}-\frac{1}{4\mu}{\rm
Tr}(S^{2}),\label{4.27}
\end{eqnarray}
where $C(k)$ denotes the $k$-th order Casimir expression built of
generators, i.e., $C(k):={\rm Tr}(\Sigma^{k})={\rm
Tr}(\hat{\Sigma}^{k})$.

By analogy with the physical 3-dimensional case the quantity
$-{\rm Tr}(S^{2})$ may be interpreted as the doubled squared norm
of the internal angular momentum, $-{\rm Tr}(S^{2})=2\|S\|^{2}$.
Similarly, $-{\rm Tr}(V^{2})=2\|V\|^{2}$. Therefore, the formulas
(\ref{4.26}), (\ref{4.27}) may be respectively written in the
following intuitive and suggestive way:
\begin{eqnarray}
\mathcal{T}_{\rm int}&=&\frac{1}{2\alpha}C(2)+
\frac{1}{2\beta}C(1)^{2}+\frac{1}{2\mu}\|V\|^{2},\label{4.28}\\
\mathcal{T}_{\rm int}&=&\frac{1}{2\alpha}C(2)+
\frac{1}{2\beta}C(1)^{2}+\frac{1}{2\mu}\|S\|^{2}.\label{4.29}
\end{eqnarray}
Obviously, for the model (\ref{4.3}) invariant under the left and
right action of linear groups we have
\begin{equation}\label{4.30}
\mathcal{T}_{\rm int}=\frac{1}{2A}C(2)+
\frac{1}{2A(n+A/B)}C(1)^{2}.
\end{equation}

When performing computations, it is convenient to use orthogonal
coordinates, $\eta_{KL}=\delta_{KL}$, $g_{ij}=\delta_{ij}$, and
rewrite some of the above formulas in terms of purely matrix
operations. Thus, (\ref{4.14}), (\ref{4.15}) become, respectively,
\begin{eqnarray}
T_{\rm int}&=&\frac{I}{2}{\rm Tr}(\hat{\Omega}^{T}\hat{\Omega})+
\frac{A}{2}{\rm Tr}(\hat{\Omega}^{2})+\frac{B}{2}({\rm
Tr}\hat{\Omega})^{2}, \\
T_{\rm int}&=&\frac{I}{2}{\rm Tr}(\Omega^{T}\Omega)+
\frac{A}{2}{\rm Tr}(\Omega^{2})+\frac{B}{2}({\rm Tr}\Omega)^{2}.
\end{eqnarray}
Obviously, the second and third terms in these formulas are
pairwise equal because ${\rm Tr}(\hat{\Omega}^{2})={\rm
Tr}(\Omega^{2})$ and ${\rm Tr}(\hat{\Omega})={\rm Tr}(\Omega)$.

Legendre transformations (\ref{4.18}), (\ref{4.20}) and their
inverses (\ref{4.19}), (\ref{4.21}) are as follows:
\[
\hat{\Sigma}=I\hat{\Omega}^{T}+A\hat{\Omega}+B({\rm
Tr}\hat{\Omega})I_{n}, \qquad \Sigma=I\Omega^{T}+A\Omega+B({\rm
Tr}\Omega)I_{n},
\]
\[
\hat{\Omega}=\frac{1}{\widetilde{I}}\hat{\Sigma}^{T}+
\frac{1}{\widetilde{A}}\hat{\Sigma}+\frac{1}{\widetilde{B}}({\rm
Tr}\hat{\Sigma})I_{n}, \qquad
\Omega=\frac{1}{\widetilde{I}}\Sigma^{T}+
\frac{1}{\widetilde{A}}\Sigma+\frac{1}{\widetilde{B}}({\rm
Tr}\Sigma)I_{n},
\]
where $I_{n}$ denotes the $n$-th order identity matrix.

Similarly, for the kinetic Hamiltonians (\ref{4.22}), (\ref{4.23})
we have, respectively,
\begin{eqnarray}
\mathcal{T}_{\rm int}&=&\frac{1}{2\widetilde{I}}{\rm
Tr}(\hat{\Sigma}^{T}\hat{\Sigma})+ \frac{1}{2\widetilde{A}}{\rm
Tr}(\hat{\Sigma}^{2})+\frac{1}{2\widetilde{B}}({\rm
Tr}\hat{\Sigma})^{2}, \\
\mathcal{T}_{\rm int}&=&\frac{1}{2\widetilde{I}}{\rm
Tr}(\Sigma^{T}\Sigma)+ \frac{1}{2\widetilde{A}}{\rm
Tr}(\Sigma^{2})+\frac{1}{2\widetilde{B}}({\rm Tr}\Sigma)^{2}.
\end{eqnarray}

This matrix representation is very lucid and useful in
calculations. Nevertheless, in comparison with the systematic
tensor language, it may be risky and misleading because it
obscures the geometric meaning of symbols and concepts. And this
is worse than the lack of aesthetics; when no care is taken, this
may lead simply to logical and numerical mistakes.

We finish this section with some geometric remarks.

Kinetic energy $T$ of a non-relativistic mechanical system is
equivalent to some Riemannian structure $\Gamma$ on its
configuration space $Q$. In terms of generalized coordinates and
velocities, we have that
\[
T=\frac{1}{2}\Gamma_{\alpha\beta}(q)\frac{dq^{\alpha}}{dt}
\frac{dq^{\beta}}{dt}, \qquad \Gamma=\Gamma_{\alpha\beta}(q)
dq^{\alpha}\otimes dq^{\beta}.
\]
Although usually somehow related to the metric tensor $g$ of the
physical space $M$, $\Gamma$ need not be directly interpretable in
terms of geometrical distances in $M$. As a rule, it depends not
only on $g$ but also on certain parameters characterizing inertial
properties of the system, i.e., masses, inertial moments, etc.

It is instructive to describe explicitly in a bit more geometric
form the metric tensors $\Gamma$ on $Q=M\times{\rm LI}(U,V)$
underlying the kinetic energies defined above. For this purpose it
is convenient to introduce auxiliary geometric objects.

Let $E_{A}$, $e_{i}$ denote the basic vectors in $U$, $V$
underlying our Lagrange and Euler coordinates $a^{K}$, $x^{j}$.
The corresponding dual basic covectors in $U^{\ast}$, $V^{\ast}$
will be denoted as usual by $E^{A}$, $e^{i}$. Generalized
coordinates in $Q$ will be, as previously, denoted by $x^{i}$,
$\varphi^{i}{}_{A}$, and no misunderstandings just simplifications
follow from using the same symbol $x^{i}$ for coordinates in $M$
and for their pull-backs to $Q$. Now, we introduce two families of
Pfaff forms (differential one-forms) on $Q$, i.e.,
$\hat{\omega}^{A}{}_{B}:=(\varphi^{-1})^{A}{}_{i}d\varphi^{i}{}_{B}$,
$\omega^{i}{}_{j}:=d\varphi^{i}{}_{A}(\varphi^{-1})^{A}{}_{j}$.
These basic systems depend on the choice of bases $E$, $e$, but
this is, so to speak, a covariant dependence. In other words the
L$(U)$- and L$(V)$-valued one-forms
$\hat{\omega}:=\hat{\omega}^{A}{}_{B}E_{A}\otimes E^{B}$,
$\omega:=\omega^{i}{}_{j}e_{i}\otimes e^{j}$ are base-independent.

In addition, we shall need the following two families of Pfaff
forms: $\hat{\theta}^{A}=(\varphi^{-1})^{A}{}_{i}dx^{i}$,
$\theta^{i}=\varphi^{i}{}_{A}\hat{\theta}^{A}=dx^{i}$. Just as
previously, they give rise to the objective base-independent $U$-
and $V$-valued differential one-forms:
$\hat{\theta}=\hat{\theta}^{A}E_{A}$, $\theta=\theta^{i}e_{i}$.

The base-independent objects $\hat{\omega},\hat{\theta}$ and
$\omega,\theta$ could be defined without any use of bases,
however, the above definitions are technically simplest. The above
objects are closely related to the concept of affine velocity in
co-moving (Lagrange) and spatial (Euler) representations. Namely,
for any history $\mathbb{R}\ni t\mapsto (x(t),\varphi(t))$, the
quantities $\hat{\Omega}^{A}{}_{B}$, $\Omega^{i}{}_{j}$ are
evaluations of $\hat{\omega}^{A}{}_{B}$, $\omega^{i}{}_{j}$ on the
tangent vectors (general velocities) given analytically by
$dx^{i}/dt,d\varphi^{i}{}_{A}/dt$. Roughly speaking, we could say
that
\[
\hat{\Omega}^{A}{}_{B}=\frac{\hat{\omega}^{A}{}_{B}}{dt},\qquad
\Omega^{i}{}_{j}=\frac{\omega^{i}{}_{j}}{dt}.
\]
This is obviously a kind of joke, but fully justified on the basis
of infinitesimal Leibniz notation. Similarly, the co-moving and
spatial components of translational velocity are given by
evaluations of $\hat{\theta}^{A}$ and $\theta^{i}$ on tangent
vectors, and using the same trick we could say that
\[
\hat{v}^{A}=\frac{\hat{\theta}^{A}}{dt},\qquad
v^{i}=\frac{\theta^{i}}{dt}.
\]

At any point of the configuration space the systems
$\theta^{i},\omega^{k}{}_{l}$ and
$\hat{\theta}^{A},\hat{\omega}^{K}{}_{L}$ provide two bases in the
corresponding space of covariant vectors. We could ask for the
corresponding bases of contravariant vector spaces. It is
convenient to use the language of contemporary differential
geometry, where vector fields $X$ with components $X^{i}$ (meant
in the sense of some local coordinates $q^{i}$) are identified
with first-order differential operators of directional derivatives
$\nabla_{X}$, i.e., $X=X^{i}\partial/\partial q^{i}$. One can
easily show that the bases $\hat{H}_{K},\hat{E}^{A}{}_{B}$ and
$H_{k},E^{a}{}_{b}$, dual respectively to
$\hat{\theta}^{K},\hat{\omega}^{A}{}_{B}$ and
$\theta^{k},\omega^{a}{}_{b}$, are given by the following
differential operators:
\[
\hat{H}_{K}=\varphi^{i}{}_{K}\frac{\partial}{\partial
x^{i}},\qquad H_{k}=\frac{\partial}{\partial x^{k}},\qquad
\hat{E}^{A}{}_{B}=\varphi^{k}{}_{B}\frac{\partial}{\partial
\varphi^{k}{}_{A}},\qquad
E^{a}{}_{b}=\varphi^{a}{}_{K}\frac{\partial}{\partial
\varphi^{b}{}_{K}}.
\]

The general $\mathcal{L}$-models (\ref{4.1}), (\ref{4.4}) are
based on metric tensors of the following form:
\[
\Gamma=m\eta_{AB}\hat{\theta}^{A}\otimes\hat{\theta}^{B}+
\mathcal{L}^{B}{}_{A}{}^{D}{}_{C}\hat{\omega}^{A}{}_{B}\otimes
\hat{\omega}^{C}{}_{D}.
\]
Similarly, for $\mathcal{R}$-models (\ref{4.2}), (\ref{4.5}) we
have
\[
\Gamma=mg_{ij}\theta^{i}\otimes\theta^{j}+
\mathcal{R}^{j}{}_{i}{}^{l}{}_{k}\omega^{i}{}_{j}\otimes
\omega^{k}{}_{l}.
\]
The corresponding contravariant metrics underlying the kinetic
Hamiltonians are given, respectively, for $\mathcal{L}$-models by
\[
\widetilde{\Gamma}=\frac{1}{m}\eta^{AB}\hat{H}_{A}\otimes
\hat{H}_{B}+
\widetilde{\mathcal{L}}^{B}{}_{A}{}^{D}{}_{C}\hat{E}^{A}{}_{B}\otimes
\hat{E}^{C}{}_{D}
\]
and for $\mathcal{R}$-models by
\[
\widetilde{\Gamma}=\frac{1}{m}g^{ij}H_{i}\otimes H_{j}+
\widetilde{\mathcal{R}}^{j}{}_{i}{}^{l}{}_{k}E^{i}{}_{j}\otimes
E^{k}{}_{l}.
\]

If the kinetic energy of internal motion (\ref{4.3}) is invariant
simultaneously under GL$(V)$ and GL$(U)$, then the corresponding
metric tensor on LI$(U,V)$ is given by
\[
\Gamma_{\rm
int}^{0}=A\hat{\omega}^{K}{}_{L}\otimes\hat{\omega}^{L}{}_{K}+
B\hat{\omega}^{K}{}_{K}\otimes\hat{\omega}^{L}{}_{L}=
A\omega^{k}{}_{l}\otimes\omega^{l}{}_{k}+
B\omega^{k}{}_{k}\otimes\omega^{l}{}_{l}
\]
and its inverse by
\[
\widetilde{\Gamma}_{\rm
int}^{0}=\frac{1}{A}\hat{E}^{K}{}_{L}\otimes\hat{E}^{L}{}_{K}-
\frac{B}{A(A+nB)}\hat{E}^{K}{}_{K}\otimes\hat{E}^{L}{}_{L}
\]
\[
=\frac{1}{A}E^{k}{}_{l}\otimes E^{l}{}_{k}-
\frac{B}{A(A+nB)}E^{k}{}_{k}\otimes E^{l}{}_{l}.
\]
Here $\Gamma_{\rm int}^{0}$ is a linear combination of two
Casimir-like objects built of Pfaff forms $\omega$ in a quadratic
way. As already mentioned, $\Gamma_{\rm int}^{0}$ becomes the
group-theoretic Killing tensor when $A=2n$, $B=-2$. This is just
the pathological situation to be excluded because for $A/B=-n$ the
tensor $\Gamma_{\rm int}^{0}$ is singular.

For our models affine in space and metrical in the body we have
that
\[
\Gamma=m\eta_{KL}\hat{\theta}^{K}\otimes\hat{\theta}^{L}+
I\eta_{KL}\eta^{MN}\hat{\omega}^{K}{}_{M}\otimes\hat{\omega}^{L}{}_{N}+
\Gamma_{\rm int}^{0}.
\]
Similarly, for models metrical in space and affine in the body:
\[
\Gamma=mg_{ij}\theta^{i}\otimes\theta^{j}+
Ig_{ik}g^{jl}\omega^{i}{}_{j}\otimes\omega^{k}{}_{l}+\Gamma_{\rm
int}^{0}.
\]
The corresponding contravariant (reciprocal) metrics are given by
\[
\widetilde{\Gamma}=\frac{1}{m}\eta^{KL}\hat{H}_{K}\otimes\hat{H}_{L}+
\frac{1}{\widetilde{I}}\eta_{KL}\eta^{MN}\hat{E}^{K}{}_{M}\otimes\hat{E}^{L}{}_{N}+
\frac{1}{\widetilde{A}}\hat{E}^{K}{}_{L}\otimes\hat{E}^{L}{}_{K}+
\frac{1}{\widetilde{B}}\hat{E}^{K}{}_{K}\otimes\hat{E}^{L}{}_{L}
\]
for spatially affine models and
\[
\widetilde{\Gamma}=\frac{1}{m}g^{ij}H_{i}\otimes H_{j}+
\frac{1}{\widetilde{I}}g_{ik}g^{jl}E^{i}{}_{j}\otimes E^{k}{}_{l}+
\frac{1}{\widetilde{A}}E^{i}{}_{j}\otimes E^{j}{}_{i}+
\frac{1}{\widetilde{B}}E^{k}{}_{k}\otimes E^{l}{}_{l}
\]
for materially affine models. Obviously, the last two terms in
both expressions coincide. The kinetic (geodetic) terms of
Hamiltonians for geodetic and potential systems are based on
$\widetilde{\Gamma}$-tensors, namely,
\[
\mathcal{T}=\frac{1}{2}\widetilde{\Gamma}^{\mu\nu}(q)p_{\mu}p_{\nu},
\]
where
$\widetilde{\Gamma}^{\mu\kappa}\Gamma_{\kappa\nu}=\delta^{\mu}{}_{\nu}$
and $p_{\mu}=\partial L/\partial \dot{q}^{\mu}=\partial T/\partial
\dot{q}^{\mu}$ are canonical momenta conjugate to $q^{\mu}$ (dual
objects to generalized velocities $\dot{q}^{\mu}$).

The above objects $\omega$, $\theta$ and $E$, $H$ possess natural
generalizations to curved manifolds with affine connection. They
appear there, respectively, as the connection form, canonical
form, fundamental vector fields, and standard horizontal vector
fields on F$M$ (the principle fibre bundle of linear frames in a
manifold $M$) \cite{Kob-Nom63}. Such a formalism is used in
mechanics of infinitesimal affinely-rigid bodies, when affine
degrees of freedom are considered as internal ones, i.e., attached
to material points moving in manifolds with curvature and torsion
\cite{JJS03}.

\section{Without translational motion}

It is instructive to consider the simplified situation when the
centre of mass is at rest and the covariant translational forces
do vanish, i.e, $Q_{i}=0$. For the "usual" d'Alembert model such a
situation, characteristic for practical elastic problems, was
discussed detailly in our earlier papers
\cite{JJS73_2,JJS75_1,JJS75_2,JJS82_2,JJS82_1,JJS86,JJS88_1}. And
from the purely geometric symmetry point of view nothing
particularly interesting happened there due to this
simplification. In our model, based on dynamical affine
symmetries, the translationless situation is an important step of
the general analysis.

We have mentioned that in spatially affine $\mathcal{L}$-models
with $Q_{i}=0$, in particular in geodetic ones, canonical linear
momentum is a constant of motion but translational velocity is not
(except some very special solutions). This violates our ideas
about Galileian symmetry, at least in the form developed in the
"usual" d'Alembert mechanics. Nevertheless, the concept of
translationless motion is well-defined because in the usual
potential models equations $v^{i}=0$, $p_{i}=0$ are equivalent;
this is one of exceptional cases when the constancy of velocity
does not contradict the constancy of linear momentum. One must
only remember that the Galilei transforms (in the usual sense) of
such space-resting solutions will not be solutions any longer.

In $\mathcal{L}$-models without translational motion the evolution
is ruled by the second of the balance laws (\ref{4.9}) with the
simplified right-hand side, i.e.,
\[
\frac{d\Sigma^{i}{}_{j}}{dt}=Q^{i}{}_{j}.
\]
Affine invariance in $M$ implies that in the completely geodetic
case this becomes simply the Noether conservation law:
\[
\frac{d\Sigma^{i}{}_{j}}{dt}=0,
\]
i.e., affine spin in spatial representation is a constant of
motion. As mentioned, to obtained a closed system of equations,
one must consider the above balance (conservation) jointly with
the Legendre transformation and the definition of affine velocity.
Unfortunately, the nice form (\ref{4.6}) with constant
coefficients cannot be used because $\hat{\Sigma}^{A}{}_{B}$ is
not a constant of motion in the geodetic case. And in general,
(\ref{4.12}) is too complicated to be effectively used. But it
turns out that something may be done for our simplified model
(\ref{4.14}), affine in $M$ and $\eta$-metrical in $U$.

Something similar may be said about $\mathcal{R}$-models,
moreover, they are in some respects simpler. The balance equations
(\ref{4.11}) reduce to their internal parts, i.e.,
\[
\frac{d\hat{\Sigma}^{A}{}_{B}}{dt}=\hat{Q}^{A}{}_{B},
\]
and become conservation laws for the co-moving affine spin in the
geodetic case, i.e.,
\[
\frac{d\hat{\Sigma}^{A}{}_{B}}{dt}=0.
\]
As previously, the simplicity is only seeming one because the
laboratory components $\Sigma^{i}{}_{j}$ fail to be constants of
motion and Legendre transformation expressed in co-moving terms
(\ref{4.13}), in general, is rather complicated. Fortunately, for
our models (\ref{4.15}), metrical in space and centro-affine in
the material, also something may be done.

Let us begin with the over-simplified model with $I=0$, affine
both in the spatial and material sense. It is easily seen that the
general solution for translation-free geodetic motion is then
given by the system of orbits of one-parameter subgroups of
GL$(V)$ or, equivalently, one-parameter subgroups of GL$(U)$,
i.e.,
\begin{equation}\label{5.1}
\varphi(t)=\exp(Et)\varphi_{0}=\varphi_{0}\exp(\hat{E}t),
\end{equation}
where $\varphi_{0}$ is an arbitrary element of LI$(U,V)$, $E$ is
an arbitrary element of L$(V)=$ GL$(V)^{\prime}$, and
$\hat{E}=\varphi^{-1}_{0}E\varphi_{0}$ is the corresponding
element of L$(U)=$ GL$(U)^{\prime}$ obtained by the
$\varphi_{0}$-similarity. If we identify formally $U$ and $V$ with
$\mathbb{R}^{n}$ (by the particular choice of bases), then the
phase portrait consists of all one-parameter subgroups of
GL$(n,\mathbb{R})$ and of all their left cosets or, equivalently,
of all their right cosets. One must only remember that although
the sets of left and right cosets coincide, they are parameterized
in a different way by the corresponding generators and initial
shifting elements. The reason is that GL$(n,\mathbb{R})$ is
non-Abelian and, in general, its one-parameter subgroups are not
normal. If we write the group-theoretical version of (\ref{5.1}),
i.e.,
\[
g(t)=\exp(at)h=h\exp(h^{-1}aht),
\]
it is seen that the coinciding left and right cosets usually refer
to different generators $a$ and $h^{-1}ah$, thus, to different
subgroups. If some left and right cosets refer to the same
subgroup, i.e., the same generator $a$, and have non-empty
intersection, then, as a rule, they are different subsets, i.e.,
\[
g_{1}(t)=\exp(at)h\neq h\exp(at)=g_{2}(t).
\]
Only the dilatational subgroup is exceptional because, consisting
of central elements, it is a normal subgroup, and $h^{-1}ah=a$ for
any dilatation generator $a$; the previous inequality becomes
equality for any $h\in$ GL$(n,\mathbb{R})$.

Let us notice that in the solution (\ref{5.1}) the pairs
$\varphi_{0},E$ and $\varphi_{0},\hat{E}$ play the role of
differently represented initial conditions; in this sense they
label the general solution. Thus, $\varphi_{0}=\varphi(0)$ is an
initial configuration, whereas $E=\Omega(0)$,
$\hat{E}=\hat{\Omega}(0)$ are initial and at the same time
permanently constant values of the laboratory and co-moving affine
velocities. Therefore, the initial values of generalized
velocities are given by
$\dot{\varphi}(0)=E\varphi_{0}=\varphi_{0}\hat{E}$.

It is seen that for $I=0$ the structure of general solution
resembles that of the spherical rigid body. It is so for every
geodetic model on a semisimple group or its trivial central
extension if the kinetic energy is doubly (left and right)
invariant \cite{Arn78}. But we should remember that even in the
simple case of a free anisotropic rigid body situation changes
drastically. Kinetic energy is invariant under left regular
translations on SO$(3,\mathbb{R})$ (identified with the
configuration space) but no longer under right translations. As a
rule, one-parameter subgroups and their cosets fail to be
solutions, i.e., they are not geodetics of left-invariant metric
tensors on SO$(3,\mathbb{R})$. There are some exceptions, however,
namely stationary rotations
\cite{Arn78,Mars-Rat94,Mars-Rat99,Rat82}. They happen when one of
main axes of inertia has a fixed orientation in space and the
remaining two perform a uniform rotation about it with a fixed
angular velocity. Thus, there is a subset of general solution
given by three one-parameter subgroups and all their left cosets
(the non-moving axis of inertia may be arbitrarily oriented in
space). This is the special case of what is known as relative
equilibria \cite{Abr-Mars78,Mars-Rat94,Mars-Rat99}. They
correspond to critical points of geodetic Hamiltonians restricted
to co-adjoint orbits in the dual space of the Lie algebra
SO$(3,\mathbb{R})^{\prime}\simeq$ SO$(3,\mathbb{R})$
\cite{Abr-Mars78,Mars-Rat94,Mars-Rat99}. Such particular
solutions, although do not exhaust the phase portrait, contain an
important information about its structure.

Something similar happens in our affine models when $I\neq 0$. The
general solution is not any longer given by orbits of
one-parameter subgroups. Nevertheless, there exist geometrically
interesting orbits which are particular solutions, i.e.,
generalized equilibria.

Let us begin with geodetic $\mathcal{L}$-models (\ref{4.14})
left-invariant under GL$(V)$ and right-invariant under
O$(U,\eta)$. One can show after some calculations that there exist
solutions of the following form:
\begin{equation}\label{5.2}
\varphi(t)=\varphi_{0}\exp(Ft),
\end{equation}
where the initial configuration $\varphi_{0}\in$ LI$(U,V)$ is
arbitrary just as in (\ref{5.1}). But now $F\in$ L$(U)\simeq$
GL$(U)^{\prime}$ is not arbitrary any longer, instead it must be
$\eta$-normal in the sense of commuting with its $\eta$-transpose,
i.e., $F^{A}{}_{C}\eta^{CD}F^{E}{}_{D}\eta_{EB}-
\eta^{AD}F^{E}{}_{D}\eta_{EC}F^{C}{}_{B}=0$. Introducing the
$\eta$-transpose symbol,
\begin{equation}\label{5.3}
(F^{\eta T})^{A}{}_{B}:=\eta^{AC}F^{D}{}_{C}\eta_{DB},
\end{equation}
we can write the above condition in the following concise form:
\begin{equation}\label{5.4}
[F,F^{\eta T}]=FF^{\eta T}-F^{\eta T}F=0.
\end{equation}
It is obvious that for such solutions affine velocities are
constant and given by
\[
\Omega=\varphi_{0}F\varphi^{-1}_{0},\qquad \hat{\Omega}=F.
\]
The initial data $\varphi_{0}$, $F$ are independent of each other.
The only restriction is that of $\eta$-normality imposed on $F$
alone. This holds, in particular, in two extremely opposite
special cases when $F$ is $\eta$-skew-symmetric or
$\eta$-symmetric, i.e.,
\[
F^{\eta T}=-F,\qquad F^{\eta T}=F.
\]
In the skew-symmetric case the one-parameter group generated by
$F$ consists of $\eta$-rotations, i.e.,
\[
\exp(Ft)\in {\rm SO}(U,\eta)\subset {\rm GL}(U).
\]
If $F$ is $\eta$-symmetric, then so are transformations
$\exp(Ft)$; they describe pure deformations in $U$ in the sense of
$\eta$-polar decomposition.

In calculations one identifies usually $U$ and $V$ with
$\mathbb{R}^{n}$ and their metrics $\eta$, $g$ with the Kronecker
delta. Then the solutions (\ref{5.2}) become all possible left
cosets of one-parameter subgroups of GL$(n,\mathbb{R})$ generated
by all possible normal matrices $F\in$ L$(n,\mathbb{R})$, i.e.,
$[F,F^{T}]=FF^{T}-F^{T}F=0$ (in this formula we mean the usual
matrix transposition).

Following (\ref{5.1}) we can try to rewrite (\ref{5.2}) in terms
of the left-acting one-parameter subgroups. It is easy to see that
\begin{equation}\label{5.5}
\varphi(t)=\varphi_{0}\exp(Ft)=\exp(\check{F}t)\varphi_{0},
\end{equation}
where $\check{F}=\varphi_{0}F\varphi_{0}^{-1}\in$ L$(V)=$
GL$(V)^{\prime}$.

In this representation $\varphi_{0}$ is still arbitrary but
$\check{F}$ is subject to some restrictions following from
(\ref{5.4}) and depending on $\varphi_{0}$. Namely, $\check{F}$ is
normal in the sense of the Cauchy deformation tensor
$C[\varphi_{0}]$ used as a kind of metric in $V$, i.e.,
\[
\check{F}^{i}{}_{a}C[\varphi_{0}]^{ak}\check{F}^{l}{}_{k}C[\varphi_{0}]_{lj}
-C[\varphi_{0}]^{ik}\check{F}^{l}{}_{k}C[\varphi_{0}]_{la}\check{F}^{a}{}_{j}=0.
\]
Introducing in analogy to (\ref{5.3}) the
$C[\varphi_{0}]$-transpose of $\check{F}$, i.e.,
\[
(\check{F}^{C[\varphi_{0}]T})^{i}{}_{j}:=
\widetilde{C}[\varphi_{0}]^{ik}\check{F}^{l}{}_{k}C[\varphi_{0}]_{lj},
\]
we can write simply
\begin{equation}\label{5.6}
[\check{F},\check{F}^{C[\varphi_{0}]T}]=0.
\end{equation}

Therefore, in the right-cosets representation the initial
configuration $\varphi_{0}$ and the generator $\check{F}$ are
mutually interrelated. Namely, if $\varphi_{0}$ is not subject to
any restrictions, then $\check{F}$ satisfies the condition
(\ref{5.6}) explicitly depending on $\varphi_{0}$. And conversely,
if $\check{F}$ is arbitrary, then the initial conditions of
$\varphi_{0}$ must be so suited to any particular choice of
$\check{F}$ that the commutator condition (\ref{5.6}) is
non-violated.

Let us observe that in all $\mathcal{L}$-models the spatial metric
$g$ does not occur in expressions for the kinetic energy at all.
Thus, as a matter of fact it does not need to exist at all and the
physical space $M$ may be purely affine. Only the material metric
$\eta$ in the body is essential for (\ref{4.14}). Let us notice,
however, that if both $g\in V^{\ast}\otimes V^{\ast}$ and $\eta
\in U^{\ast}\otimes U^{\ast}$ are fixed, then some family of
special solutions may be distinguished, for which the relationship
between initial configurations and infinitesimal generators is
simpler and more symmetric. Namely, we can start from the very
beginning with the representation
\[
\varphi(t)=\exp(Et)\varphi_{0},
\]
where $E$ and $\varphi_{0}$ are respectively some elements of
L$(V)$ and LI$(U,V)$. It is easy to see that, when some metric $g$
is fixed in $V$, then we have at disposal a very natural family of
solutions assuming that $\varphi_{0}\in$ LI$(U,V)$ is an isometry
and $E$ is $g$-normal,
\begin{equation}\label{5.7}
[E,E^{gT}]=EE^{gT}-E^{gT}E=0,
\end{equation}
where in a full analogy to the previous expression we use the
definition
\[
\left(E^{gT}\right)^{i}{}_{j}:=g^{ik}E^{l}{}_{k}g_{lj}.
\]
Obviously, such solutions form a submanifold of the family
(\ref{5.5}), (\ref{5.6}) because then $C[\varphi_{0}]=g$.

Now let us consider geodetic $\mathcal{R}$-models (\ref{4.15}),
which are left-invariant only under O$(V,g)$ but right-invariant
under the total GL$(U)$. Now, as expected, the situation will be
reversed. Let us assume solutions in the right-coset form:
\[
\varphi(t)=\exp(Et)\varphi_{0},
\]
where $\varphi_{0}\in$ LI$(U,V)$, $E\in$ L$(V)$. It is easy to
show that such a curve (right coset) satisfies, in fact, equations
of geodetic motion if $E$ is $g$-normal, just as in (\ref{5.7}),
but now $\varphi_{0}$ may be quite arbitrary isomorphism of $U$
onto $V$. And if we write the above curve as a left coset, i.e.,
\[
\varphi(t)=\varphi_{0}\exp(\widetilde{E}t),\qquad
\widetilde{E}=\varphi_{0}^{-1}E\varphi_{0}\in {\rm L}(U),
\]
then it is easy to see that, with a still arbitrary $\varphi_{0}$,
$\widetilde{E}$ will be $G[\varphi_{0}]$-normal in the sense of
Green deformation tensor $G[\varphi_{0}]\in U^{\ast}\otimes
U^{\ast}$, i.e.,
\[
[\widetilde{E},\widetilde{E}^{G[\varphi_{0}]T}]=\widetilde{E}\widetilde{E}^{G[\varphi_{0}]T}-
\widetilde{E}^{G[\varphi_{0}]T}\widetilde{E}=0,
\]
where the $G[\varphi_{0}]$-transpose is defined in a full analogy
to the above $\eta$- and $g$-transposes,
\[
(\widetilde{E}^{G[\varphi_{0}]T})^{A}{}_{B}:=
G[\varphi_{0}]^{AC}\widetilde{E}^{D}{}_{C}G[\varphi_{0}]_{DB}.
\]

Just as previously, we can distinguish an interesting submanifold
of such solutions when some material metric tensor $\eta\in
U^{\ast}\otimes U^{\ast}$ is fixed (we know it does not exist in
(\ref{4.15})). They are given by curves of the following form:
\[
\varphi(t)=\varphi_{0}\exp(Ft),
\]
where $\varphi_{0}\in$ O$(U,\eta;V,g)$ is an isometry and $F\in$
L$(U)$ is $\eta$-normal in the sense of (\ref{5.3}), (\ref{5.4}).
For such solutions we have $G[\varphi_{0}]=\eta$. Manipulating
with $\eta$ we introduce some kind of parametrization, ordering in
our manifold of relative equilibria.

The above particular solutions are very special, nevertheless very
important. Their position in our model is analogous to that of
stationary rotations in rigid body mechanics. They provide a kind
of skeleton for the general solution. Nevertheless, some, at least
qualitative, rough knowledge of the phase portrait would be mostly
welcome. The crucial question is to what extent the purely
geodetic models may predict bounded motions. Obviously, this is
impossible for compressible bodies, when the configuration space
is identical with the total LI$(U,V)$. To see this it is
sufficient to consider the special case $n=1$, when
compressibility is the only degree of freedom of internal motion.
There is only one affinely-invariant model of $T_{\rm int}$. The
resulting trivial geodetic model predicts, depending on the sign
of the initial internal velocity, either the infinite expansion or
contraction, although in the latter case the object shrinks to a
single point after infinite time. The only bounded (and
non-stable) solution is the rest state. Something similar occurs
in $n$-dimensional geodetic problems. Namely, degrees of freedom
of the isochoric motion are orthogonal to the pure dilatations and
completely independent of them. Some purely geometric comments are
necessary here. Namely, if $N$ and $M$ are purely amorphous affine
spaces, in particular no metrics $\eta$, $g$ are fixed in $U$,
$V$, then their volume measures are defined only up to
multiplicative constant factors. They are Lebesgue measures, i.e.,
special cases of Haar measures invariant under additive Abelian
translations in $U$, $V$ (in $N$, $M$). Let us denote some
particular choices respectively by $\nu_{U}$, $\nu_{V}$.
Obviously, for any measurable domain $Y\subset U$ and for any
configuration $\varphi\in$ LI$(U,V)$ we have
\[
\nu_{V}(\varphi(Y))=\Delta(\varphi)\nu_{U}(Y).
\]
The scalar multiplicator $\Delta(\varphi)$ depends on $\varphi$
and on non-correlated normalizations of $\nu_{V}$, $\nu_{U}$ but
does not depend on $Y$. Obviously, for any $A\in$ GL$(U)$, $B\in$
GL$(V)$ we have
\[
\Delta(A\varphi B)=(\det A)\Delta(\varphi)\det B.
\]
The motion is isochoric if $\Delta$ is constant in the course of
evolution. Obviously, this definition is independent of particular
normalizations of $\nu_{V}$, $\nu_{U}$. The manifold LI$(U,V)$
becomes then foliated by $(n^{2}-1)$-dimensional leaves consisting
of mutually non-compressed configurations. Every such leaf
establishes holonomic constraints, and the total foliation is what
is sometimes referred to as semi-holonomic or quasi-holonomic
constraints. If some metric tensors $\eta\in U^{\ast}\otimes
U^{\ast}$, $g\in V^{\ast}\otimes V^{\ast}$ are fixed, then the
measures $\nu_{U}$, $\nu_{V}$ may be fixed respectively as
$\nu_{\eta}$, $\nu_{g}$, and in terms of coordinates
\[
d\nu_{\eta}=\sqrt{\det[\eta_{KL}]}da^{1}\cdots da^{n},\qquad
d\nu_{g}=\sqrt{\det[g_{ij}]}dx^{1}\cdots dx^{n}.
\]
Using Euclidean coordinates we can simply put
$\Delta(\varphi)=\det[\varphi^{i}{}_{K}]$ but, obviously, this
convention fails for general coordinates. For non-Euclidean affine
coordinates we have
\[
\Delta(\varphi)=\frac{\sqrt{\det[g_{kl}]}}
{\sqrt{\det[\eta_{AB}]}}\det[\varphi^{i}{}_{K}].
\]
Let us remind that the corresponding curvilinear formula reads
\cite{Erin62}
\begin{equation}\label{5.8}
\frac{d\nu_{g}(x(a))}{d\nu_{\eta}(a)}=\frac{\sqrt{\det[g_{kl}(x(a))]}}
{\sqrt{\det[\eta_{AB}(a)]}}\det\left[\frac{\partial
x^{i}}{\partial a^{K}}\right].
\end{equation}

If some volumes are fixed in $U$ and $V$, e.g., due to some
choices of metrics $\eta$, $g$, then the volume extension ratio
$\Delta(\varphi)$ is uniquely fixed. In certain formulas it may be
convenient to use the additive parameter $\alpha(\varphi)$ instead
of the multiplicative one, i.e.,
$\Delta(\varphi)=\exp\left[\alpha(\varphi)\right]$. Another
convenient dilatation measures are those describing the linear
size extension ratio,
\[
D(\varphi)=\sqrt[n]{\Delta(\varphi)}=\exp\left[\frac{\alpha(\varphi)}{n}\right]=
\exp\left[q(\varphi)\right].
\]

The only possibility of stabilizing dilatations is to include some
extra potential preventing the unlimited expansion to the infinite
size and asymptotic contraction to the point-like object. There is
plenty of such phenomenological modelling potentials, e.g., \[
V_{\rm dil}=\frac{\kappa}{8}\left(D^{2}+D^{-2}-2\right)=
\frac{\kappa}{8}\left({\rm ch}2q-1\right),\qquad \kappa >0.
\]
Obviously, this potential is positively infinite at $q=\mp \infty$
($D=0$, $D=+\infty$) and has the global stable equilibrium at
$q=0$ ($D=1$), where it behaves as the harmonic oscillator:
$V_{\rm dil}(q)\approx \kappa q^{2}/2$ for $q\approx 0$. For
strongly extended bodies it also behaves harmonically in the
$D$-variable sense. Another phenomenological model would be just
the global form $\kappa q^{2}/2$. One can also try to use some toy
models predicting "dissociation" of the body (its unlimited
size-expansion), unlimited collapse, or both of them above some
threshold of the total dilatational energy, e.g., \[ V_{\rm
dil}(q)=\frac{\kappa}{2}\left({\rm th}^{2}q-1\right).
\]
In certain problems it may be reasonable to use phenomenological
models preventing contraction but admitting dissociation.

In quantized version of the theory one can stabilize dilatations
in an easy way with the use of the $q$-variable potential well
(perhaps with the infinite walls) concentrated around $q=0$
($D=1$).

If we identify analytically $U$ and $V$ with $\mathbb{R}^{n}$ and
LI$(U,V)$ with GL$(n,\mathbb{R})$, then it is clear that the
connected component of unity GL$^{+}(n,\mathbb{R})$ becomes the
direct product GL$^{+}(n,\mathbb{R})\simeq$
SL$(n,\mathbb{R})\times \exp(\mathbb{R})=$ SL$(n,\mathbb{R})\times
\mathbb{R}^{+}$; the second group factor is obviously meant in the
multiplicative sense, as GL$^{+}(1,\mathbb{R})$. It describes pure
dilatations, whereas SL$(n,\mathbb{R})$ refers to the isochoric
motion. Without this identification, LI$(U,V)$ may be represented
as the Cartesian product of any of the aforementioned leaves (of
mutually non-compressed configurations) and the multiplicative
group $\mathbb{R}\backslash \{0\}$. If some volume-standards
$\nu_{U}$, $\nu_{V}$ (e.g., metric-based ones $\nu_{\eta}$,
$\nu_{g}$) and orientations are fixed in $U$, $V$, then
LI$^{+}(U,V)$, i.e., the manifold of orientation-preserving
isomorphisms, is identified with the product
SLI$(U,\nu_{U};V,\nu_{V})\times \exp(\mathbb{R})$, where,
obviously, the first term consists of transformations $\varphi$
for which $\Delta(\varphi)=1$, i.e., $q(\varphi)=0$. Such a
formulation is more correct from the point of view of geometrical
purity, however, for our purposes (qualitative discussion of the
general solution), the analytical identification of LI$^{+}(U,V)$
with GL$^{+}(n,\mathbb{R})\simeq$ SL$(n,\mathbb{R})\times
\exp(\mathbb{R})$ is sufficient and, as a matter of fact, more
convenient. In any case, qualitative analysis of the general
solution (bounded and non-bounded trajectories) is not then
obscured by cosmetical aspects of geometry. Thus, from now on the
internal configuration space $Q_{\rm int}=$ LI$(U,V)$ will be
identified with GL$^{+}(n,\mathbb{R})\simeq$
SL$(n,\mathbb{R})\times \exp(\mathbb{R})$. Any matrix $\varphi\in$
GL$^{+}(n,\mathbb{R})$ will be uniquely represented as
$\varphi=l\Psi=\exp(q)\Psi$, where $\Psi\in {\rm
SL}(n,\mathbb{R})$. It is convenient to introduce the following
shear velocities:
\[
\omega:=\frac{d\Psi}{dt}\Psi^{-1},\qquad
\hat{\omega}:=\Psi^{-1}\frac{d\Psi}{dt}.
\]
Obviously, $\omega,\hat{\omega}\in$ SL$(n,\mathbb{R})^{\prime}$,
i.e., they are trace-less. Then affine velocities may be expressed
as follows:
\[
\Omega=\omega+\frac{dq}{dt}I,\qquad \hat{\Omega}=\hat{\omega}
+\frac{dq}{dt}I,
\]
where, obviously, $I$ denotes the identity matrix.

Analogously, the affine spin splits as follows:
\[
\Sigma=\sigma+\frac{p}{n}I,\qquad
\hat{\Sigma}=\hat{\sigma}+\frac{p}{n}I,\qquad
\sigma,\hat{\sigma}\in {\rm SL}(n,\mathbb{R})^{\prime},
\]
where $p$ denotes the dilatational canonical momentum. The
velocity-momentum pairing becomes ${\rm Tr}(\Sigma\Omega)={\rm
Tr}(\hat{\Sigma}\hat{\Omega})= {\rm
Tr}(\sigma\omega)+p\dot{q}={\rm
Tr}(\hat{\sigma}\hat{\omega})+p\dot{q}$.

Poisson-bracket relations for $\sigma$-components are based on the
structure constants of SL$(n,\mathbb{R})$. The same is obviously
true for $\hat{\sigma}$ with the proviso that the signs are
reversed. The mixed $\{\sigma,\hat{\sigma}\}$ brackets do vanish.
Obviously, $\{q,p\}=1$, and the quantities $q$, $p$ (dilatation)
Poisson-commute with $\Psi$, $\sigma$, $\hat{\sigma}$ (shear).

The doubly-invariant "kinetic energy" (\ref{4.3}) is a
superposition of the isochoric and dilatational terms,
\[
T=\frac{A}{2}{\rm Tr}(\omega^{2})+\frac{n(A+nB)}{2}\dot{q}^{2}=
T_{\rm sh}+T_{\rm dil}.
\]
Performing the Legendre transformation, $\sigma=A\omega$,
$p=n(A+nB)\dot{q}$, we obtain the following geodetic Hamiltonian:
\begin{equation}\label{5.9}
\mathcal{T}=\frac{1}{2A}{\rm
Tr}(\sigma^{2})+\frac{1}{2n(A+nB)}p^{2}= \mathcal{T}_{\rm
sh}+\mathcal{T}_{\rm dil}.
\end{equation}
In these expressions the quantities $\omega$, $\sigma$ may be
replaced by their co-moving representations $\hat{\omega}$,
$\hat{\sigma}$. Lagrangians and Hamiltonians of systems with
stabilized (controlled) dilatations have the following forms:
\[
L=L_{\rm sh}+L_{\rm dil}=T_{\rm sh}+T_{\rm dil}-V(q), \qquad
H=H_{\rm sh}+H_{\rm dil}=\mathcal{T}_{\rm sh}+\mathcal{T}_{\rm
dil}+V(q).
\]
There is a complete separability of shear and dilatation degrees
of freedom; they are mutually independent. This property would not
be violated if we included also a shear potential $V_{\rm
sh}(\Psi)$, i.e., if $V(\Psi,q)=V_{\rm sh}(\Psi)+V_{\rm dil}(q)$.
The question arises as to the structure of general solution for
the geodetic SL$(n,\mathbb{R})$-model, i.e., for $V_{\rm sh}=0$. A
superficial reasoning based on the analogy with d'Alembert models
might suggest that the general solution consists only of unbounded
motions (and the rest states), because there is no potential and
the configuration space is non-compact. However, it is not the
case; there is an open subset consisting of bounded orbits.

Indeed, let us assume that some trace-less matrix $\alpha\in$
SL$(n,\mathbb{R})^{\prime}$ is similar to an anti-symmetric matrix
$\lambda\in$ SO$(n,\mathbb{R})^{\prime}$, i.e., there exists such
$\chi\in$ SL$(n,\mathbb{R})$ that $\alpha=\chi\lambda\chi^{-1}$.
Then every motion $\Psi(t)=\exp(\alpha t)\Psi_{0}$ is bounded.
Indeed, $\exp(\lambda t)$ is a bounded subgroup of
SO$(n,\mathbb{R})\subset$ SL$(n,\mathbb{R})$ and so is
$\exp(\alpha t)=\chi\exp(\lambda t)\chi^{-1}$. But similarities
are globally defined continuous mappings, therefore, they
transform bounded subsets onto bounded ones. Let us observe that
for physical dimensions $n=2,3$ motions of this type are periodic.
For higher dimensions periodicity is not necessary, although
obviously possible. To see this, let us consider the simplest
situation $n=4$ and represent $\mathbb{R}^{4}$ as the direct sum
of two complementary $\mathbb{R}^{2}$-subspaces. Now, let
$\lambda\in$ SO$(n,\mathbb{R})^{\prime}$ be a block matrix
consisting of two skew-symmetric blocks
$\nu_{1}\left[\begin{tabular}{cc}
  0 & $-1$ \\
  1 & 0
\end{tabular}\right]$ and
$\nu_{2}\left[\begin{tabular}{cc}
  0 & $-1$ \\
  1 & 0
\end{tabular}\right]$. Rotations generated by separate blocks are
obviously periodic, but the total motion is periodic if and only
if $\nu_{1}/\nu_{2}\in\mathbb{Q}$, i.e., the ratio of angular
velocities is a rational number. If it is irrational, the subgroup
obtained by exponentiation of $t$-multiples of the above block
matrix is not a periodic function of the parameter $t$. It is not
a closed subset at all; its closure is a two-dimensional
submanifold of SO$(4,\mathbb{R})\subset$ SL$(4,\mathbb{R})$.
Therefore, being an algebraic subgroup of
SO$(4,\mathbb{R})\subset$ SL$(4,\mathbb{R})$ it is not its Lie
subgroup in the literal sense. The same concerns any subgroup of
SL$(4,\mathbb{R})$ obtained from the above one by a similarity
transformation. For an arbitrary $n$, solutions of this kind are
matrix-valued almost periodic functions of the time variable $t$.

If $\alpha\in$ SL$(n,\mathbb{R})^{\prime}$ is similar to a
symmetric matrix $\kappa\in$ SO$(n,\mathbb{R})^{\prime}$,
$\alpha=\chi\kappa\chi^{-1}$, then, obviously, the motion given by
$\Psi(t)=\exp(\alpha t)\Psi_{0}$ is unbounded. But one can show
that the previously described bounded almost periodic solutions
are "stable" in such a sense that for any skew-symmetric $\lambda$
there exists some open range of symmetric $\kappa$-s such that for
$\alpha=\lambda+\kappa$, or, more generally, for similar matrices
$\alpha=\chi(\lambda+\kappa)\chi^{-1}$, the corresponding
solutions $\Psi(t)=\exp(\alpha t)\Psi_{0}$ are also bounded,
although not necessarily almost periodic \cite{RWL02,Wul-Rob02}
(and not necessarily periodic in dimensions $n=2,3$). The
arbitrariness of pairs $(\alpha,\Psi_{0})$ is sufficient for the
corresponding family of bounded solutions to be open in the
general solution manifold, thus, $2(n^{2}-1)$-dimensional
(topological and differential concepts like openness and dimension
are meant in the sense of the manifold of initial conditions). Let
us observe that this statement would be false for solutions with
generators $\alpha$ similar to skew-symmetric matrices. At first
look, this might seem strange, because the structure of
SL$(n,\mathbb{R})^{\prime}$ implies that the family of such
$\alpha$-s is $(n^{2}-1)$-dimensional and so is the set of initial
configurations $\Psi_{0}$. But these data are not independent and
mutually interfere in the formula $\Psi(t)=\exp(\alpha
t)\Psi_{0}$. Therefore, the very interesting subfamily of
almost-periodic solutions is a proper subset of the manifold of
all bounded solutions.

By "anti-analogy", for symmetric matrices $\lambda=\lambda^{T}\in$
SL$(n,\mathbb{R})^{\prime}$ the corresponding solutions
$\Psi(t)=\exp(\lambda t)\Psi_{0}$ are non-bounded (escaping in
SL$(n,\mathbb{R})$) and, obviously, so are the solutions generated
by matrices similar to symmetric ones, $\Psi(t)=\chi\exp(\lambda
t)\chi^{-1}\Psi_{0}=\exp(\chi\lambda\chi^{-1}t)\Psi_{0}$. And
again this property is stable with respect to small perturbations
of $\lambda$ by skew-symmetric matrices
$\epsilon=-\epsilon^{T}\in$ SO$(n,\mathbb{R})^{\prime}$.
Therefore, the general solution of the geodetic doubly-invariant
model contains also an open subset of non-bounded (escaping)
solutions.

Roughly speaking, using analogy with the Kepler or attractive
Coulomb problem we may tell here about motions below and above
dissociation threshold, however, without any potential, just in
purely geodetic models on the non-compact manifold
SL$(n,\mathbb{R})$.

In a quantized version of this model the family of bounded
classical trajectories is replaced by the discrete energy spectrum
and the $L^{2}$-class wave functions of stationary states. And
similarly, the manifold of non-bounded orbits is a classical
counterpart of the continuous spectrum and non-normalized wave
functions (scattering situations).

Obviously, the above description in terms of groups
GL$(n,\mathbb{R})$, SL$(n,\mathbb{R})$ is an analytical
simplification used for computational purposes. To use
systematically a more correct geometrical language we should
replace the terms "skew-symmetric" and "symmetric" by $g$- or
$\eta$-skew-symmetric and symmetric: $\lambda^{i}{}_{j}=\mp
g^{ik}\lambda^{m}{}_{k}g_{mj}$, $\hat{\lambda}^{A}{}_{B}=\mp
\eta^{AC}\hat{\lambda}^{D}{}_{C}\eta_{DB}$.

Finally, we can conclude that if dilatations are stabilized by
some potential $V_{\rm dil}(q)$, then for the model with the
kinetic energy invariant under spatial and material affine
transformations, there exists a $2n^{2}$-dimensional family of
bounded solutions even if the shear component of motion is purely
geodetic. If the stabilizing dilatational potential has an upper
bound, there exists also a $2n^{2}$-dimensional family of
unbounded, escaping motions. The above arguments are based on
properties of one-parameter subgroups and their cosets in
SL$(n,\mathbb{R})$. Therefore, they do not apply directly to
affine-metrical and metrical-affine models. Indeed, as we have
seen, if the spatial or material symmetry of the kinetic energy is
restricted to the rotation group, then, except some special
solutions (relative equilibria), one-parameter subgroups and their
cosets fail to be solutions. Nevertheless, our arguments may be
used then in a non-direct way.

In analogy to (\ref{5.9}) we can rewrite the kinetic Hamiltonians
(\ref{4.24}), (\ref{4.25}), i.e., (\ref{4.28}), (\ref{4.29}), as
follows:
\begin{eqnarray}
\mathcal{T}_{\rm int}&=&\frac{1}{2(I+A)}{\rm
Tr}(\hat{\sigma}^{2})+
\frac{1}{2n(I+A+nB)}p^{2}+\frac{I}{2(I^{2}-A^{2})}\|V\|^{2}
\label{5.10}\\
&=&\frac{1}{2(I+A)}C_{{\rm SL}(n)}(2)+
\frac{1}{2n(I+A+nB)}p^{2}+\frac{I}{2(I^{2}-A^{2})}\|V\|^{2},
\nonumber\\
\mathcal{T}_{\rm int}&=&\frac{1}{2(I+A)}{\rm Tr}(\sigma^{2})+
\frac{1}{2n(I+A+nB)}p^{2}+\frac{I}{2(I^{2}-A^{2})}\|S\|^{2}
\label{5.11}\\
\nonumber &=&\frac{1}{2(I+A)}C_{{\rm SL}(n)}(2)+
\frac{1}{2n(I+A+nB)}p^{2}+\frac{I}{2(I^{2}-A^{2})}\|S\|^{2},
\end{eqnarray}
where $C_{{\rm SL}(n)}(2)={\rm Tr}(\sigma^{2})={\rm
Tr}(\hat{\sigma}^{2})$.

The formulas (\ref{4.28}), (\ref{4.29}) or, equivalently,
(\ref{5.10}), (\ref{5.11}) imply that for
dilatatio\-nally-stabilized models $H=\mathcal{T}_{\rm int}+V_{\rm
dil}(q)$ with the affine-metrical and metrical-affine kinetic
terms $\mathcal{T}_{\rm int}$, all the above statements concerning
bounded and unbounded solutions of affine-affine models
(\ref{5.9}), (\ref{4.3}) remain true. In particular, for the
purely geodetic incompressible models with $\mathcal{T}_{\rm int}$
invariant under SL$(V)\times$ O$(U,\eta)$ or under O$(V,g)\times$
SL$(U)$, there exists an open subset of bounded solutions
(vibrations) and an open subset of non-bounded ones. What concerns
spatially affine and materially metrical models, the very rough
argument is that the evolution of quantities $\Sigma$,
$\mathcal{K}$ is exactly the same as it was for Hamiltonians $H$
with $\mathcal{T}_{\rm int}$ affinely-invariant both in the
physical and in the material spaces, in this case in (\ref{5.9})
$A$ is replaced by $I+A$. This is a direct consequence of
equations of motion written in terms of Poisson brackets,
\[
\frac{dF}{dt}=\{F,H\}.
\]
In fact, $\|V\|$ is a constant of motion for both types of
Hamiltonians (affine-affine and affine-metrical). In addition to
Lie-algebraic relations of GL$(V)^{\prime}\simeq$ L$(V)$ satisfied
by $\Sigma^{i}{}_{j}$, we have the following obvious Poisson
rules:
\[
\{\Sigma^{i}{}_{j},C(2)\}=\{\Sigma^{i}{}_{j},C(1)\}=0, \qquad
\{\Sigma^{i}{}_{j},\|V\|^{2}\}=0,\qquad
\{\mathcal{K}_{a},\|V\|^{2}\}=0.
\]
The first equations express an obvious property of $C(k)$ as
Casimir invariants of $\Sigma^{i}{}_{j}$ (and
$\hat{\Sigma}^{A}{}_{B}$). The second formula follows from the
obvious relationship
$\{\Sigma^{i}{}_{j},\hat{\Sigma}^{A}{}_{B}\}=0$, because
$\|V\|^{2}$ is an algebraic function of $\hat{\Sigma}^{A}{}_{B}$.
And the third equation is due to the fact that the deformation
invariants $\mathcal{K}_{a}$ are invariant under the group of
material isometries generated by $V^{A}{}_{B}$.

Therefore, the time evolution of variables $\Sigma^{i}{}_{j}$,
$\mathcal{K}_{a}$ is identical in both types of models, i.e.,
(\ref{5.9}) and (\ref{5.10}); the former with $A$ replaced by
$I+A$. As a matter of fact, for geodetic models with
dilatation-stabilizing potentials $V(q)$, the deviator
$\sigma^{i}{}_{j}=\Sigma^{i}{}_{j}-(1/n)\Sigma^{a}{}_{a}\delta^{i}{}_{j}$
is a constant of motion and, obviously, it is so for the purely
geodetic incompressible models. The only difference occurs in
degrees of freedom ruled by SO$(V,g)$, SO$(U,\eta)$, describing
the orientation of principal axes of deformation tensors $C\in
V^{\ast}\otimes V^{\ast}$, $G\in U^{\ast}\otimes U^{\ast}$. But,
roughly speaking, these degrees of freedom have compact topology
and their evolution does not influence the bounded or non-bounded
character of the total orbits.

The same reasoning applies to dilatationally stabilized geodetic
models invariant under O$(V,g)\times$ GL$(U)$ or purely geodetic
incompressible models invariant under O$(V,g)\times$ SL$(U)$
(spatially metrical and materially affine models). Then everything
follows from Poisson brackets
\[
\{\hat{\Sigma}^{A}{}_{B},C(2)\}=\{\hat{\Sigma}^{A}{}_{B},C(1)\}=0,
\qquad \{\hat{\Sigma}^{A}{}_{B},\|S\|^{2}\}=0,\qquad
\{\mathcal{K}_{a},\|S\|^{2}\}=0.
\]
Now on the level of state variables $\hat{\Sigma}^{A}{}_{B}$,
$\mathcal{K}_{a}$ the time evolution is exactly identical with
that based on the affine-affine model of $\mathcal{T}_{\rm int}$
(again with $A$ in (\ref{5.9}) replaced by $I+A$).

Let us stress an important point that it is the time evolution of
deformation invariants that decides whether the total motion is
bounded or not. This is a purely geometric fact independent on any
particular dynamical model. There is an analogy with the material
point motion in $\mathbb{R}^{n}$. An orbit is bounded if and only
if the range of the radial variable $r$ is bounded.

The above point plays an essential role in the qualitative
discussion of deformative motion. It suggests one to use
analytical descriptions of degrees of freedom based on deformation
invariants.

\section{Analytical description}

In section 3 some fundamental facts concerning deformation tensors
and deformation invariants were summarized. Below we continue this
subject and present some natural descriptions of affine degrees of
freedom well-adapted to the study of isotropic problems.

The material and physical spaces are endowed with fixed metric
tensors, $\eta\in U^{\ast}\otimes U^{\ast}$, $g\in V^{\ast}\otimes
V^{\ast}$, and any configuration $\varphi\in$ LI$(U,V)$ gives rise
to the symmetric positively definite tensors $G[\varphi]\in
U^{\ast}\otimes U^{\ast}$, $C[\varphi]\in V^{\ast}\otimes
V^{\ast}$, i.e., Green and Cauchy deformation tensors. Raising
their first indices respectively with the help of $\eta$ and $g$,
we obtain the mixed tensors $\hat{G}[\varphi]\in U\otimes
U^{\ast}$, $\hat{C}[\varphi]\in V\otimes V^{\ast}$ with
eigenvalues $\lambda_{a}$, $\lambda^{-1}_{a}$, $a=\overline{1,n}$.
It is also convenient to use the quantities $Q^{a}$, $q^{a}$,
where $Q^{a}=\exp(q^{a})=\sqrt{\lambda_{a}}$. The diagonal matrix
$D={\rm diag}(Q^{1},\ldots,Q^{n})$ is identified with the linear
mapping $D:\mathbb{R}^{n}\rightarrow \mathbb{R}^{n}$.

The configuration $\varphi\in$ LI$(U,V)$ may be characterized by
$D$, i.e., by the system of fundamental stretchings
$Q^{a}=\exp(q^{a})$, and by the systems of eigenvectors $R_{a}\in
U$, $L_{a}\in V$ of $\hat{G}$, $\hat{C}$ normalized, respectively,
in the sense of $\eta$ and $g$,
\[
\hat{G}R_{a}=\lambda_{a}R_{a}=\exp(2q^{a})R_{a},\qquad
\hat{C}L_{a}=\lambda^{-1}_{a}L_{a}=\exp(-2q^{a})L_{a}.
\]
Obviously, when the spectrum is non-degenerate, then $R_{a}$,
$L_{a}$ are uniquely defined (up to re-ordering) and pair-wise
orthogonal,
$\eta(R_{a},R_{b})=\eta_{CD}R^{C}{}_{a}R^{D}{}_{b}=\delta_{ab}=g(L_{a},L_{a})=
g_{ij}L^{i}{}_{a}L^{j}{}_{b}$. Such a situation is generic, thus,
when at some time instant $t\in \mathbb{R}$ $\varphi(t)$
corresponds to degenerate situation, then $L_{a}(t)$, $R_{a}(t)$
may be also uniquely defined due to the continuity demand.

The elements of the corresponding dual bases will be denoted
respectively by $R^{a}\in U^{\ast}$, $L^{a}\in V^{\ast}$. When
necessary, to avoid misunderstandings, we shall indicate
explicitly the dependence of the above quantities on $\varphi\in$
LI$(U,V)$: $q^{a}[\varphi]$, $R_{a}[\varphi]$, $L_{a}[\varphi]$,
etc.

Green and Cauchy deformation tensors may be respectively expressed
as follows:
\[
G[\varphi]=\sum_{a}\lambda_{a}[\varphi]R^{a}[\varphi]\otimes
R^{a}[\varphi]=\sum_{a}\exp\left(2q^{a}[\varphi]\right)R^{a}[\varphi]\otimes
R^{a}[\varphi],
\]
\[
C[\varphi]=\sum_{a}\lambda^{-1}_{a}[\varphi]L^{a}[\varphi]\otimes
L^{a}[\varphi]=\sum_{a}\exp\left(-2q^{a}[\varphi]\right)L^{a}[\varphi]\otimes
L^{a}[\varphi].
\]

In this way $\varphi$ has been identified with the triple of
fictitious objects: two rigid bodies in $U$ and $V$ with
configurations represented, respectively, by orthonormal frames
$R\in$ F$(U,\eta)$, $L\in$ F$(V,g)$ and a one-dimensional
$n$-particle system with coordinates $q^{a}$ (or $Q^{a}$). Even
for non-degenerate spectra of $\hat{G}[\varphi]$,
$\hat{C}[\varphi]$ this representation is not unique because the
labels $a$ under the summation signs may be simultaneously
permuted without affecting $\varphi$ itself. For degenerate
spectra this representation becomes continuously non-unique in a
similar (although much stronger) way as, e.g., spherical
coordinates at $r=0$.

Let us observe that the linear frames $L=(\ldots,L_{a},\ldots)$
and $R=(\ldots,R_{a},\ldots)$ may be, as usual, identified with
linear isomorphisms $L:\mathbb{R}^{n}\rightarrow V$ and
$R:\mathbb{R}^{n}\rightarrow U$. Similarly, their dual co-frames
$\widetilde{L}=(\ldots,L^{a},\ldots)$ and
$\widetilde{R}=(\ldots,R^{a},\ldots)$ are equivalent to
isomorphisms $L^{-1}:V\rightarrow \mathbb{R}^{n}$ and
$R^{-1}:U\rightarrow \mathbb{R}^{n}$. Identifying the diagonal
matrix diag$(\ldots,Q_{a},\ldots)$ with a linear isomorphism
$D:\mathbb{R}^{n}\rightarrow \mathbb{R}^{n}$, we may finally
represent
\[
\varphi=LDR^{-1},
\]
this is a geometric description of what is sometimes referred to
as the two-polar decomposition
\cite{JJS82_1,JJS88_1,JJS91_5,JJS98_2,JJS02_1,JJS-AKS98}.

Strictly speaking, in continuum mechanics, when the orientation of
the body is constant during any admissible motion (no
mirror-reflections), one has to fix some pattern orientations in
$U$, $V$ and admit only orientation-preserving mappings $\varphi$.
And then the non-connected sets of all orthonormal frames
F$(U,\eta)$, F$(V,g)$ are to be replaced by their connected
submanifolds F$^{+}(U,\eta)$, F$^{+}(V,g)$ of positively oriented
frames.

Obviously, the spatial and material orientation-preserving
isometries $A\in$ SO$(V,g)$, $B\in$ SO$(U,\eta)$ affect only the
$L$- and $R$-gyroscopes on the left. Indeed, $L\mapsto AL$,
$R\mapsto BR$ result in $\varphi\mapsto A\varphi B^{-1}$. Their
Hamiltonian generators, spin and minus-vorticity (i.e.,
respectively $V$- and $U$-spatial canonical spins) have identical
Poisson-commutation rules.

For any of the mentioned rigid bodies, one can define in the usual
way the angular velocity in two representations. One should stress
that both $V$ and $U$ are from this point of view interpreted as
"physical spaces". The "material" ones are both identified with
$\mathbb{R}^{n}$. The "co-moving" and "current" representations
$\hat{\chi}\in$ SO$(n,\mathbb{R})^{\prime}$, $\chi\in$
SO$(V,g)^{\prime}$ for the $L$-top are respectively given by
\[
\hat{\chi}^{a}{}_{b}:=
<L^{a},\frac{dL_{b}}{dt}>=L^{a}{}_{i}\frac{dL^{i}{}_{b}}{dt},
\qquad \chi:=\hat{\chi}^{a}{}_{b}L_{a}\otimes L^{b},\ {\rm i.e.,}\
\chi^{i}{}_{j}=\frac{dL^{i}{}_{a}}{dt}L^{a}{}_{j}.
\]
The corresponding objects $\hat{\vartheta}\in$
SO$(n,\mathbb{R})^{\prime}$, $\vartheta\in$ SO$(U,\eta)^{\prime}$
for the $R$-top are defined by analogous formulas:
\[
\hat{\vartheta}^{a}{}_{b}:=<R^{a},\frac{dR_{b}}{dt}>=R^{a}{}_{K}\frac{dR^{K}{}_{b}}{dt},
\qquad \vartheta:=\hat{\vartheta}^{a}{}_{b}R_{a}\otimes R^{b},\
{\rm i.e.,}\
\vartheta^{K}{}_{L}=\frac{dR^{K}{}_{a}}{dt}R^{a}{}_{L}.
\]
In certain problems it is convenient to use non-holonomic
velocities $\dot{q}^{a}$, $\hat{\chi}^{a}{}_{b}$,
$\hat{\vartheta}^{a}{}_{b}$ or $\dot{q}^{a}$, $\chi^{i}{}_{j}$,
$\vartheta^{A}{}_{B}$. Similarly, non-holonomic conjugate momenta
$p_{a}$, $\hat{\rho}^{a}{}_{b}$, $\hat{\tau}^{a}{}_{b}$ or
$p_{a}$, $\rho^{i}{}_{j}$, $\tau^{A}{}_{B}$ are used, where again
$\hat{\rho},\hat{\tau}\in{\rm SO}(n,\mathbb{R})^{\prime}$,
$\rho\in{\rm SO}(V,g)^{\prime}$, $\tau\in{\rm
SO}(U,\eta)^{\prime}$. The pairing between non-holonomic momenta
and velocities is given by
\[
<(\rho,\tau,p),(\chi,\vartheta,\dot{q})>=
<(\hat{\rho},\hat{\tau},p),(\hat{\chi},\hat{\vartheta},\dot{q})>
\]
\[
=p_{a}\dot{q}^{a}+\frac{1}{2}{\rm Tr}(\rho\chi)+\frac{1}{2}{\rm
Tr}(\tau\vartheta)=p_{a}\dot{q}^{a}+\frac{1}{2}{\rm
Tr}(\hat{\rho}\hat{\chi})+\frac{1}{2}{\rm
Tr}(\hat{\tau}\hat{\vartheta}).
\]

\noindent {\bf Remark:} Our system of notations is slightly
redundant, because $\rho$ and $\tau$ coincide, respectively, with
spin and negative vorticity, $\rho=S$, $\tau=-V$. The reason is
that they are Hamiltonian generators of transformations
$\varphi\mapsto A\varphi$, $\varphi\mapsto \varphi B^{-1}$, $A\in
{\rm SO}(V,g)$, $B\in {\rm SO}(U,\eta)$.

The objects $\hat{\rho}$, $\hat{\tau}$ generate transformations
\begin{equation}\label{trrt71}
L\mapsto LA,\quad R\mapsto RB,\quad A,B\in {\rm SO}(n,\mathbb{R})
\end{equation}
and express the quantities $\rho$, $\tau$ in terms of the
reference frames given, respectively, by the principal axes of the
Cauchy and Green deformation tensors,
\[
\rho=\hat{\rho}^{a}{}_{b}L_{a}\otimes L^{b},\qquad
\tau=\hat{\tau}^{a}{}_{b}R_{a}\otimes R^{b}.
\]

\noindent {\bf Remark:} In dynamical models based on the
d'Alembert principle the quantities $Q^{a}$ and their conjugate
momenta $P_{a}$ are more convenient that $q^{a}$ and $p_{a}$. The
latter ones are useful in models with affinely-invariant kinetic
energy.

If $V$ and $U$ both are identified with $\mathbb{R}^{n}$ and
LI$(U,V)$ with GL$(n,\mathbb{R})$, then $L$ and $R$ in the
two-polar splitting $\varphi=LDR^{-1}$ become elements of
SO$(n,\mathbb{R})$ and $D$, as previously, is a diagonal matrix
with positive elements. The two-polar decomposition is a
by-product of the polar decomposition of GL$^{+}(n,\mathbb{R})$,
\[
\varphi=UA,
\]
where $U\in$ SO$(n,\mathbb{R})$, thus, $U^{T}=U^{-1}$, and
$A=A^{T}$ is a symmetric positively-definite matrix. It is
well-known that this decomposition is unique, whereas the
two-polar one is charged with some multivaluedness. Green and
Cauchy deformation tensors are then represented as follows:
$G=\varphi^{T}\varphi=A^{2}$, $C=(\varphi^{-1})^{T}\varphi^{-1}=
UA^{-2}U^{-1}$. One can also use the reversed polar decomposition
\[
\varphi=BU,\qquad U\in {\rm SO}(n,\mathbb{R}),\qquad
B=UAU^{-1}=B^{T}.
\]
Then $G=U^{-1}B^{2}U$, $C=B^{-2}$. The two-polar decomposition is
achieved by the orthogonal diagonalization of the matrix $A$,
$A=VDV^{-1}$, $V\in {\rm SO}(n,\mathbb{R})$. Then $L=UV$, $R=V$.

The polar splitting was described above in an over-simplified
standard way, namely, $U$ and $V$ were identified with
$\mathbb{R}^{n}$ and LI$(U,V)$ with GL$(n,\mathbb{R})$. Let us
remind that in continuum mechanics the connected components of
LI$(U,V)$ and GL$(n,\mathbb{R})$ are used as configuration spaces,
LI$^{+}(U,V)$, GL$^{+}(n,\mathbb{R})$, where the first symbol
denotes the manifold of orientation-preserving isomorphisms (it is
assumed here that some orientations in $U$, $V$ are fixed). It is
instructive to see what the both polar splittings are from the
geometric point of view, when $U$ and $V$ are distinct linear
spaces, non-identified with $\mathbb{R}^{n}$.

As mentioned above, when metric tensors $\eta\in U^{\ast}\otimes
U^{\ast}$, $g\in V^{\ast}\otimes V^{\ast}$ are fixed, then any
$\varphi\in$ LI$(U,V)$ with non-degenerate spectra of deformation
tensors gives rise to the pair of orthonormal bases
$(L_{a}[\varphi]\in V,\ a=\overline{1,n})$, $(R_{a}[\varphi]\in
U,\ a=\overline{1,n})$. There exists exactly one isometry
$U[\varphi]:U\rightarrow V$ such that $U[\varphi]\cdot
R_{a}[\varphi]=L_{a}[\varphi]$. Obviously, the isometry property
is meant in the sense that $\eta=U[\varphi]^{\ast}\cdot g$, i.e.,
analytically
$\eta_{AB}=g_{ij}U[\varphi]^{i}{}_{A}U[\varphi]^{j}{}_{B}$.
Geometric meaning of the polar decomposition is as follows:
\[
\varphi=U[\varphi]A[\varphi]=B[\varphi]U[\varphi],
\]
where the automorphisms $A[\varphi]\in$ GL$(U)$, $B[\varphi]\in$
GL$(V)$ are symmetric, respectively, in the $\eta$- and $g$-sense,
i.e., $\eta(A[\varphi]x,y)=\eta(x,A[\varphi]y)$,
$g(B[\varphi]w,z)=g(w,B[\varphi]z)$ for arbitrary $x,y\in U$,
$w,z\in V$. They are also positively definite,
$\eta(A[\varphi]x,x)>0$, $g(B[\varphi]w,w)>0$ for arbitrary
non-null $x\in U$, $w\in V$.

In spite of the non-uniqueness contained in $L[\varphi]$,
$R[\varphi]$, the mappings $U[\varphi]$, $A[\varphi]$,
$B[\varphi]$ are unique. And the symmetric parts are obtained from
each other by the $U[\varphi]$-intertwining,
$B[\varphi]=U[\varphi]A[\varphi]U[\varphi]^{-1}$.

In mechanics of discrete affine systems we are free to admit
orientation-reversing isometries $U$ or symmetric mappings $A$,
$B$ not necessarily positively-definite.

The non-uniqueness of the two-polar decomposition mentioned above
is important in certain computational and also principal problems,
so some comments are necessary here. The problem is technically
complicated, thus, only necessary facts will be quoted here, some
of them formulated in a rather brief, rough way.

The subgroup of O$(n,\mathbb{R})$ consisting of matrices which
have exactly one non-vanishing entry in every row and column will
be denoted by $K$. Obviously, $K$ is finite and the mentioned
entries are $\pm 1$, reals with absolute value $1$. The subgroup
of proper $K$-rotations will denoted by $K^{+}:=K\cap{\rm
SO}(n,\mathbb{R})$. Obviously, the orders (numbers of elements) of
$K$, $K^{+}$ equal respectively $2n\cdot n!$ and $n\cdot n!$. Let
$W\in K$ be a corresponding similarity transformation preserving
the group of diagonal matrices, ${\rm Diag}(\mathbb{R}^{n})\ni D
\mapsto W^{-1}DW\in {\rm Diag}(\mathbb{R}^{n})$, and resulting in
permutation of the diagonal elements of $D={\rm
diag}(Q^{1},\ldots,Q^{n})$, i.e., we have
$(Q^{1},\ldots,Q^{n})\mapsto (Q^{\pi_{W}(1},\ldots,Q^{n)})$ or
$(q^{1},\ldots,q^{n})\mapsto (q^{\pi_{W}(1},\ldots,q^{n)})$, where
$Q^{a}=\exp(q^{a})$. Obviously, the mapping $K\ni W\mapsto
\pi_{W}\in {\rm S}^{(n)}$ is a $2n:1$ epimorphism of $K$ onto the
permutation group S$^{(n)}$. Its restriction to $K^{+}$ has an
$n$-element kernel. The non-uniqueness of representation of
$\varphi\in {\rm GL}^{+}(n,\mathbb{R})$ through elements of
SO$(n,\mathbb{R})\times\mathbb{R}^{n}\times{\rm SO}(n,\mathbb{R})$
depends strongly on the degeneracy of spectra of deformation
tensors. The multi-valuedness is discrete, thus, simplest in the
case of simple spectra.

Let GL$^{+(n)}(n,\mathbb{R})\subset {\rm GL}^{+}(n,\mathbb{R})$ be
the subset of $\varphi$-s with non-degenerate spectra of $C$, $G$.
The corresponding subset $M^{(n)}$ of
SO$(n,\mathbb{R})\times\mathbb{R}^{n}\times{\rm SO}(n,\mathbb{R})$
consists of such triplets $(L;q^{1},\ldots,q^{n};R)$ that all
$q^{i}$-s are pairwise distinct. The group $K^{+}$ may be
faithfully realized by the following transformation group
$H^{(n)}$ of $M^{(n)}$:
\[
(L;q^{1},\ldots,q^{n};R)\mapsto(LW;q^{\pi_{W}(1},\ldots,q^{n)};RW).
\]
Obviously, this transformation does not affect $\varphi=LDR^{-1}$.
Therefore, we have a diffeomorphism
\[
{\rm GL}^{+(n)}(n,\mathbb{R})\simeq M^{(n)}/H^{(n)}.
\]
Non-degenerate spectrum is a generic one, nevertheless the
coincidence case must be also taken into account because some new
qualities appear then and they are relevant for qualitative
analysis of classical phase portraits and for quantum conditions
on admissible wave functions.

Let GL$^{+(k;p_{1},\ldots,p_{k})}\subset{\rm
GL}^{+}(n,\mathbb{R})$ consist of $\varphi$-s for which
deformation tensors have $k\leq n$ different principal values,
every one of them with the corresponding multiplicity
$p_{\sigma}$, $\sum_{\sigma=1}^{k}p_{\sigma}=n$. And similarly,
let $M^{(k;p_{1},\ldots,p_{k})}$ be the set of such triplets
$(L;q^{1},\ldots,q^{n};R)\in{\rm
SO}(n,\mathbb{R})\times\mathbb{R}^{n}\times{\rm SO}(n,\mathbb{R})$
that there are only $k$ different $q^{i}$-s with the same
conditions concerning multiplicity. And now let us consider the
transformation group $H^{(k;p_{1},\ldots,p_{k})}$ acting on
$M^{(k;p_{1},\ldots,p_{k})}$ as follows:
\[
(L;q^{1},\ldots,q^{n};R)\mapsto(LW;q^{\pi_{W}(1},\ldots,q^{n)};RW),
\]
where $W$ runs over the subgroup of SO$(n,\mathbb{R})$ that is
generated by $K^{+}$ and the subgroup
$H^{(k;p_{1},\ldots,p_{k})}\subset{\rm SO}(n,\mathbb{R})$ composed
of $k$ blocks $p_{\sigma}\times p_{\sigma}$, every one given by
the corresponding SO$(p_{\sigma},\mathbb{R})$. Then we have that
\[
{\rm GL}^{+(k;p_{1},\ldots,p_{k})}\simeq
M^{(k;p_{1},\ldots,p_{k})}/H^{(k;p_{1},\ldots,p_{k})}.
\]
When $k<n$, then at least one of multiplicities is non-trivial and
the resulting group $H^{(k;p_{1},\ldots,p_{k})}$ is continuous.
The resulting quotient is lower-dimensional because of this
continuity of the divisor transformation group.

In the physical case $n=3$, we have obviously only two
possibilities of the non-trivial blocks, namely the total
SO$(3,\mathbb{R})$ and SO$(2,\mathbb{R})\times{\rm
SO}(1,\mathbb{R})$ (respectively, all three $q^{i}$'s equal or two
of them); obviously SO$(1,\mathbb{R})=\{1\}$.

In the extreme case $k=1$, $D$ is proportional to the $n\times n$
identity matrix and it is only the total $LR^{-1}$ that is
well-defined; on the other hand $L$, $R$ separately are
meaningless.

It is very convenient and instructive to express our Hamiltonians,
kinetic energies and configuration metrics in terms of the
two-polar splitting. The previous statements concerning the phase
pictures become then much more lucid. Let us introduce some
auxiliary quantities $M:=-\hat{\rho}-\hat{\tau}$,
$N:=\hat{\rho}-\hat{\tau}$. One can easily show that the
second-order Casimir invariant $C(2)$ occurring in the main terms
of our affine-affine, affine-metrical and metrical-affine kinetic
Hamiltonians has the following form:
\begin{equation}\label{6.1}
C(2)=\sum_{a}p^{2}_{a}+\frac{1}{16}\sum_{a,b}\frac{(M^{a}{}_{b})^{2}}{{\rm
sh}^{2}\frac{q^{a}-q^{b}}{2}}-\frac{1}{16}\sum_{a,b}\frac{(N^{a}{}_{b})^{2}}{{\rm
ch}^{2}\frac{q^{a}-q^{b}}{2}}.
\end{equation}
Obviously, $M$ and $N$ are antisymmetric in the Kronecker-delta
sense, $M^{a}{}_{b}=-M_{b}{}^{a}=-g_{bk}g^{al}M^{k}{}_{l}$,
$N^{a}{}_{b}=-N_{b}{}^{a}=-g_{bk}g^{al}N^{k}{}_{l}$. The first
term in (\ref{6.1}) may be suggestively decomposed into the
"relative" and the "over-all" ("centre of mass") parts:
\[
\frac{1}{2n}\sum_{a,b}(p_{a}-p_{b})^{2}+\frac{p^{2}}{n}.
\]
Obviously, $C(1)=p$.

For geodetic systems and for more general systems with potentials
$V$ depending only on deformation invariants, spin $S=\rho$ and
vorticity $V=\tau$ are constants of motion and may be used for
extracting from equations of motion some information concerning
the general solution. Unlike this the quantities $\hat{\rho}$,
$\hat{\tau}$, thus, also $M$, $N$, fail to be constants of motion
except the special case $n=2$, when the rotation group is Abelian.
However, on the level of qualitative analysis, the expression
(\ref{6.1}) based on $\hat{\rho}$, $\hat{\tau}$ is more convenient
because it does not involve $L$, $R$-variables, i.e., rotational
degrees of freedom of deformation tensors. Therefore, our Poisson
bracket relations imply that on the level of variables $q^{a}$,
$p_{a}$, $M^{a}{}_{b}$, $N^{a}{}_{b}$ equations of motion based on
(\ref{5.10}) (equivalently (\ref{4.28})), (\ref{5.11})
(equivalently (\ref{4.29})), and (\ref{5.9}) with $A$ replaced by
$I+A$ are identical. In particular, for geodetic incompressible
models and for compressible models with stabilized dilatations
there exists an open family of bounded (vibrating) solutions and
an open family of non-bounded (decaying) solutions. The reason is
that it is so for (\ref{5.9}) with $A$ replaced by $I+A$, and the
additional terms proportional to $S^{2}$ or $V^{2}$ do not
influence anything because they have vanishing Poisson brackets
with $q^{a}$, $p_{a}$, $M^{a}{}_{b}$, $N^{a}{}_{b}$ and only those
variables occur in $H$. The only difference appears when the
evolution of $L$- and $R$-variables is taken into account.
However, the corresponding configuration spaces F$(V,g)$,
F$(U,\eta)$ are compact (they are manifolds of orthonormal frames)
and do not influence the boundedness of orbits.

Let us observe that after substituting (\ref{6.1}), the first main
term of (\ref{5.10}) (equivalently (\ref{4.28})), (\ref{5.11})
(equivalently (\ref{4.29})), and (\ref{5.9}) with $A$ replaced by
$I+A$ acquires the characteristic lattice structure,
\[
\mathcal{T}_{\rm
latt}=\frac{1}{2\alpha}\sum_{a}p^{2}_{a}+\frac{1}{32\alpha}
\sum_{a,b}\frac{(M^{a}{}_{b})^{2}}{{\rm
sh}^{2}\frac{q^{a}-q^{b}}{2}}-\frac{1}{32\alpha}\sum_{a,b}\frac{(N^{a}{}_{b})^{2}}{{\rm
ch}^{2}\frac{q^{a}-q^{b}}{2}}.
\]
This expression resembles structurally the hyperbolic Sutherland
$n$-body system on the straight line. Positions of the fictitious
material points are given by deformation invariants $q^{a}$. The
"particles" have identical masses and are indistinguishable.
Unlike the hyperbolic Sutherland system, the coupling amplitudes
$M^{a}{}_{b}$, $N^{a}{}_{b}$ are non-equal and non-constant;
rather they are dynamical variables on the equal footing with
$q^{a}$, $p_{a}$. The negative $N$-contribution to
$\mathcal{T}_{\rm latt}$ describes the attractive forces between
lattice points, whereas the positive $M$-term corresponds to
repulsion. Under the appropriate initial conditions we have stable
bounded vibrations without any use of the potential energy term.
Therefore, the non-definiteness of $\mathcal{T}_{\rm latt}$ is not
only non-embarrassing, but just desirable as a tool for describing
"elastic" vibrations on the basis of purely geodetic models. Let
us observe that the purely affine-affine part of (\ref{5.10}),
(\ref{5.11}) (equivalently (\ref{4.28}), (\ref{4.29})), i.e.,
(\ref{5.9}) with $A$ replaced by $I+A$ (composed of its first two
Casimir terms), splits in the following suggestive way into the
binary SL$(n,\mathbb{R})$-part and dilatational contribution:
\begin{eqnarray}
\mathcal{T}^{\rm aff}_{\rm
int}&=&\frac{1}{2\alpha}C(2)+\frac{1}{2\beta}C(1)^{2}=
\frac{1}{4\alpha n}\sum_{a,b}(p_{a}-p_{b})^{2}
\nonumber\\
&+&\frac{1}{32\alpha}\sum_{a,b}\frac{(M^{a}{}_{b})^{2}}{{\rm
sh}^{2}\frac{q^{a}-q^{b}}{2}}-\frac{1}{32\alpha}\sum_{a,b}\frac{(N^{a}{}_{b})^{2}}{{\rm
ch}^{2}\frac{q^{a}-q^{b}}{2}}+\frac{n\alpha+\beta}{2n\alpha\beta}p^{2},\nonumber
\end{eqnarray}
or, in a more explicit form,
\begin{eqnarray}
\mathcal{T}^{\rm aff}_{\rm int}&=&\frac{1}{4(I+A)
n}\sum_{a,b}(p_{a}-p_{b})^{2}+\frac{1}{32(I+A)}\sum_{a,b}\frac{(M^{a}{}_{b})^{2}}{{\rm
sh}^{2}\frac{q^{a}-q^{b}}{2}}\nonumber
\\
&-&\frac{1}{32(I+A)}\sum_{a,b}\frac{(N^{a}{}_{b})^{2}}{{\rm
ch}^{2}\frac{q^{a}-q^{b}}{2}}+\frac{1}{2n(I+A+nB)}p^{2}.\label{6.2}
\end{eqnarray}
Obviously, for (\ref{5.10}) (equivalently (\ref{4.28})) and
(\ref{5.11}) (equivalently (\ref{4.29})) we have, respectively,
\begin{eqnarray}
\mathcal{T}^{\rm aff-metr}_{\rm int}&=&\mathcal{T}^{\rm aff}_{\rm
int}+\frac{I}{2(I^{2}-A^{2})}\|V\|^{2},\label{6.3}
\\
\label{6.4} \mathcal{T}^{\rm metr-aff}_{\rm
int}&=&\mathcal{T}^{\rm aff}_{\rm
int}+\frac{I}{2(I^{2}-A^{2})}\|S\|^{2}.
\end{eqnarray}
Copmparing this with (\ref{5.10}), (\ref{5.11}), we conclude that
\[
C_{{\rm SL}(n)}(2)={\rm Tr}(\sigma^{2})={\rm
Tr}(\hat{\sigma}^{2})
\]
\[
=\frac{1}{2n}\sum_{a,b}(p_{a}-p_{b})^{2}+
\frac{1}{16}\sum_{a,b}\frac{(M^{a}{}_{b})^{2}}{{\rm
sh}^{2}\frac{q^{a}-q^{b}}{2}}
-\frac{1}{16}\sum_{a,b}\frac{(N^{a}{}_{b})^{2}}{{\rm
ch}^{2}\frac{q^{a}-q^{b}}{2}}.
\]
This expression is very suggestive because it expresses the
quantity $C_{{\rm SL}(n)}(2)$ and the corresponding contribution
to $\mathcal{T}_{\rm int}$, i.e., the metric tensor on the
manifold of incompressible motions, as the sum of $n(n-1)/2$
two-dimensional clusters, i.e., $\mathbb{R}^{2}$-coordinate planes
in $\mathbb{R}^{n}$. Incompressibility is expressed by the fact
that the invariants $q^{a}$ and their conjugate momenta $p_{a}$
enter the above formula through the shape-describing differences
$(q^{a}-q^{b})$ (ratios $Q^{a}/Q^{b}$) and $p_{a}-p_{b}$. This
expression may be very convenient when studying invariant geodetic
models on the projective group Pr$(n,\mathbb{R})$, i.e., when
dealing with the mechanics of projectively-rigid bodies (bodies
subject to such constraints that all geometric relationships of
projective geometry are preserved, in particular, the material
straight lines remain straight lines). The point is that
Pr$(n,\mathbb{R})$ may be identified in a standard way with
SL$(n+1,\mathbb{R})$.

For the d'Alembert model the two-polar splitting leads to the
following kinetic Hamiltonian term:
\begin{equation}\label{6.5}
\mathcal{T}_{\rm int}=\frac{1}{2I}\sum_{a}P^{2}_{a}+
\frac{1}{8I}\sum_{a,b}\frac{(M^{a}{}_{b})^{2}}{(Q^{a}-Q^{b})^{2}}
+\frac{1}{8I}\sum_{a,b}\frac{(N^{a}{}_{b})^{2}}{(Q^{a}+Q^{b})^{2}}.
\end{equation}
It is purely repulsive on the level of $Q$-variables, thus,
without any potential term it is non-realistic as a model of
elastic vibrations. It is related to the Calogero-Moser lattices
similarly as the previous models show some kinship with the
hyperbolic Sutherland lattices
\cite{Cal-Mar74,Mos75_1,Mos75_2,JJS91_5,JJS98_2,JJS02_1,JJS-AKS98,Toda81}.

What concerns affine models, we can compactify deformation
invariants $q^{a}$ by taking them modulo $2\pi$ ($n$-dimensional
torus), i.e., by putting formally $Q^{a}=\exp(iq^{a})$. This is
equivalent to replacing GL$(n,\mathbb{R})$ by U$(n)$, i.e.,
another and completely opposite real form of GL$(n,\mathbb{C})$.
The Lie algebra U$(n)^{\prime}$ consists of anti-Hermitian
matrices, and the positively definite kinetic energy may be
postulated in the following form:
\[
T_{\rm int}=-\frac{A}{2}{\rm Tr}(\Omega^{2})-\frac{B}{2}({\rm
Tr}\Omega)^{2} =\frac{A}{2}{\rm
Tr}(\Omega^{+}\Omega)+\frac{B}{2}{\rm Tr}(\Omega^{+}){\rm
Tr}(\Omega),
\]
where $\Omega=(d\varphi/dt)\varphi^{-1}$, $A>0$, $B>0$. Obviously,
in this expression for $T$, $\Omega$ may be as well replaced by
$\hat{\Omega}=\varphi^{-1}(d\varphi/dt)$.

Using again the "two-polar" decomposition $\varphi=LDR^{-1}$,
where $L,R\in$ SO$(n,\mathbb{R})$, $D=$
diag$(\ldots,\exp(iq^{a}),\ldots)$, one obtains for the geodetic
Hamiltonian:
\begin{equation}\label{89}
\mathcal{T}_{\rm int}=\frac{1}{2A}\sum_{a}p_{a}^{2}+
\frac{1}{32A}\sum_{a,b}\frac{(M^{a}{}_{b})^{2}}{\sin^{2}\frac{q^{a}-q^{b}}{2}}
+\frac{1}{32A}\sum_{a,b}\frac{(N^{a}{}_{b})^{2}}{\cos^{2}
\frac{q^{a}-q^{b}}{2}}-\frac{B}{2A(A+nB)}p^{2}.
\end{equation}
The first three terms, corresponding to the $C(2)$-Casimir,
resemble the usual Sutherland lattice for $q$-particles with the
same provisos as previously. Geodetic motion is bounded, because
U$(n)$ is compact. Just as previously, it may be convenient to use
the splitting into SU$(n)$- and U$(1)$-terms,
\[
\mathcal{T}_{\rm int}=\frac{1}{4An}\sum_{a,b}(p_{a}-p_{b})^{2}+\frac{1}{2n(A+nB)}p^{2}
\]
\[
+\frac{1}{32A}\sum_{a,b}\frac{(M^{a}{}_{b})^{2}}{\sin^{2}\frac{q^{a}-q^{b}}{2}}
+\frac{1}{32A}\sum_{a,b}\frac{(N^{a}{}_{b})^{2}}{\cos^{2}
\frac{q^{a}-q^{b}}{2}}.
\]
And, in particular,
\[
C_{{\rm SU}(n)}(2)=\frac{1}{2n}\sum_{a,b}(p_{a}-p_{b})^{2}+
\frac{1}{16}\sum_{a,b}\frac{(M^{a}{}_{b})^{2}}{\sin^{2}\frac{q^{a}-q^{b}}{2}}
+\frac{1}{16}\sum_{a,b}\frac{(N^{a}{}_{b})^{2}}{\cos^{2}
\frac{q^{a}-q^{b}}{2}}.
\]

The binary structures of $C_{SL(n,\mathbb{R})}(2)$ and
$C_{SU(n)}(2)$ and their dependence on the variables $q^{a}$,
$p_{a}$ through their differences $q^{a}-q^{b}$, $p_{a}-p_{b}$ is
geometrically interesting in itself. The splitting into
SL$(2,\mathbb{R})$- and $SU(2)$-clusters corresponding to all
possible coordinate planes $\mathbb{R}^{2}$ in $\mathbb{R}^{n}$
may be also analytically helpful. However, some sophisticated
mathematical techniques would be necessary then, like, e.g., the
Dirac procedure for degenerate/constrained system. The point is
that, in general, different clusters are not analytically
independent. And any procedure based on some ordering of variables
destroys the explicit binary structure and makes the structure of
$\mathcal{T}$ rather obscure.

It is interesting that the general solution of $C(2)$-based
geodetic models contains as a particular subfamily the general
solution of the mentioned Calogero-Moser and Sutherland models. It
is obtained by putting $N^{a}{}_{b}=0$, and all $M^{a}{}_{b}$ with
$b\neq a$ equal to some fixed constant $M$.

As mentioned, we are particularly interested in geodetic affine
models. Nevertheless, it is instructive to admit a wider class of
Hamiltonians:
\begin{equation}\label{6.6}
H=\mathcal{T}+V(q^{1},\ldots,q^{n}),
\end{equation}
where $\mathcal{T}$ is any of the kinetic energy models described
above, and the potential $V$ depends on $\varphi$ only through the
deformation invariants $q^{a}$. This means that it is isotropic
both in the physical and material space. The mentioned
non-uniqueness of the two-polar decomposition implies that $V$ as
a function on $\mathbb{R}^{n}$ must be permutation-invariant to
represent a well-defined function on the configuration space. When
the extra potential, e.g., elastic one, is admitted, then also the
"usual" model of $\mathcal{T}$ based on the d'Alembert principle
may be sensibly used for describing bounded elastic vibrations.
Therefore, from now on all the above models of $\mathcal{T}_{\rm
int}$ (\ref{6.2}), (\ref{6.3}), (\ref{6.4}), (\ref{6.5})  are
admitted, although of course the "non-usual" affine models
(\ref{6.2}), (\ref{6.3}), (\ref{6.4}) are still particularly
interesting for us.

As mentioned, the most convenient way of discussing and solving
equations of motion is that based on Poisson brackets,
\[
\frac{dF}{dt}=\{F,H\},
\]
where $F$ runs over some maximal system of (functionally)
independent functions on the phase space. The most convenient and
geometrically distinguished choice is $q^{a}$, $p_{a}$,
$M^{a}{}_{b}$, $N^{a}{}_{b}$, $L$, $R$ or, more precisely, some
coordinates on SO$(n,\mathbb{R})$ parameterizing $L$ and $R$. In
d'Alembert models $Q^{a}$, $P_{a}$ are more convenient the
$q^{a}$, $p_{a}$.

An important point is that $q^{a}$, $p_{a}$, $M^{a}{}_{b}$,
$N^{a}{}_{b}$ generate some Poisson subalgebra, because their
Poisson brackets may be expressed by them alone without any use of
$L$, $ R$-variables. And Hamiltonians also depend only on $q^{a}$,
$p_{a}$, $M^{a}{}_{b}$, $N^{a}{}_{b}$, whereas $L$, $R$ are
non-holonomically cyclic variables. This enables one to perform a
partial reduction of the problem. In fact, the following subsystem
of equations is closed:
\[
\frac{dq^{a}}{dt}=\{q^{a},H\}=\frac{\partial H}{\partial p_{a}},
\]
\[
\frac{dM^{a}{}_{b}}{dt}=\{M^{a}{}_{b},H\}=
\{M^{a}{}_{b},M^{c}{}_{d}\}\frac{\partial H}{\partial
M^{c}{}_{d}}+\{M^{a}{}_{b},N^{c}{}_{d}\}\frac{\partial H}{\partial
N^{c}{}_{d}},
\]
\[
\frac{dp_{a}}{dt}=\{p_{a},H\}=-\frac{\partial H}{\partial q^{a}},
\]
\[
\frac{dN^{a}{}_{b}}{dt}=\{N^{a}{}_{b},H\}=
\{N^{a}{}_{b},M^{c}{}_{d}\}\frac{\partial H}{\partial
M^{c}{}_{d}}+\{N^{a}{}_{b},N^{c}{}_{d}\}\frac{\partial H}{\partial
N^{c}{}_{d}}.
\]
Obviously, $\{q^{a},p_{b}\}=\delta^{a}{}_{b}$,
$\{q^{a},M^{c}{}_{d}\}=\{p_{a},M^{c}{}_{d}\}=
\{q^{a},N^{c}{}_{d}\}=\{p_{a},N^{c}{}_{d}\}=0$. Poisson brackets
of $M$, $N$-quantities follow directly from those for
$\hat{\rho}$, $\hat{\tau}$, and the latter ones correspond exactly
to the structure constants of SO$(n,\mathbb{R})$, thus,
\[
\{\hat{\rho}_{ab},\hat{\rho}_{cd}\}=\hat{\rho}_{ad}\delta_{cb}-
\hat{\rho}_{cb}\delta_{ad}+\hat{\rho}_{db}\delta_{ac}-\hat{\rho}_{ac}\delta_{db},
\]
\[
\{\hat{\tau}_{ab},\hat{\tau}_{cd}\}=\hat{\tau}_{ad}\delta_{cb}-
\hat{\tau}_{cb}\delta_{ad}+\hat{\tau}_{db}\delta_{ac}-\hat{\tau}_{ac}\delta_{db},
\qquad \{\hat{\rho}_{ab},\hat{\tau}_{cd}\}=0,
\]
where the raising and lowering of indices are meant in the
Kronecker-delta sense. From these Poisson brackets we obtain the
following ones:
\[
\{M_{ab},M_{cd}\}=\{N_{ab},N_{cd}\}=M_{cb}\delta_{ad}-M_{ad}\delta_{cb}
+M_{db}\delta_{ac}-M_{ac}\delta_{db},
\]
\[
\{M_{ab},N_{cd}\}=N_{cb}\delta_{ad}-N_{ad}\delta_{cb}
-N_{ac}\delta_{db}+N_{db}\delta_{ac}.
\]

The subsystem for $(q^{a},p_{a},M^{a}{}_{b},N^{a}{}_{b})$ may be
in principle autonomously solvable. When the time dependence of
$\hat{\rho}=(N-M)/2$ and $\hat{\tau}=-(N+M)/2$ is known, then
performing the inverse Legendre transformation we can obtain the
time dependence of angular velocities $\hat{\chi}$,
$\hat{\vartheta}$:
\[
\hat{\chi}^{a}{}_{b}=\frac{\partial H}{\partial
\hat{\rho}^{b}{}_{a}},\qquad
\hat{\vartheta}^{a}{}_{b}=\frac{\partial H}{\partial
\hat{\tau}^{b}{}_{a}}
\]
(some care must be taken when differentiating with respect to
skew-symmetric matrices). And finally the evolution of $L$, $R$ is
given by the following time-dependent systems:
\[
\frac{dL}{dt}=L\hat{\chi},\qquad \frac{dR}{dt}=R\hat{\vartheta}.
\]

There is some very important consequence of this reduction
procedure, i.e., in doubly-isotropic models spin $S$, vorticity
$V$, and their magnitudes $\|S\|$, $\|V\|$ are constants of
motion. Moreover, $\|S\|$ and $\|V\|$ have vanishing Poisson
brackets with all quantities $q^{a}$, $p_{a}$, $M^{a}{}_{b}$,
$N^{a}{}_{b}$. Therefore, on the level of these variables, all
Hamiltonian systems (\ref{6.6}) with the same doubly isotropic
potential $V$ and with three affine models of the kinetic energy
(\ref{6.2}), (\ref{6.3}), (\ref{6.4}) are identical. In
particular, the solutions for variables $q^{a}$, $p_{a}$,
$M^{a}{}_{b}$, $N^{a}{}_{b}$ coincide with those for the
affine-affine model (\ref{6.2}). And this applies, in particular,
to the geodetic model (when $V=0$) and to the geodetic shear model
with extra imposed dilatations stabilized by $V_{\rm dil}(q)$,
where $q=(q^{1}+\cdots+q^{n})/n$. And then, as mentioned, the
argument about one-parameter subgroups and their cosets decides
about the existence of open subsets of bounded and non-bounded
trajectories. The only difference between various
$\mathcal{T}$-models appears only on the level of $L$, $R$-degrees
of freedom. But the compactness of the corresponding configuration
spaces F$(V,g)$, F$(U,\eta)$ implies that this part of motion does
not influence the property of the total orbits in $Q=$ LI$(U,V)$
to be bounded or non-bounded. One should stress that for the
affine-metrical and metrical-affine geodetic models (\ref{6.3}),
(\ref{6.4}) only exceptional solutions are given by one-parameter
subgroups and their cosets (relative equilibria). Nevertheless,
extracting from all possible one-parameter subgroups and their
cosets their $(q^{a},p_{a},M^{a}{}_{b},N^{a}{}_{b})$-content, we
obtain true statements concerning all three geodetic models
(\ref{6.2}), (\ref{6.3}), (\ref{6.4}).

Our affine geodetic models (\ref{6.2}), (\ref{6.3}), (\ref{6.4})
have a nice binary structure with an additional degree of freedom
related to the motion of the centre $q$ of deformation invariants
$q^{a}$, $a=\overline{1,n}$. In practical applications this term
in $\mathcal{T}_{\rm int}$ should be stabilized by some extra
introduced dilatational potential. If we perturb geodetic models
by admitting more general doubly-isotropic potentials, then it
follows from the mentioned structure of $\mathcal{T}_{\rm int}$
that the most natural and computationally effective potentials
will be those somehow adapted to the above splitting into shear
and dilatation parts, i.e., \[ V(q^{1},\ldots,q^{n})=V_{\rm
dil}(q)+\frac{1}{2}\sum_{i,j}V_{\rm sh}{}^{ij} (|q^{i}-q^{j}|).
\]
Here the additional shear part is not only binary but, just as it
should be, it is depending only on the relative positions of
deformation invariants $|q^{i}-q^{j}|$ on $\mathbb{R}$ or on the
circle U$(1)$ when the group U$(n)$ is used. Obviously, the model
of $V_{\rm sh}=(1/2)\sum_{i,j}V_{\rm sh}{}^{ij} (|q^{i}-q^{j}|)$
will be computationally effective only when the structure of
functions $V_{\rm sh}{}^{ij}$ will have something to do with ${\rm
sh}\left([q^{i}-q^{j}]/2\right)$, ${\rm
ch}\left([q^{i}-q^{j}]/2\right)$,
$\sin\left([q^{i}-q^{j}]/2\right)$,
$\cos\left([q^{i}-q^{j}]/2\right)$.

\section*{Acknowledgements}

The paper is the first in the series which is thought on as a
draft of some future monograph. It contains results obtained
during our work within the framework of the research project
8T07A04720 of the Committee of Scientific Research (KBN)
"Mechanical Systems with Internal Degrees of Freedom in Manifolds
with Nontrivial Geometry". Also results obtained within the KBN
Supervisor Programme 5T07A04824 "Group-Theoretical Models of
Internal Degrees of Freedom" are included here. Authors are
greatly indebted to KBN for the financial support.

Included are also some results obtained or inspired during the
stay of one of us (Jan Jerzy S\l awianowski) in Pisa in 2001, at
the Istituto Nazionale di Alta Matematica "Francesco Severi",
Universit\`{a} di Pisa. It is a pleasure to express the deep
gratitude to Professor Gianfranco Capriz for his hospitality,
discussions, and inspiration. They contributed in an essential way
to the results presented here. The same concerns discussions with
Professor Carmine Trimarco, Dipartimento di Matematica Applicata
"U. Dini", Universit\`{a} di Pisa. Also our "three-body
interactions" ("three-soul interactions"), i.e., common
discussions in our small group (Gianfranco Capriz, Carmine
Trimarco, Jan Jerzy S\l awianowski) contributed remarkably to
results presented in this treatise. One of us (Jan Jerzy S\l
awianowski) is deeply indebted to Istituto Nazionale di Alta
Matematica "Francesco Severi", Gruppo Nazionale per la Fizica
Matematica Firenze, for the fellowship which made this
collaboration possible.

Discussions with Professor Paolo Maria Mariano during his stay at
the Institute of Fundamental Technological Research of Polish
Academy of Sciences in Warsaw also influenced this paper and are
cordially acknowledged.

\end{document}